\documentclass[10pt,twocolumn,journal,compsoc]{IEEEtran}

\usepackage{fancyhdr, epsfig, epsf, amsthm, amsmath, amssymb, amsfonts, subfigure,color}
\usepackage{threeparttable}
\usepackage[noadjust]{cite}
\usepackage{dsfont}
\usepackage{enumerate}
\usepackage{comment}
\usepackage{multirow,booktabs}
\usepackage{bigstrut}
\usepackage[most]{tcolorbox}
\tcbuselibrary{breakable} 
\usepackage{enumitem}
\usepackage{url}

\def\E{\mathsf{E}}

\def\l{\left}
\def\r{\right}
\def\({\left(}
\def\){\right)}

\def\[{\left[}
\def\]{\right]}

\newcommand{\nn}{\nonumber}

\newcommand{\set}[1]{\mathcal{#1}}
\newcommand{\sample}{Q}
\newcommand{\sampleSet}{\set{\sample}}
\newcommand{\gpdCombined}{\vect{d}}
\newcommand{\gpdScale}{\sigma}
\newcommand{\gpdShape}{\xi}

\newcommand{\gpdML}{G^{\gpdCombined}}
\newcommand{\vue}{u}
\newcommand{\VUE}{U}
\newcommand{\vueSet}{\set{\VUE}}
\newcommand{\setSize}[1]{|#1|}
\newcommand{\loglikelihood}{f^{\gpdCombined}}
\newcommand{\sampleSizeRatio}{\kappa_{\vue}}
\newcommand{\grad}[2]{\nabla_{#1}#2}
\newcommand{\loglikelihoodSP}[1]{f^{#1}}
\newcommand{\localmodel}{( \grad{\gpdCombined}{\loglikelihoodSP{\gpdCombined_{\vue}}_{\vue}}, \gpdCombined_{\vue}, \setSize{\sampleSet_{\vue}} )}
\newcommand{\globalmodel}{( \grad{\gpdCombined}{\loglikelihoodSP{\gpdCombined}}, \gpdCombined, \sum_{\vue} \setSize{\sampleSet_{\vue}} )}

\newcommand{\queue}{q}
\newcommand{\queueTH}{\queue_0}

\newcommand{\optimal}{^\star}

\newcommand{\vect}[1]{\boldsymbol{#1}}

\makeatletter
\newtcolorbox[auto counter]{bluebox}[2][]{%
	enhanced, 
	breakable,
	colback=white,
	colframe=blue!30!black,
	attach boxed title to top left={yshift=-2pt}, title={#2},
	boxed title size=standard,
	boxrule=0pt,
	left=3pt,right=3pt,top=3pt,bottom=3pt,
	boxed title style={%
		sharp corners, 
		rounded corners=northwest, 
		colback=tcbcol@frame, 
		boxrule=0pt, size = small},
	sharp corners=north,
	overlay unbroken={%
		\path[fill=tcbcol@back] 
		([xshift=-2pt]title.south east) 
		to[out=0, in=180] ([xshift=1.5cm]title.east)--
		(title.east-|frame.east) |- 
		([xshift=-2pt]title.south east)--cycle;
		\path[fill=tcbcol@frame] (title.south east) 
		to[out=0, in=180] ([xshift=1.5cm]title.east)--
		(title.east-|frame.east)
		[rounded corners=\kvtcb@arc] |- 
		(title.north-|frame.north) 
		[sharp corners] -| (title.south east);
		\draw[line width=.2mm, rounded corners=\kvtcb@arc, 
		tcbcol@frame] 
		(title.north west) rectangle 
		(frame.south east);
	}, 
	overlay first={%
		\path[fill=tcbcol@back] 
		([xshift=2pt]title.south east) 
		to[out=0, in=180] ([xshift=1.5cm]title.east)--
		(title.east-|frame.east) |- 
		([xshift=2pt]title.south east)--cycle;
		\path[fill=tcbcol@frame] (title.south east) 
		to[out=0, in=180] ([xshift=1.5cm]title.east)--
		(title.east-|frame.east)
		[rounded corners=\kvtcb@arc] |- 
		(title.north-|frame.north) 
		[sharp corners] -| (title.south east);
		\draw[line width=.5pt, rounded corners=\kvtcb@arc, 
		tcbcol@frame] 
		(frame.south west) |- (title.north) -| 
		(frame.south east);
	}, 
	overlay middle={%
		\draw[line width=.5pt, tcbcol@frame] 
		(frame.north west)--(frame.south west) 
		(frame.north east)--(frame.south east);
	}, 
	overlay last={%
		\draw[line width=.5pt, rounded corners=\kvtcb@arc, 
		tcbcol@frame] 
		(frame.north west) |- (frame.south) -|
		(frame.north east);
	}, 
	#1
}
\makeatother

\definecolor{myyellow}{RGB}{226, 139, 39}
\makeatletter
\newtcolorbox[auto counter]{yellowbox}[2][]{%
	enhanced, 
	breakable,
	colback=white,
	colframe=myyellow,
	attach boxed title to top left={yshift=-2pt}, title={#2},
	boxed title size=standard,
	boxrule=0pt,
	left=3pt,right=3pt,top=3pt,bottom=3pt,
	boxed title style={%
		sharp corners, 
		rounded corners=northwest, 
		colback=tcbcol@frame, 
		boxrule=0pt, size = small},
	sharp corners=north,
	overlay unbroken={%
		\path[fill=tcbcol@back] 
		([xshift=-2pt]title.south east) 
		to[out=0, in=180] ([xshift=1.5cm]title.east)--
		(title.east-|frame.east) |- 
		([xshift=-2pt]title.south east)--cycle;
		\path[fill=tcbcol@frame] (title.south east) 
		to[out=0, in=180] ([xshift=1.5cm]title.east)--
		(title.east-|frame.east)
		[rounded corners=\kvtcb@arc] |- 
		(title.north-|frame.north) 
		[sharp corners] -| (title.south east);
		\draw[line width=.2mm, rounded corners=\kvtcb@arc, 
		tcbcol@frame] 
		(title.north west) rectangle 
		(frame.south east);
	}, 
	overlay first={%
		\path[fill=tcbcol@back] 
		([xshift=2pt]title.south east) 
		to[out=0, in=180] ([xshift=1.5cm]title.east)--
		(title.east-|frame.east) |- 
		([xshift=2pt]title.south east)--cycle;
		\path[fill=tcbcol@frame] (title.south east) 
		to[out=0, in=180] ([xshift=1.5cm]title.east)--
		(title.east-|frame.east)
		[rounded corners=\kvtcb@arc] |- 
		(title.north-|frame.north) 
		[sharp corners] -| (title.south east);
		\draw[line width=.5pt, rounded corners=\kvtcb@arc, 
		tcbcol@frame] 
		(frame.south west) |- (title.north) -| 
		(frame.south east);
	}, 
	overlay middle={%
		\draw[line width=.5pt, tcbcol@frame] 
		(frame.north west)--(frame.south west) 
		(frame.north east)--(frame.south east);
	}, 
	overlay last={%
		\draw[line width=.5pt, rounded corners=\kvtcb@arc, 
		tcbcol@frame] 
		(frame.north west) |- (frame.south) -|
		(frame.north east);
	}, 
	#1
}
\makeatother

\def\papertitle{Wireless Network Intelligence at the Edge}

\IEEEoverridecommandlockouts

\begin{document}
\title{ \fontsize{24}{28}\selectfont  \papertitle}

\author{Jihong~Park, Sumudu~Samarakoon, Mehdi~Bennis, and $^\dagger$M\'{e}rouane Debbah
 \thanks{J.~Park, S.~Samarakoon, and M.~Bennis are with the Centre for Wireless Communications, University of Oulu, Oulu 90014, Finland (email: \{jihong.park, sumudu.samarakoon, mehdi.bennis\}@oulu.fi). }
\thanks{$^\dagger$M.~Debbah is with CentraleSup\'{e}lec, Universit\'{e} Paris-Saclay, 91190 Gif-sur-Yvette, France (email: merouane.debbah@centralesupelec.fr).}
}

\maketitle \thispagestyle{empty}

\begin{abstract} 
Fueled by the availability of more data and computing power, recent breakthroughs in cloud-based machine learning (ML) have transformed every aspect of our lives from face recognition and medical diagnosis to natural language processing. However, classical ML exerts severe demands in terms of energy, memory and computing resources, limiting their adoption for resource constrained edge devices. The new breed of intelligent devices and high-stake applications (drones, augmented/virtual reality, autonomous systems, etc.), requires a novel paradigm change calling for distributed, low-latency and reliable ML at the wireless network edge (referred to as \emph{edge ML}). In \emph{edge ML}, training data is unevenly distributed over a large number of edge nodes, which have access to a tiny fraction of the data. Moreover training and inference are carried out collectively over wireless links, where edge devices communicate and exchange their learned models (not their private data). In a first of its kind, this article explores key building blocks of edge ML, different neural network architectural splits and their inherent tradeoffs, as well as theoretical and technical enablers stemming from a wide range of mathematical disciplines. Finally, several case studies pertaining to various high-stake applications are presented demonstrating the effectiveness of edge ML in unlocking the full potential of 5G and beyond.
\end{abstract}

\section{Significance and Motivation}\label{sec:significance}

This research  endeavor sits at the confluence of two transformational technologies, namely the fifth generation of wireless communication systems, known as 5G~\cite{tech:3gppTR38801}, and machine learning (ML) or artificial intelligence. On the one hand while the evolutionary part of 5G, enhanced mobile broadband (eMBB), focusing mainly on millimeter-wave transmissions has made significant progress~\cite{ITU5G:15}, fundamentals of ultra-reliable and low-latency communication (URLLC)~\cite{NGMNKPI:16,PetarURLLC:17}, one of the major tenets of the 5G revolution, are yet to be fully understood. In essence, URLLC warrants a departure from average-based system design towards a clean-slate design centered on tail, risk, and scale~\cite{MehdiURLLC:18}. While risk is encountered when dealing with decision making under uncertainty, scale is driven by the sheer amount of devices, antennas, sensors, and actuators, all of which pose unprecedented challenges in network design, optimization, and scalability.

On the other hand in just a few years, breakthroughs in ML and particularly deep learning have transformed every aspect of our lives from face recognition~\cite{aiman:2017:face,wan:2017:face}, medical diagnosis~\cite{maity:2017:diag,ker:2018:diag}, and natural language processing~\cite{lakhanpal:2015:lang,yang:2017:lang}. 
This progress has been fueled mainly by the availability of more data and more computing power. 
However, the current premise in classical ML is based on a single node in a centralized and remote data center with full access to a global dataset and a massive amount of storage and computing power, sifting through this data for inference. 
Nevertheless the advent of a new breed of intelligent devices and high-stake applications ranging from drones to augmented/virtual reality (AR/VR) applications, and self-driving vehicles, makes cloud-based ML inadequate. 
These applications are real-time, cannot afford latency, and must operate under high reliability, even when network connectivity is lost. 

Indeed, an autonomous vehicle that needs to apply its brakes, cannot allow even a millisecond of latency that might result from cloud processing, requiring split second decisions for safe operation~\cite{Lin:2018:auto,mohammad:2018:aoi}. 
A user enjoying visuo-haptic perceptions requires not only minimal individual perception delays but also minimal delay variance, to avoid motion sickness~\cite{ParkGC:18,ABIQualcommVR:17}. 
A remotely controlled drone or a robotic assembler in a smart factory should  always be operational even when  network connection is temporarily unavailable~\cite{kagawa:2017:uav,Mozaffari:2016:UAV2,fotouhi:2018:uav}, by sensing and reacting rapidly to local (and possibly hazardous) environments. 

These new applications have sparked a huge interest in distributed, low-latency and reliable ML calling for a major departure from cloud-based and centralized training and inference toward a novel system design coined \emph{edge ML}, in which: 
(i) training data is unevenly distributed over a large number of edge devices, such as network base stations (BSs) and/or mobile devices including phones, cameras, vehicles, and drones; and 
(ii) every edge device has access to a tiny fraction of the data and training and inference are carried out collectively. Moreover, edge devices communicate and exchange their locally trained models (e.g., neural networks (NNs)), instead of  exchanging their private data. 

\begin{figure*}\centering
	\includegraphics[width=.9\textwidth]{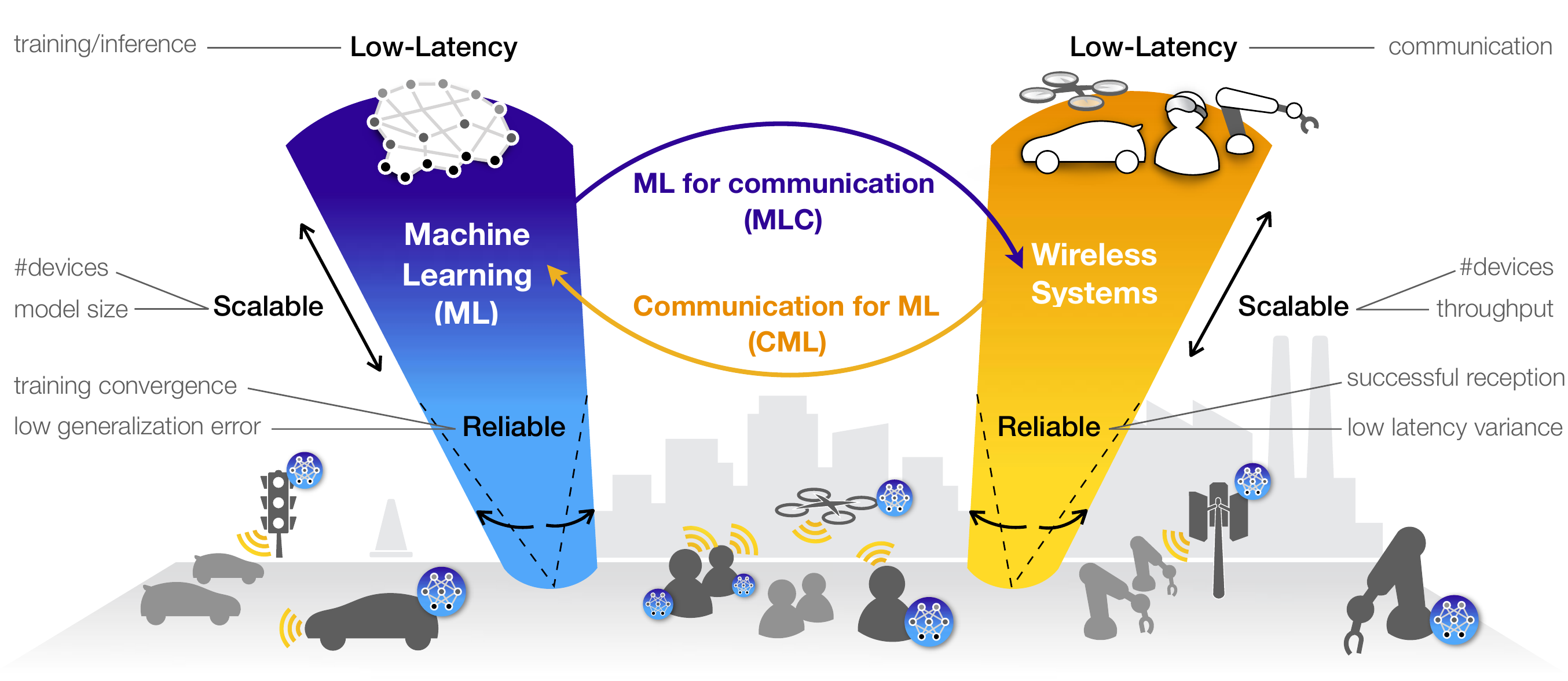}
	\caption{An illustration of \emph{edge ML} where both ML inference and training processes are pushed down into the network edge (bottom), highlighting two research directions: (1)~\emph{ML for communication} (\emph{MLC}, from left to right) and (2) \emph{communication for ML} (\emph{CML}, from right to left).}
	\label{fig:edge_ML}
\end{figure*}

There are clear advantages using edge ML: (i) performing inference locally on connected devices reduces latency and cost of sending device-generated data to the cloud for prediction; (ii) rather than sending all data to the cloud for performing ML inference, inference is run directly on the device, and data is sent to the cloud only when additional processing is required; (iii) getting inference results with very low latency is important in making mission-critical Internet of things (IoT) applications respond quickly to local events; (iv) unlike cloud-based ML, edge ML is privacy-preserving in which training data is not logged at the cloud, but is kept locally on every device, and the global shared model is learned by aggregating locally computed updates, denoted as \emph{model state information (MSI)}, in a peer-to-peer manner or via a coordinating (federating) server; and (v) higher inference accuracy can be achieved by training with a wealth of user-generated data samples that may even include privacy sensitive information, such as healthcare records, factory/network operational status, and personal location history.

Edge ML is a nascent research field whose system design is entangled with communication and on-device resource  constraints (i.e., energy, memory and computing power). In fact, the size of an NN and its energy consumption may exceed the memory size and battery level of a device, hampering decentralized inference. Moreover, the process of decentralized training involves a large number of devices that are interconnected over wireless links, hindering the training convergence due to the stale MSI exchange under poor wireless channel conditions. As such, enabling ML at the network edge introduces novel fundamental research problems in terms of jointly optimizing training, communication, and control under end-to-end (E2E) latency, reliability,  privacy, as well as  devices' hardware requirements. As depicted in Fig.~\ref{fig:edge_ML}, these research questions can be explored through the following two research directions.

\begin{figure*}\centering
\includegraphics[width=.98\textwidth]{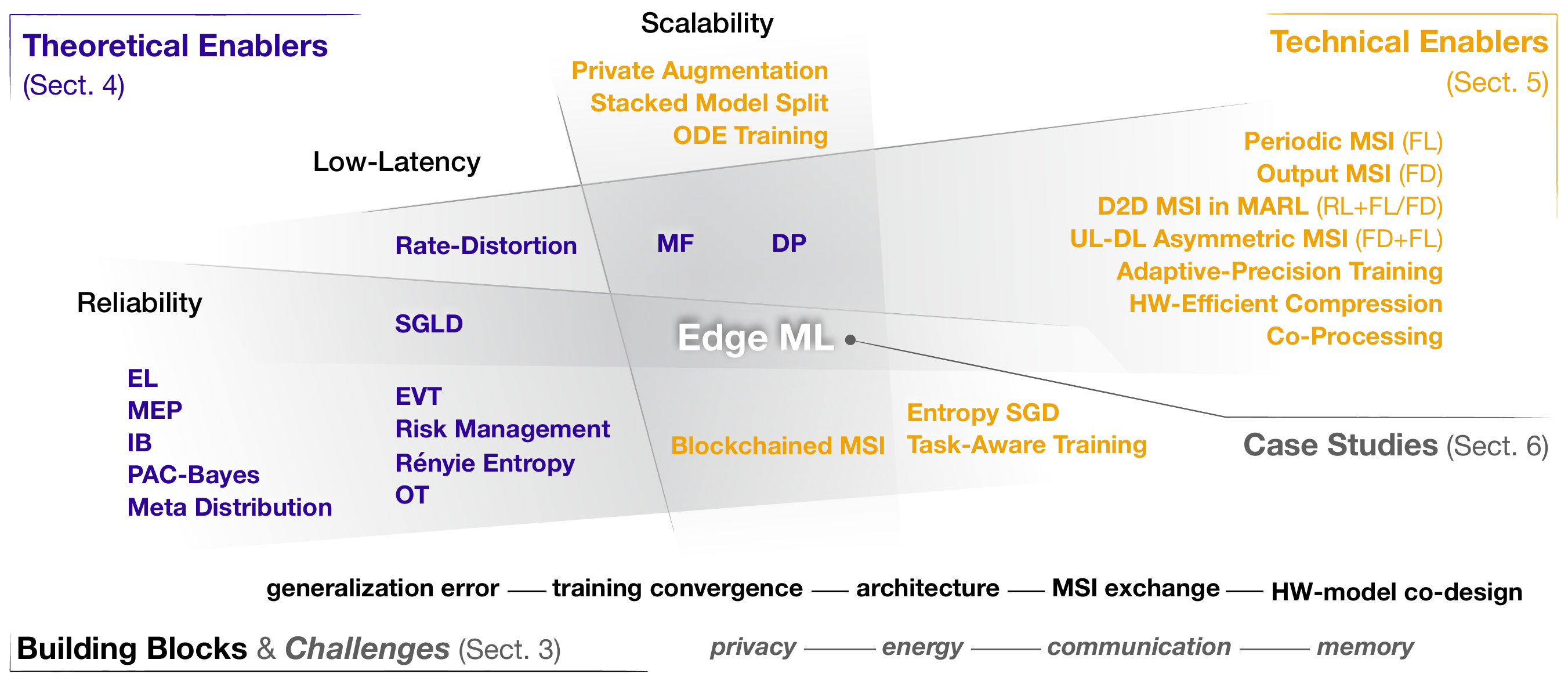}
\caption{ Overview of the key building blocks and theoretical/technical enablers of edge ML.}
\label{fig:overciew_enablers}
\end{figure*}

\vspace{5pt}
\noindent 1) \textbf{ML for Communications (MLC)}\quad 
Exploiting edge ML for improving communication, on the one hand, epitomizes the research direction of MLC. The recent groundswell  interest in ML-aided (and mostly data-driven)  wireless system design  fits into this direction. At its core, MLC leverages a large amount of data samples (e.g., radio signals) to acquire  an accurate knowledge of the RF environment to, for instance, optimize modulation coding schemes (MCS).
Towards this vision, at the physical layer an end-to-end communication framework was studied under unknown channels,  using an autoencoder (AE)~\cite{Dorner:18,Balevi2018OneBit}, recurrent neural network (RNN) \cite{Farsad2018RNN}, and generative adversarial network (GAN) \cite{Ye2018ChannelAE}. To overcome millimeter-wave's link sensitivity to blockage~\cite{Mattia:18}, a GAN-aided long-term channel estimation~\cite{Li2018GAN} and a reinforcement learning (RL) based beam alignment technique~\cite{wang2018RL} were proposed. 
%
At the network layer, an RNN-aided caching solution was proposed in~\cite{Chen2017MachineLF}, and an unsupervised clustering algorithm was used with real user traffic patterns. Basics of ML, its NN architectures and  communication designs were recently overviewed in~\cite{simeone2018brief,Chen2017MachineLF}.

While interesting, these works however focus on centralized ML  whose goal is to improve the communication performance assuming a well-trained ML model is available, and a large number of data samples,  overlooking  the additional latency induced by the prior training process and the posterior inference latency. In sharp contrast to these approaches, the latency and reliability of edge ML have to be examined with respect not only to communication but also to decentralized ML training and inference processes, calling for novel analytical methods based on studying tail distributions, a novel communication and ML co-design, and uncertainty/risk assessment.
\vspace{5pt}

\noindent2) \textbf{Communication for ML (CML)}\quad 
As alluded to, training ML at the network edge over wireless networks while taking into account latency and reliability opens up a novel research direction. In this respect, edge ML architectures and their operations should be optimized by accounting for communication overhead and channel dynamics, while coping with several problems such as straggling devices in the training process and generalization to unmodeled phenomena under limited local training data. Not only that, all these aspects need to factor in \emph{on-device constraints} including energy,  memory, and compute, not to mention privacy guarantees.

An interesting example of edge ML training architecture is \emph{federated learning (FL)} \cite{Brendan17,jnl:jakub16} wherein mobile devices periodically exchange their NN weights and gradients during local training. FL has been shown to improve communication efficiency by    MSI quantization~\cite{pap:jakub16}, adjusting the MSI update period~\cite{Brendan17}, and optimizing  devices' scheduling policy~\cite{FL_Nishio}. 
These methods are still in their infancy and need to address a myriad of fundamental challenges including the ML-communication co-design, while accounting for on-device constraints and wireless channel characteristics.

From a theoretical  standpoint, the overarching goal of this article is to explore building blocks, principles, and techniques focusing on CML. As on-device processing becomes more powerful, and ML grows more prevalent, the confluence of these two research directions will be instrumental in spearheading the vision of truly intelligent next-generation communication systems, 6G~\cite{UO:6G}.
\vspace{10pt}

\noindent\textbf{Scope and Organization}\quad Enabling ML at the network edge  hinges on investigating several fundamental questions, some of which are briefly summarized next.
\vspace{5pt}

\begin{tcolorbox}[colback=white, colframe=blue!30!black, boxrule=.5pt]
\textbf{Q1}. How do edge devices train a high-quality centralized model in a decentralized manner, under communication/on-device resource constraints and different NN architectures?
\end{tcolorbox}\vskip -3pt

Classical ML has been based on the precept of a single central entity having full access to the global dataset over ultra-fast wired connections: 
e.g., a local inter-chipset controller manipulating multiple graphics processing units (GPUs) through PCI Express intra-computer connections (supporting up to 256 Gbps) \cite{NvidiaComp}; 
InfiniBand inter-computer links (up to 100 Gbps) \cite{Infiniband}; 
or a cloud server commanding multiple computing devices via Ethernet communication (up to 25 Gbps) \cite{PandaKeyNote}. 
Using a deep NN, the central controller sifts through this global data for training and inference by exploiting the massive amount of storage and computing power: 
e.g., 11.5 Petaflops processing power supported by 256 tensor processing units (TPUs) and 4 TB high bandwidth memory~(HBM) \cite{TPUv2pod}, which is sufficient for operating the Inception V4 NN model consuming 44.3~GB~\cite{Wang:2018:inception}.

These figures are in stark contrast with the capability of devices under edge ML. 
While 5G peak rates achieve 20~Gbps \cite{tech:3gppTR38801} that is comparable only with Ethernet connections, the instantaneous rate may frequently fluctuate due to poor wireless channel conditions, hindering inter-device communication. 
Moreover, the computing power of mobile devices is a million times less powerful (e.g., Qualcomm Snapdragon 845's 16.6 Gflops~\cite{Snapdragon845}). 
Their memory size is also ten times smaller (Apple iPhone XS Max's 4 GB~\cite{Spec:iPhoneXS}) than a deep NN model size. 
Besides, the energy consumption (e.g., Google Pixel 3 XL's 2.15 W~\cite{Spec:Pixel3}), delimited by the battery capacity, is half million times smaller than a powerful centralized ML architecture (e.g., AlphaGo's 1~MW~\cite{Ceva:AlphaGo}).
Since computation and communication consume battery power, the energy consumption of edge ML operations should be flexibly optimized over time, which is not feasible under the classical ML architecture, wherein a NN model size is fixed.

Last but certainly not the least, the privacy guarantee of each device is crucial in edge ML, particularly when on-device datasets are associated with privacy sensitive information. Perturbing the information exchange can ensure privacy, which may be at odds with the goal of high accuracy and reliability. Adding redundant information can  provide  a solution, at the cost of extra communication latency. With these communication and on-device hardware/privacy requirements in mind, we explore and propose decentralized architectures and training algorithms that are suitable for edge ML. To this end, we first describe the key building blocks of edge ML (Sect. \ref{sec:building_blocks}), and then introduce suitable technical enablers (Sect. \ref{sec:tech_enablers}), as illustrated in Fig.~\ref{fig:overciew_enablers}.

\vspace{10pt}

\begin{figure*}
\subfigure[CML: Communication reliability $\rightarrow$ ML training convergence.]{
	\includegraphics[width=\columnwidth]{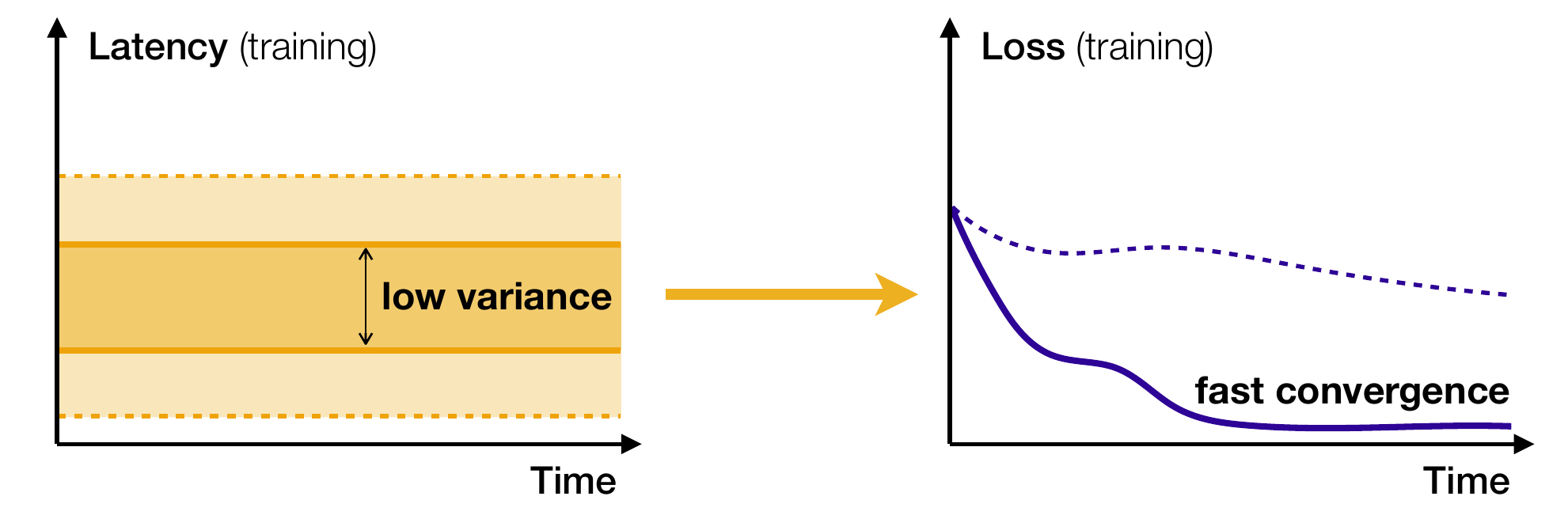}
	\label{fig:CML}
}
\subfigure[MLC: ML generalization error $\rightarrow$ communication reliability.]{
	\includegraphics[width=\columnwidth]{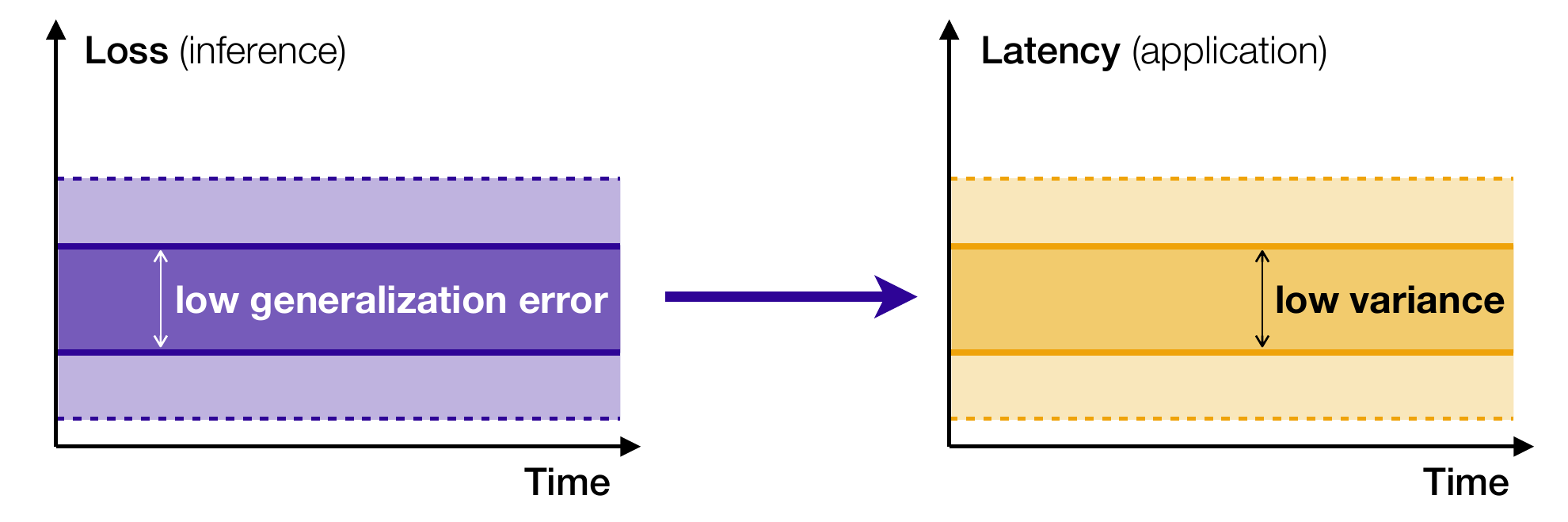}
	\label{fig:MLC}
}
\caption{Two examples illustrating CML and MLC: (a) Training latency and variance over wireless connections impact the ML training convergence; and (b) ML inference loss and variance affect the latency of a wireless system as its application.}
\end{figure*}

\begin{tcolorbox}[colback=white, colframe=blue!30!black, boxrule=.5pt]
\textbf{Q2}. How to enable reliable edge ML as opposed to best-effort cloud-based ML, subject to non-convex loss functions and unevenly dispersed and unseen data samples?

\end{tcolorbox}\vskip -3pt
A requisite for reliable edge intelligence is high inference accuracy, i.e., low loss, under not only the training dataset but also unseen data samples. Therefore, average inference accuracy is insufficient, and its credibility interval for a given training dataset ought to be considered. Credibility can be measured by the loss difference under training and the entire samples, referred to as a \emph{generalization error}, in which the credibility interval is an achievable target generalization error that can be decreased by utilizing more training samples and a proper NN model architecture. Calculating the generalization error is relatively easy in centralized ML where the central controller feeds independent and identically distributed (IID) training data samples into each device. By contrast, it becomes more challenging in edge ML where the training data samples may become non-IID across devices.

Next, decentralized training dynamics in edge ML become more complicated even under a simple gradient descent algorithm. In fact, it is difficult to characterize the convergence behavior of a decentralized training process, especially under non-IID training dataset, as well as the limited communication/computation resource budget  fluctuating over time. The situation is aggravated when a single NN model is split and shared by multiple devices due to the limited on-device memory size. 

Besides, most training algorithms intentionally insert noise, i.e., regularizers,  to cope with non-convex loss functions. However, it is challenging to optimize the regularization in edge ML, which is intertwined with wireless communication and privacy-preserving methods generating noise. To tackle these difficulties, we address Q2 by revisiting the fundamental principles of ML (Sect.~\ref{sec:building_blocks}), followed by key theoretical enablers for edge ML (Sect.~\ref{sec:theory_enablers}).
\vspace{10pt}

\begin{tcolorbox}[colback=white, colframe=blue!30!black, boxrule=.5pt]
\textbf{Q3}. How do the theoretical and technical  enablers of edge ML impact E2E latency, reliability, and scalability throughout the  training and inference processes,  under  both CML and MLC frameworks?
\end{tcolorbox}\vskip -3pt

From the standpoint of MLC, stringent URLLC applications can be empowered using edge ML, whereas in CML edge ML is enhanced via wireless connectivity under on-device constraints. In this respect, edge ML design not only enhances CML but also MLC, calling for optimizing E2E latency, reliability, and scalability.

Specifically, the worst-case E2E latency of a reference device is given by `$\text{training}+\text{inference}+\text{application}$' latency. MLC can reduce the application delays that are proportional to wireless communication latency, given as 'payload/(bandwidth$\times$spectral efficiency).' For instance, this is viable by improving spectral efficiency via enabling real-time joint source-channel coding \cite{Farasad:ICASSP18}, channel-agnostic end-to-end communications \cite{Dorner:18}, and interference management \cite{Sidiropoulos:18}. The effective amount of bandwidth can also be increased using low-complexity dynamic spectrum access \cite{Cohen:GC17} and proactive resource management \cite{ChallitaSaad:18}. Furthermore, the payload size can be minimized by exchanging semantic information rather than raw data \cite{PetarOsvaldo:19}.

Achieving these benefits in MLC entails extra training and inference delays of ML operations. Here, training latency is captured by a loss or weight convergence delay~\cite{Wang:18}, whereas inference latency refers to computation and memory access delays \cite{Zhang:Energy}. Since both training and inference processes are performed at the network edge, their computation and communication need to be jointly optimized subject to  limited energy, memory, computing power, and radio resource constraints. To further reduce  latency, as done in \cite{Liu:2018:MEC}, training process can partly be offloaded to other proximal edge nodes with higher computational power.

E2E reliability of a reference device is the probability that the E2E latency does not exceed a target latency deadline, and is determined by the reliability of both training and inference. For training reliability, the staleness of exchanged MSIs is the key bottleneck, as it disrupts the training convergence, and may  lead to an unbounded training latency,  emphasizing the importance of communication techniques to synchronize edge nodes. For inference reliability, high inference accuracy is mandatory for reducing the application latency, and needs to be ensured under both training and unseen samples, necessitating suitable mathematical tools to quantify the generalization error.

Through E2E reliability, CML and MLC are intertwined. For CML, the higher communication latency reliability during training, i.e., low communication latency variance, the higher training convergence guarantee by avoiding training devices lagging behind, known as stragglers. As illustrated in Fig.~\ref{fig:CML}, lower latency variance decreases the inference loss, thereby reducing the application (i.e., communication) latency and increasing E2E reliability. In the opposite direction, for MLC, a lower generalization error yields  less fluctuations in the application latency, as visualized in Fig.~\ref{fig:MLC}, thereby achieving higher E2E reliability.

Finally, E2E scalability is specified by the number of supported devices, NN model size, and communication throughput to ensure a target E2E reliability. In this respect, the major obstacles are on-device constraints, whereby the number of federating devices is limited by preserving their privacy. The range of federation is also determined by memory and NN model sizes, as FL requires an identical model size for all federating devices. Moreover, the federation range should take into account tasks' correlations and channel conditions, so as not to exchange redundant MSIs and to avoid straggling devices under limited wireless capacity, respectively. 

With these challenges and E2E performance definitions, to address Q3, we revisit the state-of-the-art literature in URLLC and ML (Sect.~\ref{sec:state_of_the_art}), and provide several case studies that showcase the essence of both MLC and CML frameworks (Sect.~\ref{sec:case_studies}), followed by concluding remarks (Sect.~\ref{sec:conclusion}).


\section{State-of-the-Art}\label{sec:state_of_the_art}

\subsection{From Vanilla 5G Towards URLLC Compound}

Since its inception, 5G requirements have been targeting three generic and distinct services: eMBB, mMTC, and URLLC \cite{tech:3gppTR22891,tech:3gppTR38802,tech:ITUR-M2410}. 
The prime concern of eMBB is to maximize spectral efficiency (SE) by providing higher capacity for both indoor and outdoor highly dense areas, 
enhancing seamless connectivity for everyone and everywhere,
 supporting high mobility including cars, trains, and planes.
Beyond  voice and data services, eMBB focuses on immersive AR/VR applications with high definition $360^{\circ}$ video streaming for entertainment and navigation.
Enabling communication among large number of devices is the focus of mMTC.
The success therein relies on coverage, low power consumption, longer life span of devices, and cost efficiency.
The goal of URLLC is to ensure reliability and latency guarantees. 
Therein, mission critical applications including autonomous driving, remote surgery, and factory automation require outage probabilities of the range from $10^{-5}$ to $10^{-9}$.

That said, recent field tests have shown that  packet sizes of less than 200 Bytes were supported for a single device moving within 1 km~\cite{Docomo:18}. This demonstrates the fundamental challenge in  achieving (even moderate) URLLC requirements, e.g., 99.999\% decoding success rate with 1 ms latency~\cite{3GPPRel14:16}.
On the other hand, new  emerging use cases advocate a \emph{mixture} of URLLC with eMBB and/or mMTC via slicing, in which a pool of shared resources (spectrum, computing power, and memory)  is used~\cite{jnl:pocovi18,jnl:esswie18,pap:wu17,pap:lien17,jnl:zhao18,jnl:gaun18,jnl:popovski18}.

The success behind fine-grained and dynamic network slicing hinges on the ability to identify and distinguish different services within the network. Unfortunately, such traditional slicing methods are unfit for the upcoming high-stake applications that may not only demand more stringent requirements but also give rise to a \emph{compound} of URLLC with eMBB and/or mMTC, where a single device simultaneously requests several services out of eMBB, mMTC, and URLLC. 
For instance, multi-player mobile AR/VR gaming applications demand URLLC for achieving low motion-to-photon latency (MTP) and eMBB for rendering 360 degree video frames.
Likewise, autonomous vehicles rely on sharing high-resolution real-time maps with low latency, which is an exemplary use case of ultra-High Speed Low Latency Communications (uHSLLC) \cite{tech:3gppTS22261}. 
In a similar vein,  mission-critical  factory automation applications  may require sensing, reasoning, and perception-based modalities (e.g., haptic), going beyond classification tasks ~\cite{ZHONG:2017:industry},  necessitating both eMBB and URLLC.

	Beyond communication, automated or remotely controlled vehicles, drones, and factories demand ultra-reliable and low-latency control.
	A swarm of remotely controlled drones affected by sudden gust of wind may need to report and receive  control commands within a split of a second. Likewise,
	a remote surgery robot operating a critical patient requires a high definition video signal upload while receiving real-time commands of high precision
	are two such examples.
	The reliability of the control feedback loop of the aforementioned control systems directly affects the system performance. 
	Both  lost or outdated information and commands can yield undesirable and chaotic system behaviors in which, the communication links between  controllers and devices play a pivotal role.
	In this regard, the research in URLLC and control has recently emerged focusing on 	coordination \cite{Liu:2017:platoon,mozaffari:2018:uav,YANMAZ:2018:uav},
	robustness \cite{mahar:2018:uav,Joelianto:2011:robust,Vu:2018:RSRL},
	and sensing \cite{akyol:2014:sense,LI:2014:sense}.

Existing orthogonal and non-orthogonal slicing approaches are ill-suited for supporting these compounded URLLC services. 
Compared to a non-compounded scenario, orthogonal slicing under compounded URLLC consumes the resource amount multiplied by the number of compounded links per device. 
Due to the use of multiple links by a single device, non-orthogonal slicing  induces severe multi-service self-interference \cite{ParkGC:18}, negating the effectiveness of slicing. 
These limitations call for the aid from a new dimension, namely edge ML as elaborated in the next section.

\subsection{From Centralized ML Towards Edge ML}

\subsubsection{Types of ML}

The training process of ML is categorized into supervised, unsupervised, and reinforcement learning as follows. 
\vspace{5pt}

\noindent1) \textbf{Supervised Learning}\quad 
By feeding an input data sample, the goal of supervised learning is to predict a target quantity, e.g., regression, or classification of the category within pre-defined labels. 
This ability can be obtained by optimizing the NN parameters by feeding training data samples, referred to as a training process. 
In supervised learning, the input training samples are paired with the ground-truth output training samples. 
These output samples `supervise' the NN to infer the correct outputs for the actual input samples after the training process completes. 
\vspace{5pt}

\noindent2) \textbf{Unsupervised Learning}\quad 
The training process of unsupervised learning is performed using only the input training samples. 
In contrast to supervised learning, unsupervised learning has no target to predict, yet aims at inferring a model that may have generated the training samples. 
Clustering of un-grouped data samples and generating new data samples by learning the true data distribution, i.e., a generative model, belong to this category.
\vspace{5pt}
\begin{figure*}\centering
	
\subfigure[Classical Q-learning.]{
	\includegraphics[width=.32\textwidth]{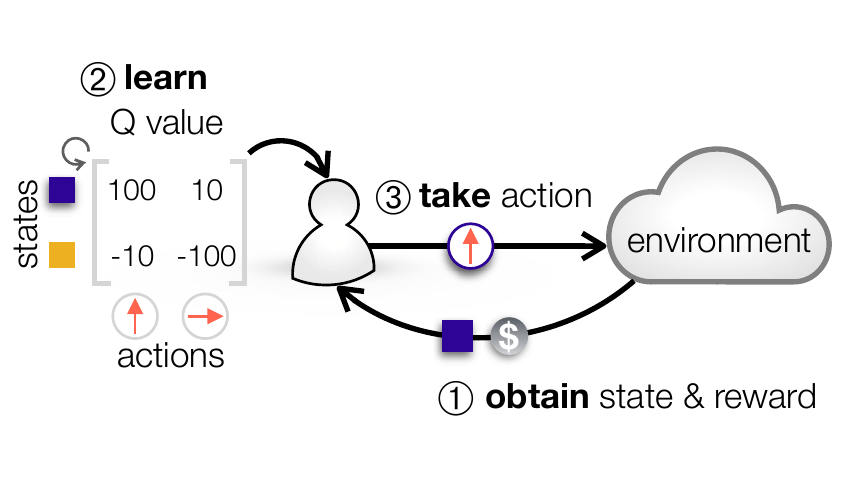}
	\label{fig:RL_Q}
}%
\subfigure[Deep Q-learning.]{
	\includegraphics[width=.32\textwidth]{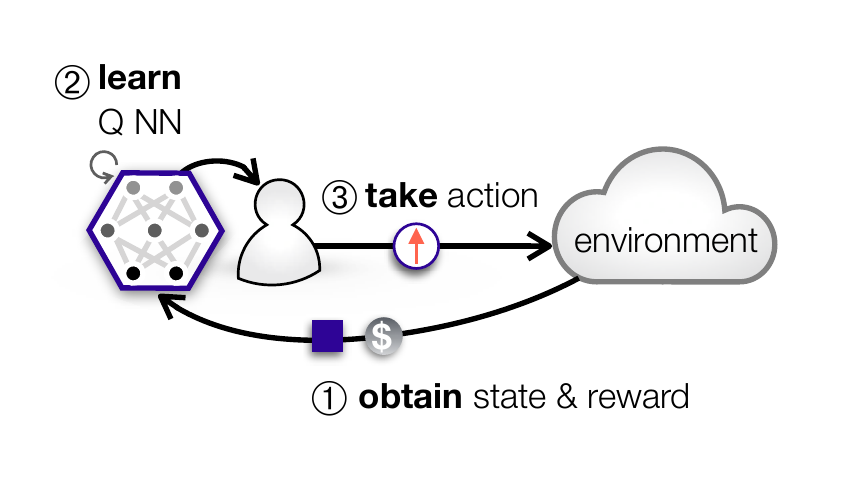}
	\label{fig:RL_DQN}
}%
\subfigure[Actor-critic RL.]{
	\includegraphics[width=.32\textwidth]{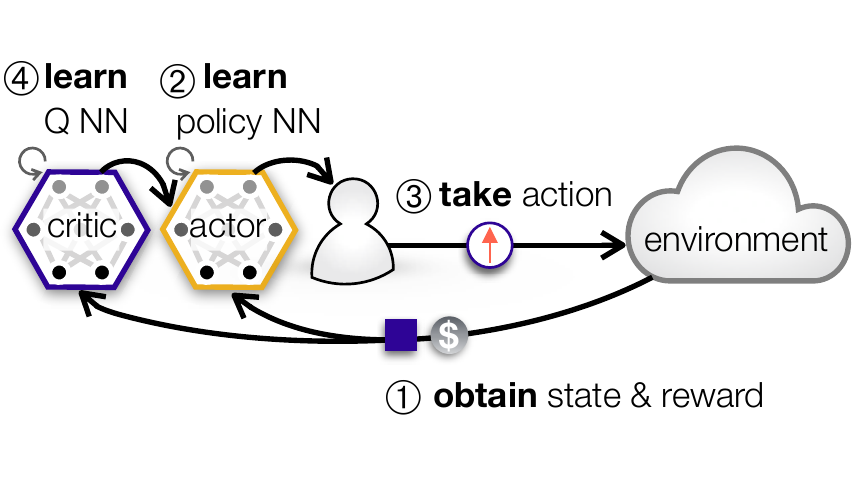}
	\label{fig:RL_AC}
}%

\caption{Examples of RL: (a) classical Q-learning without any NN; (b) deep Q-learning with a NN, and (c) actor-critic RL with actor and critic NNs.}%
\label{fig:RL}
\end{figure*}

\noindent3) \textbf{Reinforcement Learning (RL)}\quad 
The goal of RL is to make an agent in an environment take an optimal action at a given current state, where the interaction between the agent's action and state through the environment is modeled as a Markov decision process (MDP). 
When each action is associated with a return, the agent takes an action that maximizes its predicted cumulative return, e.g., Q-learning that maximizes the Q value for each state, as illustrated in Fig.~\ref{fig:RL_Q}. 
In Q-learning, the larger state dimension, the more computation. This problem is resolved by deep Q-learning as shown in Fig.~\ref{fig:RL_DQN}, where a NN approximates the Q function and produces the Q values by feeding a state. 
These value-based RL can take actions only through Q values that are not necessarily required. 
Instead, one can directly learn a policy that maps each state into the optimal action, which is known as policy-based RL whose variance may become too large~\cite{Sutton:1999:PGM}. 
Actor-critic RL is a viable solution to both problems, comprising a NN that trains a policy (actor NN) and another NN that evaluates the corresponding Q value (critic NN), as visualized in Fig.~\ref{fig:RL_AC}.
\vspace{5pt}

For the sake of convenience, unless otherwise specified, we hereafter describe ML from the perspective of supervised learning, most of which can also be applied to unsupervised and RL algorithms with minor modifications.

\subsubsection{Types of NN Architectures}

\begin{figure*}\centering
	\subfigure[MLP.]{
		\includegraphics[width=.155\textwidth]{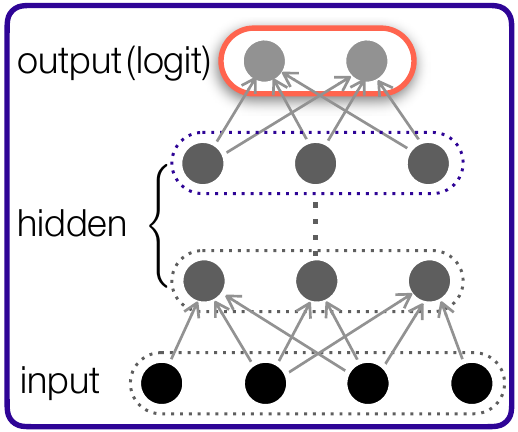}
		\label{fig:NN_arch_MLP}
	}%
	\subfigure[RNN.]{
		\includegraphics[width=.155\textwidth]{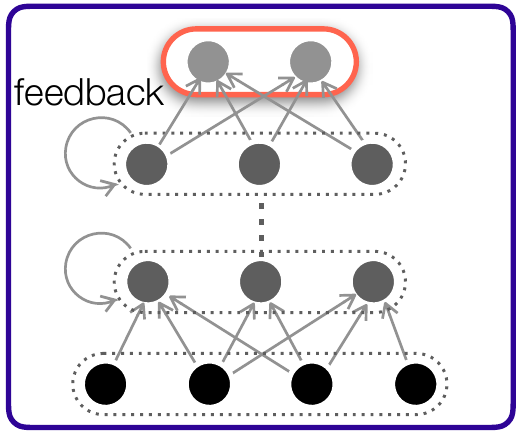}
		\label{fig:NN_arch_RNN}
	}%
	\subfigure[CNN.]{
		\includegraphics[width=.155\textwidth]{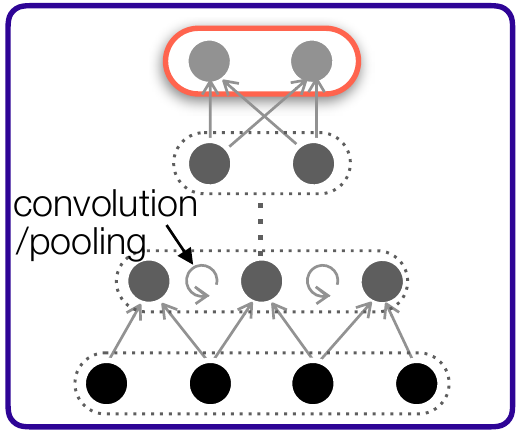}
		\label{fig:NN_arch_CNN}
	}%
	\subfigure[DBN.]{
		\includegraphics[width=.155\textwidth]{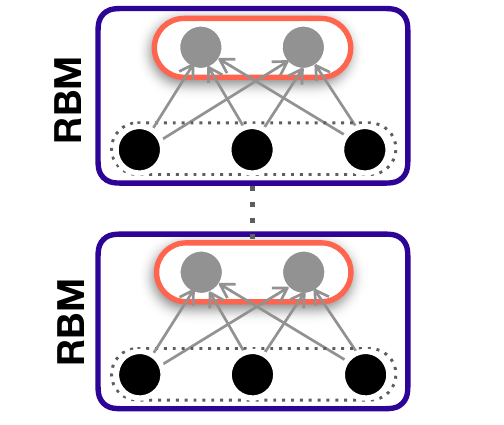}
		\label{fig:NN_arch_DBN}
	}%
	\subfigure[AE.]{
		\includegraphics[width=.155\textwidth]{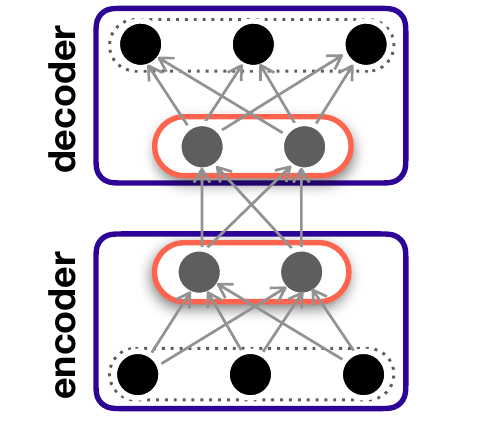}
		\label{fig:NN_arch_autoencoder}
	}%
	\subfigure[GAN.]{
		\includegraphics[width=.155\textwidth]{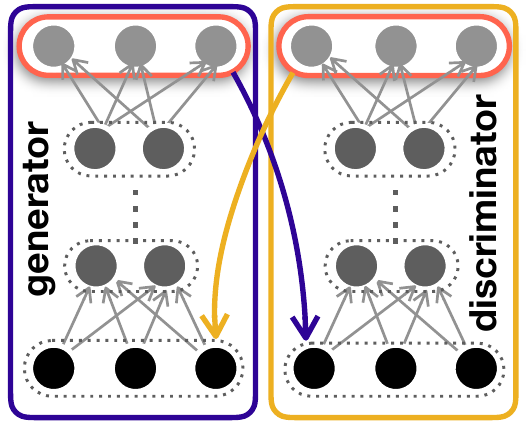}
		\label{fig:NN_arch_GAN}
	}%
	\caption{Types of NN architectures: (a) multilayer perceptron (MLP); (b) recurrent neural network (RNN); (c) convolutional neural network (CNN); (d) deep belief network (DBN) comprising a stack of pre-trained restricted Boltzmann machines (RBMs); (e) autoencoder; and (f) generative adversarial network (GAN).}%
	\label{fig:NN_architecture}
\end{figure*}

Fig.~\ref{fig:NN_architecture} visualizes an NN consisting of multiple layers.
The input layer accepts input data samples, and the output layer produces the training/inference outcomes. 
These two layers are connected through at least a single hidden layer. 
The total number of hidden layers is called the depth of the NN. An NN with a large depth is referred to as a deep NN (deep NN), otherwise, a shallow NN. 

Each of the layers comprises perceptrons connected to each other, and such connections are associated with weights, respectively. 
At a single perceptron, all the inputs are thus weighted, and aggregated, followed by the output value after passing through a non-linear activation function, e.g., sigmoid function, softmax function, or rectified linear unit (ReLU). 
The output is fed to the next layer's perceptrons until reaching the output layer. 
\vspace{5pt}

\noindent1) \textbf{Multilayer Perceptron (MLP)}\quad
If the output of each layer is fed forward to the next layer, then the NN is called a feedforward NN (FNN). 
The default (i.e., \emph{vanilla}) baseline of the FNN is an MLP. 
As visualized in Fig.~\ref{fig:NN_arch_MLP}, each perceptron's output is passed directly to the next layer's perceptrons, without any recursion and/or computation other than the activation function.
Even with this simple structure, an MLP is capable of distinguishing data that is not linearly separable, as long as the number of perceptrons (i.e., NN model size) is sufficiently large. 
Its theoretical backgrounds are elaborated in Sect.~\ref{subsec:BB_arch_split}.
Training an MLP is performed using a gradient decent optimization algorithm, called the backpropagation method \cite{werbos1975beyond}.
\vspace{5pt}

\noindent2) \textbf{Recurrent NN (RNN)}\quad 
If the output of a layer  encounters recursions within the same layer, the corresponding NN is referred to as a recurrent NN (RNN). 
As shown in Fig.~\ref{fig:NN_arch_RNN}, the vanilla RNN is a form of MLP with feedback loops at hidden layers. 
For training a RNN, the feedback loops can be unrolled in a sequential way, such that the hidden layer prior to a feedback loop is chained to the hidden layer posterior to the loop. 
This chained structure easily allows a sequence of inputs, which is plausible for instance in natural language processing (NLP). 
Vanilla RNN struggles with the vanishing gradients problem as the feedback loop iterates, and thus cannot capture a feature with long-term correlations. 
To cope with this, the vanilla RNN's hidden layer can be replaced with a long short-term memory (LSTM) unit \cite{Hochreiter:1997:LSM}. 
The LSTM unit introduces a memory cell that can store the current hidden layer's values in the memory, which is controlled by several gates that determine whether to store or forget. Similar operations can also be implemented using gates, yielding a gated recurrent unit (GRU) \cite{cho:2014:gru}.
\vspace{5pt}

\noindent3) \textbf{Convolutional NN (CNN)}\quad 
Processing image data samples through an MLP may induce a large number of perceptrons, as each image pixel needs to be associated with a single perceptron in the input layer. 
As a variant of MLPs, a CNN resolves this problem by inserting two pre-processing layers, i.e., convolutional and pooling layers, in-between the hidden layers as illustrated in Fig.~\ref{fig:NN_arch_CNN}. 
Inspired by human visual stimuli, at a convolutional layer, the input information is processed using a convolution operation, thereby extracting the features while compressing the information. 
Next, at the pooling layer, the previous layer's outputs are combined into a single perceptron in the next layer, by selecting their maximum value or taking their average value \cite{Ciresan:2011:CNN}. 
Lastly, the compressed feature information is fed back to a conventional MLP structure to obtain the final output, of which the corresponding stages are called  fully connected layers.
\vspace{5pt}

\noindent4) \textbf{Deep Belief Network (DBN)}\quad 
A notorious problem of training a deep NN is  the vanishing gradients as the depth increases. 
To resolve this, a DBN exploits a divide and conquer method: first pre-train each small part of the network, i.e., restricted Boltzmann machine (RBM), then combine all pre-trained parts, followed by fine-tuning the entire network. 
As a result, a DBN comprises a stack of pre-trained RBMs, as shown in Fig.~\ref{fig:NN_arch_DBN}. 
Each RBM has a single hidden layer and a visible layer that accepts inputs or produces outputs. 
The connections between these two layers are bidirectional, in which an RBM is not an FNN. 
Nonetheless, after the pre-training process, the pre-trained RBMs in a DBN are stacked in a way that each RBM's hidden layer connects no longer to its visible layer but to the next RBM's visible layer. 
Thus, a DBN is an FNN. 
DBN can easily be extended to a CNN by partitioning each hidden layer into several groups and applying a convolutional operation to each group \cite{Lee:2009:CDB}.
\vspace{5pt}

\noindent5) \textbf{Autoencoder (AE)}\quad
As illustrated in Fig.~\ref{fig:NN_arch_autoencoder}, an AE is a stack of two FNNs, copying the input to its output in an unsupervised learning way.
The first NN learns representative features of the input data, known as encoding. 
The second NN receives the feature as the input, and reproduces the approximation of the original input as the final output, referred to as decoding. 
The key focus of an AE is to learn useful features of the input data en route for copying the input data. 
In this respect, reproducing the output exactly the same as the original data may become too accurate to capture the latent features. 
A viable solution for mitigating this is to constrain the feature space to have a smaller dimension than the original input space, yielding an undercomplete AE whose encoding NN compresses the original input data into a short code~\cite{GoodfellowBook:16}. 
Decoding the NN then uncompresses the code so as to approximately reproduce the original data. 
These encoding and decoding NNs resemble a transmitter and its receiver in communication systems. 
Due to such an analogy, AEs have recently been used for data-driven E2E communication designs~\cite{Dorner:18,Balevi2018OneBit}, after inserting a set of extra layers that emulate the transmission power normalization and the channel propagation between the encoding and decoding NNs.
\vspace{5pt}

\noindent6) \textbf{Generative Adversarial Network (GAN)}\quad
As a generative model in the class of unsupervised learning, the goal of GAN is to generate new data samples given by the estimated distribution of the input data samples. 
This is achieved by training two NNs, referred to as a generator and a discriminator as depicted in Fig.~\ref{fig:NN_arch_GAN}, as if they play a zero-sum game. 
Here, the generator produces fake samples to fool the discriminator, while the discriminator tries to identify the fake samples. 
As the game reaches a Nash equilibrium, i.e., training completion, the generator becomes capable of producing fake-yet-realistic samples that are indistinguishable from the real samples. 
Mathematically, playing this game is identical to minimizing the Jensen-Shannon (JS) divergence \cite{book:manning:1999} between the real and  generated sample distributions. 
When these two distributions are disjoint, JS divergence goes to infinity, yielding the gradient vanishing problem. 
A naive solution is to insert noise so as to make the distributions overlapped, which may degrade the output quality. 
A better solution is to replace JS divergence with a proper loss function that is capable of measuring the distance between disjoint distributions with a finite value. Wasserstein distance satisfies this characteristics, thus prompting the recent success of Wasserstein GAN \cite{wassGAN}. 
More details on the Wasserstein distance are deferred to Sect.~\ref{subsec:T_E_reliability}. 

\subsubsection{Limitations of Centralized ML}

The classical concept of ML focuses mainly on \emph{offline and centralized ML} \cite{Chen2017MachineLF,Farsad2018RNN}. 
In this case, the entire dataset is given a priori, and is used for the training process. 
To obtain accurate and reliable inference, a central controller divides the training dataset into mini-batches, and allocates them to multiple processing devices  e.g., via a message passing interface (MPI) \cite{ben:2018:demy}, running local training operations. 
The central controller iteratively collects and aggregates their local training results, until the training loss converges. 
The said training process is separated from  inference, and hence the training cost and latency are commonly neglected. 

Unfortunately, such a one-time training process is vulnerable to initially un-modeled phenomena, as it is practically impossible to enumerate all preconditions and ensuing consequences, referred to as \emph{qualification} and \emph{ramification} problems, respectively \cite{diett:2017:robust}. 
In fact, the trained model using centralized ML is biased towards the initially fixed training dataset, which fails to capture user-generated and  the time-varying nature of data, yielding less reliable inference results. 
Besides, a user-generated dataset can be privacy sensitive, and the owners may not allow the central controller to directly  access the data.
\emph{Online} and \emph{edge ML} is able to address these problems. 
Indeed, \emph{online decentralized training} can preserve privacy by exchanging not the dataset but the model parameters with (or without) a simple parameter aggregator \cite{jnl:jakub16}, thereby reflecting a huge volume of user-generated data samples in real-time. 
In so doing, the trained models are immediately obtained at the local mobile devices, enabling low-latency \emph{inference}. 
In spite of the rich literature in ML, edge ML with a large number of mobile devices over wireless remains a nascent field of research, motivating us to explore its key building blocks and enablers as elaborated in the next section.
%

\section{Building Blocks and Challenges}\label{sec:building_blocks}

\subsection{Neural Network Training Principles}
The unprecedented success of ML has yet to be entirely demystified. Till today, it is not completely clear how to train an NN so as to achieve high accuracy even for unseen data samples. Nonetheless, recent studies on the asymptotic behaviors of a deep NN training process and on the loss landscape shed some light on providing  guideline principles, presented in the following subsections.

\subsubsection{Asymptotic Training Principles}

\noindent1) \textbf{Universal Approximation Theorem (UAT)}\quad
The theorem states that an ideally trained MLP with infinitely many perceptrons can approximate any kind of non-linear function. This holds irrespective of the number of hidden layers. Thus, both very-wide shallow NNs and very-deep NN can become ideal classifiers. The proof is provided for shallow NNs \cite{Hornik:89,Cybenko:89} and recently for deep NNs \cite{Hanin:17}. This explains that the key benefit of deep NNs compared to shallow NNs is attributed not to the inference accuracy but to its credibility, as detailed next. 
\vspace{5pt}

\noindent2) \textbf{Energy Landscape (EL)}\quad
The success of deep NNs has recently been explained through the lens of EL from statistical physics. In this approach, a fully-connected FNN with $L$ layers is first transformed into a spherical spin-glass model with  $L$ spins, and the FNN's non-convex loss function is thereby approximated as the spin-glass model's non-convex Hamiltonian \cite{Choromanska:15}, i.e., energy of the neural network model. Exploiting this connection, a recent discovery \cite{Becker:18} verifies that the number of Hamiltonian's critical points is a positively-skewed unimodal curve over $L$. For a sufficiently large $L$ in a deep NN, the number of local minima and saddle points thus monotonically decreases with $L$. At the same time, it also verifies that the Hamiltonian's local minima become more clustered in EL as $L$ increases. For these reasons, in a deep NN, any local minimum of a non-convex loss function well approximates the global minimum. This enables to train a deep NN using a simple gradient-descent approach. In other words, a deep NN training can always achieve the highest inference accuracy, so long as sufficiently large amount of data samples are fed for training, emphasizing the importance of sufficient data sample acquisition.

\begin{figure}\centering
\subfigure[Trade-off between \emph{bias} and \emph{variance} whose addition yields a generalization error.]{
	\includegraphics[width=.9\columnwidth]{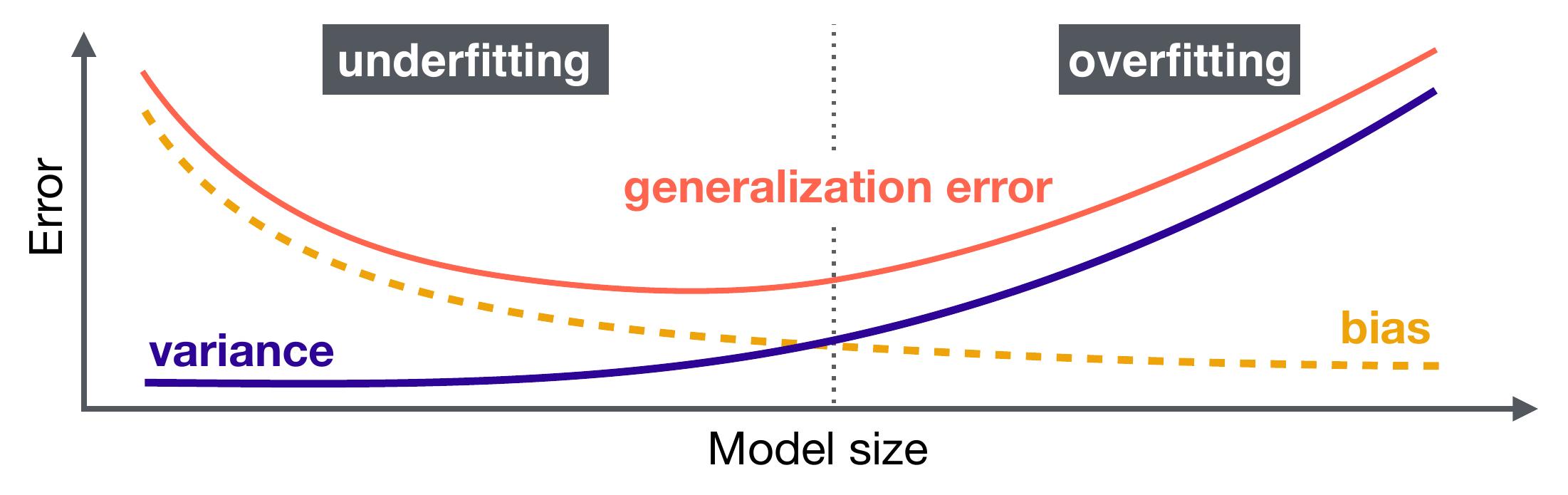}
	\label{fig:bias_var_tradeoff}
}%
\\
\subfigure[Trade-off between \emph{empirical} average and \emph{expected} average whose difference yields a generalization error.]{
	\includegraphics[width=.9\columnwidth]{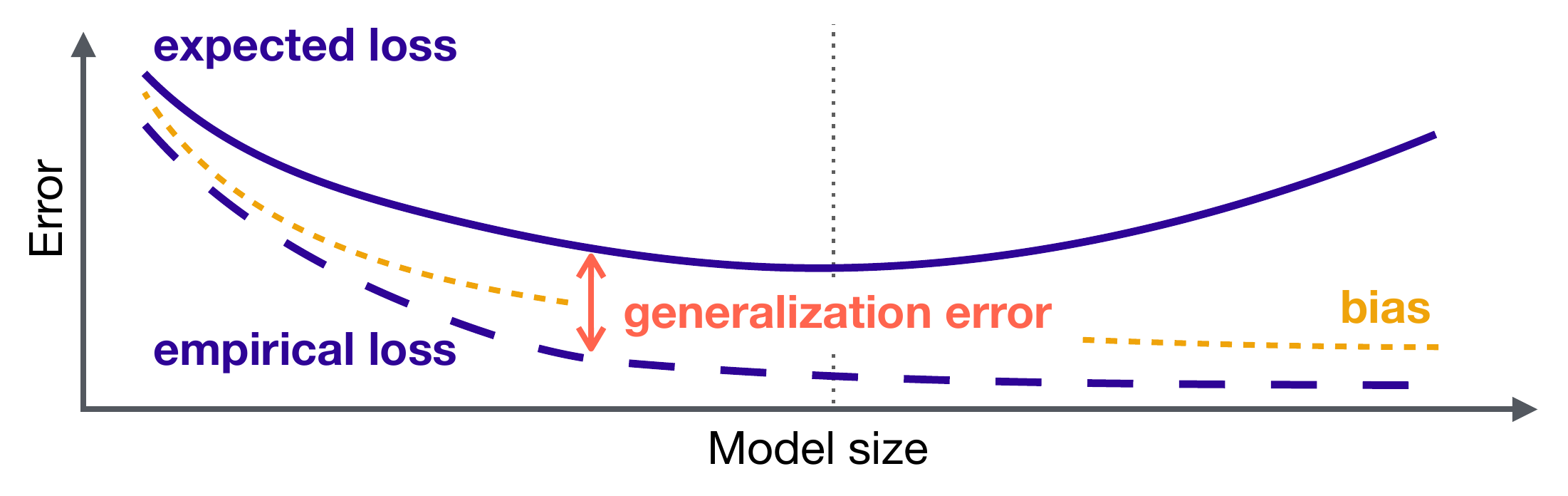}
	\label{fig:avg_risk_tradeoff}
}%
\caption{Illustration of underfitting and overfitting via (a) the bias-variance trade-off and (b) the approximation-generalization trade-off.}
\label{fig:fitting_over_under}
\end{figure}

\subsubsection{Practical Training Principles}\label{subsec:BB_training_prin_pract}

The aforementioned asymptotic characteristics of a deep NN make the non-convex training problem being in favor of simple convex optimization methods. This partly justifies the use of \emph{de-facto} stochastic gradient descent (SGD) algorithms in modern ML applications. Nevertheless, the number of layers is finite in reality, and the resultant mismatch between the non-convex problem and its convex-based training algorithm has to be taken into account in practice, as elaborated below.
\vspace{5pt}

\noindent1) \textbf{Underfitting vs. Overfitting}\quad
The objective of training an NN is two-fold. The first goal is to minimize the training loss, and the other goal is to minimize the loss difference between the values under training data samples and unseen samples, referred to as mitigating underfitting and overfitting, respectively. 
How to achieve both goals is exemplified in the following \emph{bias-variance} trade-off. Consider $f(x)$ is the true function to be estimated using an approximate function $\hat{f}(x)$ for an unseen data sample $x$. 
When the distribution of $x$ has zero mean and variance $\sigma^2$, the mean squared error of $f(x)$ and $\hat{f}(x)$ can be decomposed into bias and variance terms as follows:
\begin{align}
\E\[\(f(x) - \hat{f}(x) \)^2\] = \text{Bias}^2 + \text{Var} + \sigma^2,
\end{align}
where $\text{Bias} = \E[\hat{f}(x)-f(x)]$ and $\text{Var}= \!\E[\hat{f}(x)]^2 -(\E[\hat{f}(x)])^2$. 
For a given training dataset, as the model size grows, Bias decreases whereas Var keeps increasing, as visualized in Fig.~\ref{fig:bias_var_tradeoff}. 
The sum of Bias and Var is denoted as \emph{generalization error}, and both underfitting and overfitting can be mitigated at the minimum generalization error.

Such a behavior can also be characterized through the lens of the \emph{approximation-generalization} trade-off as follows. 
Let $\hat{L}(w)$ denote the empirical average loss under the training dataset. This empirical loss decreases as the NN better \emph{approximates} the features of the training dataset. On the other hand, $L(w)$ represents the expected loss under the entire dataset with a set $w$ of model parameters, which depends not only on the training samples but also on unseen samples, i.e., \emph{generalization}.
As illustrated in Fig.~\ref{fig:avg_risk_tradeoff}, $\hat{L}(w)$ decreases with the model size, while $L(w)$ is convex-shaped. 
In this case, both underfitting and overfitting can be avoided at the minimum $L(w)$, and the gap between $L(w)$ and $\hat{L}(w)$ implies the generalization error. 
It is noted that calculating $L(w)$ requires unseen data samples, which is unavailable in practice. 
Instead, the probability that the generalization error is less than a certain value can be evaluated using statistical learning frameworks, as  detailed in Sect.~\ref{subsec:T_E_reliability}.
\vspace{5pt}

\noindent2) \textbf{Information Bottleneck Principle (IB)}\quad
The bias-variance trade-off is well observed in the NN training process under IB \cite{Tishby,Shwartz-Ziv:17}. 
In this approach, denoting $X$ and $Y$ as a random input and its desired output, respectively, model training is interpreted as adjusting the model configuration $\hat{X}$ so that its predicted output $\hat{Y}$ can be close to $Y$. 
A convincing model training strategy is to minimize the redundant information $I(\hat{X};X)$ for predicting~$Y$, while maximizing the prediction relevant information $I(\hat{X};Y)$. This is recast by solving the following Lagrangian optimization problem:
\begin{align}
\underset{p(\hat{x}\mid x),\; p(y\mid \hat{x}),\;p(\hat{x})}{\min}\;\;  I(X;\hat{X})-\beta I(\hat{X};Y) , \label{Eq:IB}
\end{align}
such that $Y\rightarrow X\rightarrow \hat{X}\rightarrow \hat{Y}$, where $\beta>0$ is a Lagrangian multiplier. 
In the objective function \eqref{Eq:IB}, the first term $I(X;\hat{X})$ decreases with the level of model generalization, while the second term $I(\hat{X};Y)$ increases with the inference performance, thereby capturing the bias-variance trade-off. 
Furthermore, the problem can  be interpreted as passing the information in $X$ about $Y$ through a bottleneck $\hat{X}$ \cite{Huang:16}. 
To illustrate,  following the data processing inequality \cite{Cover:Book} in information theory, the second term $I(\hat{X};Y)$ is upper bounded by $I(X;\hat{X})$ that coincides with the first term $I(X;\hat{X})$. 
Such a conflict in the optimization process incurs an information bottleneck, thereby leading to two model training phases. Namely, during the first phase, both $I(X;\hat{X})$ and $I(\hat{X};Y)$ increase, whereas during the second phase, only $I(\hat{X};Y)$ increases whereas $I(X;\hat{X})$ decreases. 
This unfolds the microscopic model training behaviors: model training tends to first increase the inference performance while relying highly on its own input data samples, and then tries to generalize the model by reducing the impact of its own data samples. 
Nonetheless, the two-phase training dynamics are not always manifested since the original IB formulation is sensitive to the types of hidden layer activations \cite{Saxe:18} and the training objective function \cite{Kolchinsky:18}, which can be partly rectified by modifying the objective function and/or inserting noise \cite{Amjad:18}.
\vspace{5pt}

\begin{figure}\centering
\subfigure[Structure.]{\includegraphics[width=.41\columnwidth]{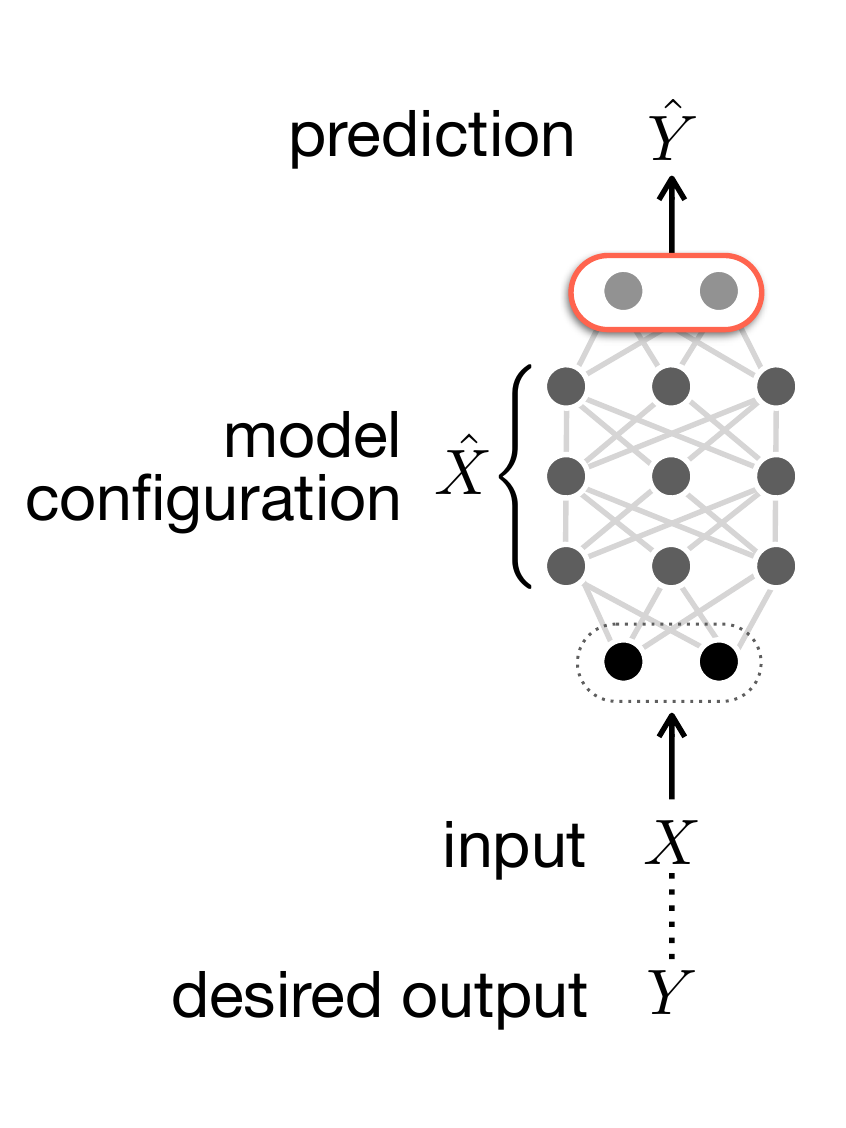}}
\subfigure[Trace of model training.]{\includegraphics[width=.58\columnwidth]{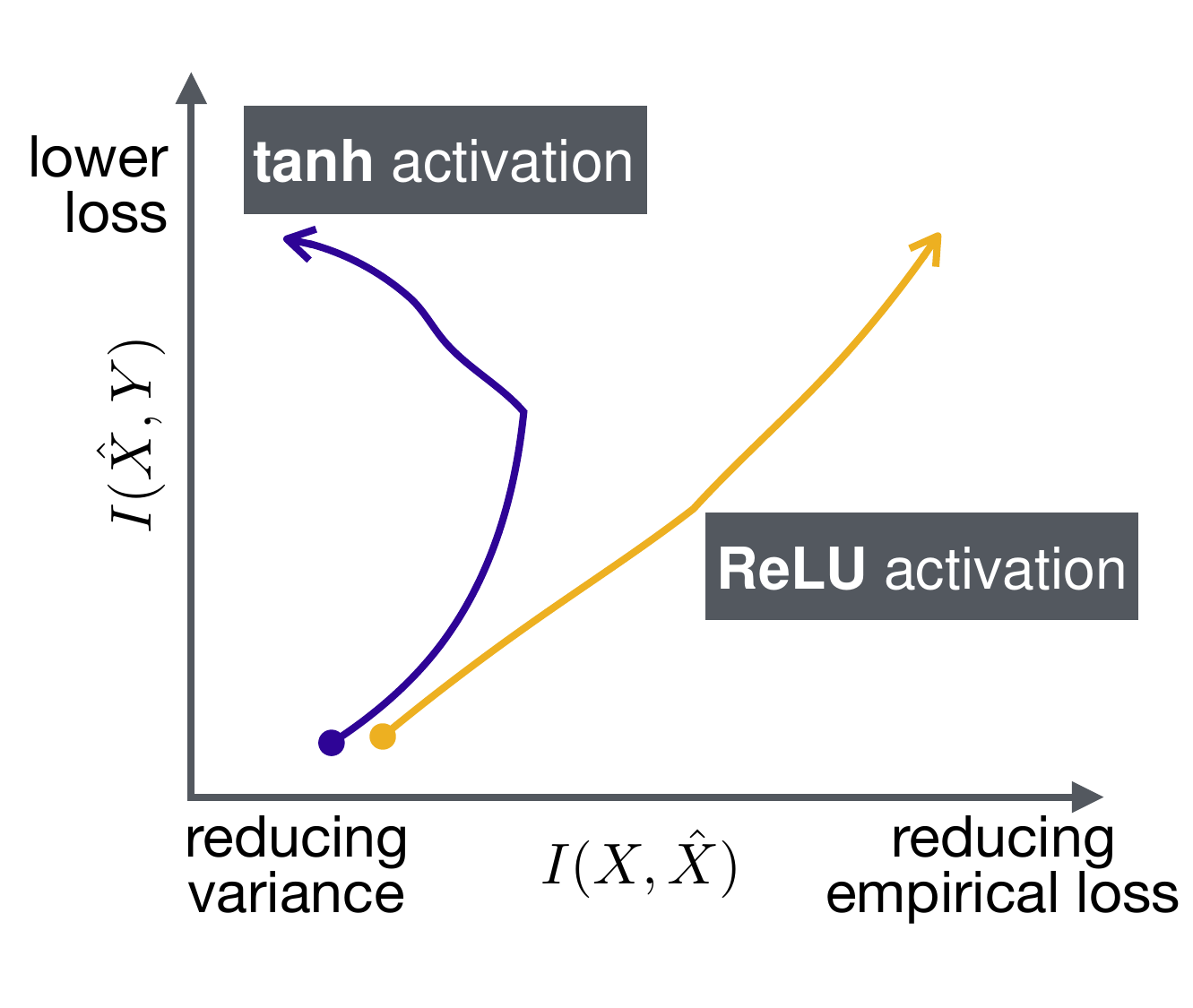}}
\caption{Illustration of (a) model training structure and (b) trace of $\{I(X,\hat{X}),I(\hat{X},Y)\}$ during model training under the information bottleneck principle (IB).}
\end{figure}

\begin{figure}\centering
\includegraphics[width= \columnwidth]{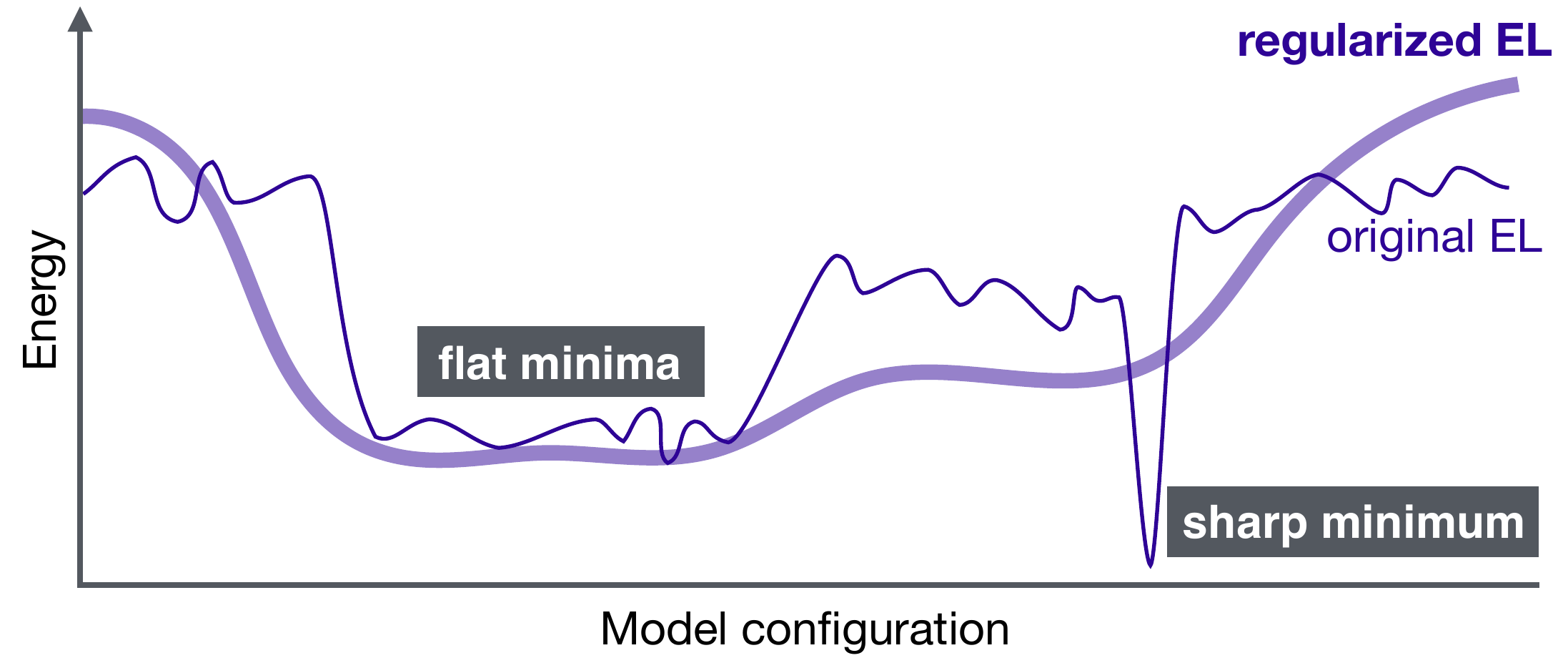}
\caption{Flat vs. sharp minima and the impact of regularization in EL.}
\label{fig:minima_flat_sharp}
\end{figure}

\noindent3) \textbf{Flat Minima}\quad
Although deep NNs asymptotically guarantee easily trainable characteristics \cite{Becker:18}, our practical concern is the deep NN with finite depth $L$ that cannot make all local minima become negligibly separated. 
In the EL shown in Fig.~\ref{fig:minima_flat_sharp}, some of the local minima are clustered in a wide valley, i.e., \emph{flat minima} \cite{Horchreiter:97}, whereas a sharp minimum \cite{Dinh:17} isolated from other minima may have lower energy than the flat minima.
At this point, the well-known problem of generalization is in order. 
Each sharp minimum represents the overfitting case that achieves the minimum energy only at the given data and/or model configuration. 
In order to train a generalized model, it is thus preferable to find a flat minimum at each training iteration \cite{Horchreiter:97,Chaudhari:ICLR17,Wu:17,LiXu:18}. 
\vspace{5pt}

\begin{figure*}\centering
\subfigure[m-d split.]{
	\includegraphics[width=.24\textwidth]{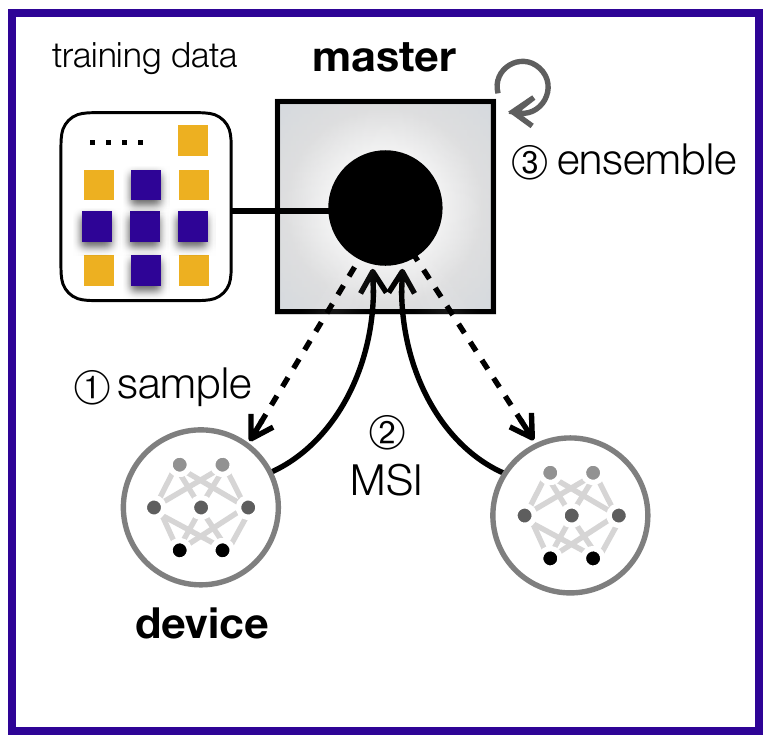}
	\label{fig:split_master_device}
}%
\subfigure[h-d split.]{
	\includegraphics[width=.24\textwidth]{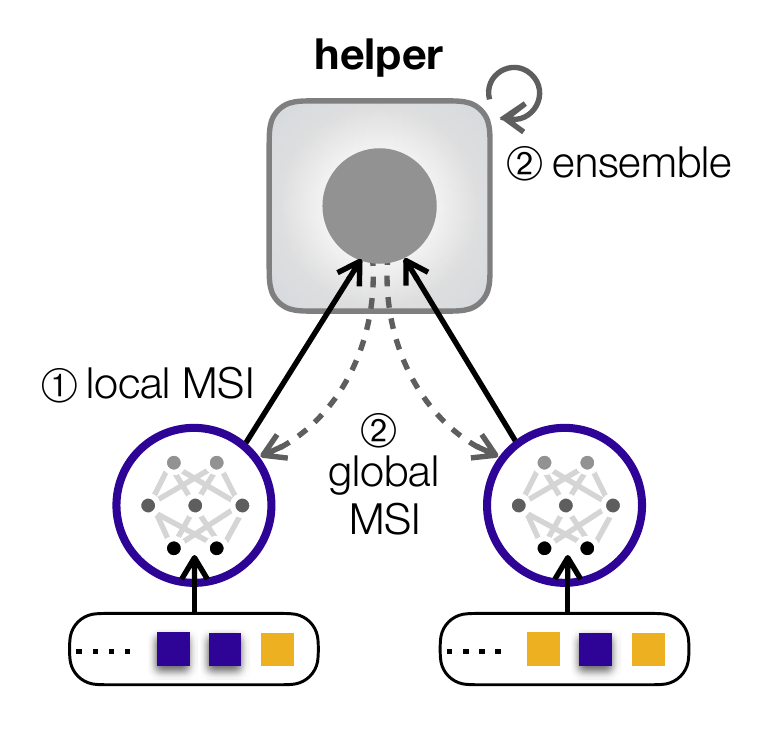}
	\label{fig:split_helper_device}
}%
\subfigure[d-d split.]{
	\includegraphics[width=.24\textwidth]{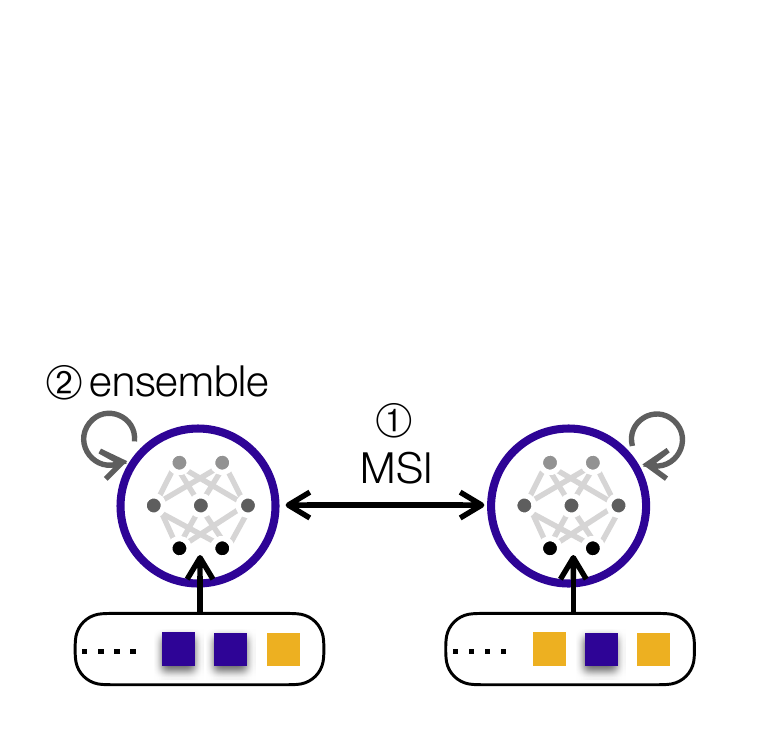}
	\label{fig:split_device_device}
}%
\subfigure[Model split.]{
	\includegraphics[width=.24\textwidth]{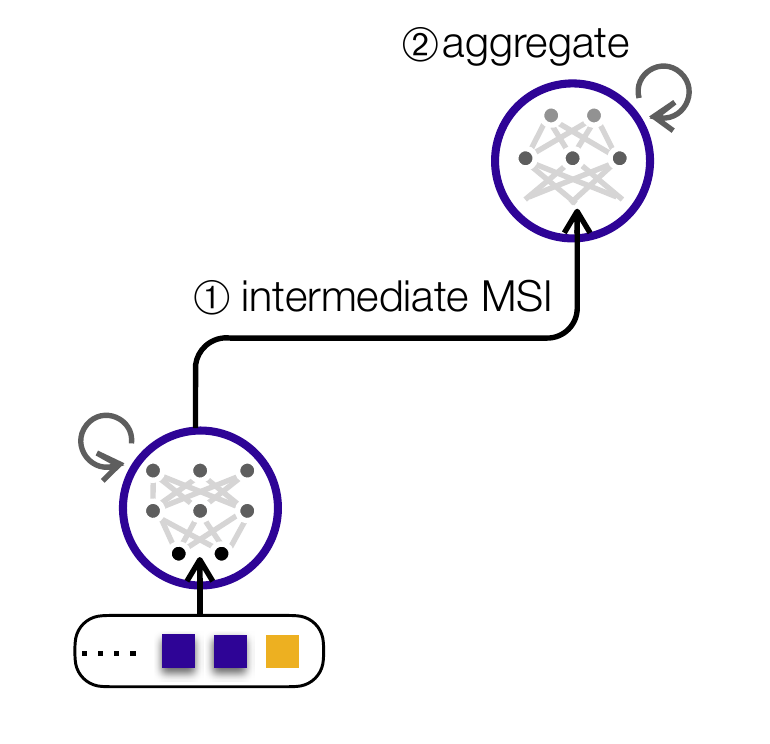}
	\label{fig:split_model}
}%
\caption{Centralized ML with (a) master-device (m-d) split, compared to edge ML with (b) helper-device (h-d) split, (c) device-device (d-d) split, and (d) model split. The master node in centralized ML has full access to the dataset and controls all devices within the blue square, whereas the helper node in edge ML can only exchange MSIs with each device having access to its local data samples.}
\label{fig:arch_split}
\end{figure*}

\noindent3) \textbf{Minimum Energy Path (MEP)}\quad
Finding flat minima by using EL has recently been tackled by a counter example \cite{Dinh:17}. 
Namely, while producing the same final output, it is possible to rescale the model weights, changing the EL. 
In this case, one can imagine a sufficiently generalized NN whose trained weights lie within flat minima. 
Rescaling this original NN can produce a set of modified weights lying not within flat minima, and both observations contradict with each other. 
This calls for a careful observation on the EL. 
On this account, a filter normalization technique \cite{LiXu:18} has been proposed, which heuristically guarantees a tight relationship between the resulting landscape and the final output. 
A recent study \cite{Draxler:18} detours this problem by using an MEP method in statistical physics. 
In this approach, from a given weight to another arbitrary weight, the goal is to find the path that follows saddle points in the energy landscape, implying the path minimizing the maximum energy. 
Such a path does not allow the weight rescaling operation, thereby negating the aforementioned counterexample. 
Furthermore, the result shows that the minimum energy paths of widely-used deep NNs are almost the same as the local minima, uncovering their generalization excellence through the lens of the energy landscape.
\vspace{5pt}

\noindent4) \textbf{Regularization and Ensembling}\quad
In practice, the goal of high inference accuracy with the low generalization error is commonly achieved by heuristically inserting noise during the training phase. 
This smoothens out the EL as observed in Fig.~\ref{fig:minima_flat_sharp}, thereby allowing a gradient descent method to achieve the desired objective. 
To this end, one can add a regularizer term to the non-convex loss function and/or insert noise into the weight update process. 
The latter can be implemented indirectly by randomly sampling the training data samples in SGD or by aggregating and averaging the weight parameters computed by multiple devices, referred to as ensembling or bootstrap aggregating (bagging)~\cite{GoodfellowBook:16}.

\subsection{Distributed Training}
The NN training process occupies the majority of the E2E latency budget, and impacts the inference reliability, while delimiting the overall scalability. Towards achieving these E2E targets, parallelizing the training process is crucial. Ideally, the distributed training process can exploit a larger amount of aggregate computing power with more training data samples. In the following subsections we discuss how to parallelize the architecture and  carry out distributed training.

\subsubsection{Architectural Split}\label{subsec:BB_arch_split}

An NN training process can be split by parallelizing the training data samples to multiple devices that have an identical NN structure, referred to as \emph{data split}. 
Alternatively, when an NN model size is too large, a single NN structure can be split into multiple segments that are distributed over multiple devices, i.e., \emph{model split}.
\vspace{5pt}

\noindent1) \textbf{Data Split}\quad
In centralized ML, a central controller, hereafter denoted as a \emph{master}, owns the entire set of training data samples, and feeds a small portion of the entire dataset, i.e., a minibatch, into each subordinated training device at each round, i.e., an epoch. 
Afterwards, the master collects and aggregates the trained model parameters, as illustrated in Fig.~\ref{fig:split_master_device}. 
This \emph{master-device} (m-d) split corresponds to the case with a cloud server fully controlling its associated devices. 
The parallel computation inside a single device also fits with this split, where a controller handles multiple training processors, e.g., GPUs, inside a single device. 
Such a centralized training operation is ill-suited for online decentralized training in edge ML, where  training data samples are generated and owned by local devices. 
Even in offline learning, some private data samples, e.g., patient records, may not be accessible to the master, obstructing the training process. 

In edge ML, one can instead consider a \emph{helper-device} (h-d) split, where each device first trains the local model using its own data samples, and then exchanges the trained local model parameters with a helper that aggregates the parameters uploaded from multiple devices.
Downloading the aggregated model parameter to each device completes a single epoch, as shown in Fig.~\ref{fig:split_helper_device}. 
A well-known example of this h-d split is the structure with a parameter server that assists the model parameter exchanges~\cite{Dean:12,Zinkevich:10,Synch_sgd,Jin:16}. 
A natural extension of this is a \emph{device-device} (d-d) without any coordinator, as visualized in Fig.~\ref{fig:split_device_device}. 
The communication among devices can be enabled by a push-pull gossip algorithm \cite{Jin:16} or by following a pre-defined communication network topology~\cite{Lian:17,Anis:2019:GC}. 
In the context of edge ML, we henceforth focus on the h-d split and d-d split, to be elaborated in Sect.~\ref{subsec:TEC_E_latency}.
\vspace{5pt}

\noindent2) \textbf{Model Split}\quad
If an NN model size is larger than the device memory, the model has to be split into segments distributed over multiple devices, as illustrated in Fig.~\ref{fig:split_model}. 
In this case, devices need to exchange intermediate model parameters during the forward and backward training operations, requiring sophisticated pipelining of the processes. 
In order to minimize the dependency among the split segments, one can first unroll the original NN model and construct its dataflow graph~\cite{WangHuang:18,Schuiki:18}. 
Then, the processing efficiency is maximized by, for example, grouping and merging common operations in the forward and backward processes, respectively. 
In this article,  we mainly focus on the data split architecture, unless otherwise specified.

\subsubsection{MSI Exchange}\label{subsec:BB_MSI_xchange}

For a given data split, each device first trains its local model, and then exchanges its current \emph{model state information (MSI)}. 
In what follows we exemplify which MSI is exchanged and how to update the local model based on the exchanged MSI in centralized ML with m-d split, followed by their limitation from the edge ML's standpoint.
\vspace{5pt}

\noindent1) \textbf{Centralized Parallel SGD (CSGD)} \quad
This baseline method, also known as full synchronous SGD or mini-batch SGD \cite{Synch_sgd,Zinkevich:10,Jin:16,Dean:12}, exchanges the gradients of the training loss function. To elaborate, every device uploads its \emph{local MSI} to the master, and subsequently downloads the master's \emph{global MSI} for the next local weight update. Local MSI is each device's local gradients, and global MSI is the mean gradients averaged over all devices. At the $k$-th epoch, the $i$-th device is fed with a randomly selected group of data samples, i.e., a mini-batch, by the master. Its local weight $w_{k}^{(i)}$ is updated~as
\begin{align}
w_{k+1}^{(i)} = w_k^{(i)} - \eta \bar{g}_k,
\end{align}
where $\bar{g}_k = 1/M \sum_{j=1}^M g(w_k^{(j)})$ is the gradient averaged over $M$ devices and $g(w_k^{(j)})$ is the $j$-th device's gradient. The parameter $\eta$ is the learning rate that should decrease with $k$ in order to compensate the weight update variance induced by the random mini-batch sampling process.
\vspace{5pt}

\noindent2) \textbf{Elastic SGD (ESGD)} \quad
Instead of exchanging gradients, the weights of an NN model during training can be exchanged. A typical example is ESGD, where the local MSI is each device's local model weights, and the global MSI is the mean weights averaged over all devices~\cite{Zhang:15}. In addition, ESGD focuses on guaranteeing the mean weight convergence, i.e., global MSI reliability, which affects all devices' local MSIs. To this end, the master in ESGD updates the global MSI not only using the current average weight but also based on the previous average weight. The local weight update of the $i$-th device is given as
\begin{align}
w_{k+1}^{(i)} &= (1-\alpha)w_k^{(i)} - \eta g(w_k^{(i)}) + \alpha \hat{w}_k\\
\hat{w}_{k}&= (1- \beta) \hat{w}_{k-1} +   \beta \bar{w}_{k},
\end{align}
where $\bar{w}_k$ is the average weight at the $k$-th epoch, and $\alpha,~\beta<1$ are constants.
\vspace{5pt}

In both CSGD and ESGD, the MSI payload size is proportional to the model size, and is exchanged every epoch for all devices. 
This becomes challenging in edge ML where the wireless communication cost is not negligible. 
In addition, their global MSI based weight update rule fundamentally relies on an ensembling technique suitable for IID training datasets of devices. 
Therefore, it becomes less effective in edge ML where the user-generated training data samples may be non-IID. 
These issues are tackled in Sect~5.4 and~6.4.

\subsection{Hardware-Model Co-Design}
Modern computational devices are equipped with multiple types of memory. On-chip cache such as static random access memory (SRAM) has the smallest size, yet is the fastest, e.g., 24 MB with tens of nanoseconds access latency \cite{N.-P.-Jouppi-et.-al.:2017aa,Skylake}. Dynamic random access memory (DRAM) is larger but slower than SRAM, e.g., 16 GB with hundreds of nanoseconds access latency \cite{Skylake}. Non-volatile memory such as solid-state drive (SSD) is the largest but the slowest, e.g., 30 TB with several microseconds access latency \cite{Anandtech}. 

Compressing the size of an NN model makes the model fit into smaller and faster memory, empowering low-latency inference and training. In addition, model compression improves energy efficiency, since the number of memory accesses is the major source of NN's energy consumption, which is proportional to the model size~\cite{Han:16,Zhang:Energy}. Lastly, in distributed training, model compression minimizes the MSI payload size, thereby reducing communication latency. Model compression needs thus to take into account communication and computation aspects, which is commonly achieved by the following approaches.
\vspace{5pt}

\noindent1) \textbf{Quantization} \quad
Training an NN accompanies a large number of simple arithmetic operations, of which the intermediate calculations need to be stored in the memory. Therefore, it is effective to properly reducing the arithmetic precision of model parameters during the training process. In this respect, mixed-precision training~\cite{S.-Narang-et.-al.:2018aa} is a viable solution, where the precision is downconverted from floating point (FP) $32$ to FP $16$ during the forward and backward propagation processes, and then upconverted back to FP $32$ when updating the master copy of model weights. In so doing, memory consumption during the training is  halved, while not compromising accuracy compared to the training with single-precision FP $32$.
\vspace{5pt}

\noindent2) \textbf{Pruning} \quad 
A deep NN commonly has a large amount of redundant weights and connections. Partially pruning them  helps compress the model while maintaining the original inference accuracy. In this direction, the simplest pruning is dropping a set of perceptrons by setting their final activations to $0$, known as DropOut~\cite{N.-Srivastava-G.-Hinton-A.-Krizhevsky-I.-Sutskever-and-R.-Salakhutdinov:2014aa}, which removes all the subordinated connections of the pruned perceptrons. Alternatively, one can only prune the connections by setting the weights to $0$, referred to as DropConnect~\cite{Wan:13}. The pruning can be performed uniformly randomly until the target accuracy is maintained, or in a more sophisticated way, e.g., based on the Fisher information of the weights~\cite{Theis:18}.
\vspace{5pt}

\begin{figure}\centering
\includegraphics[width= \columnwidth]{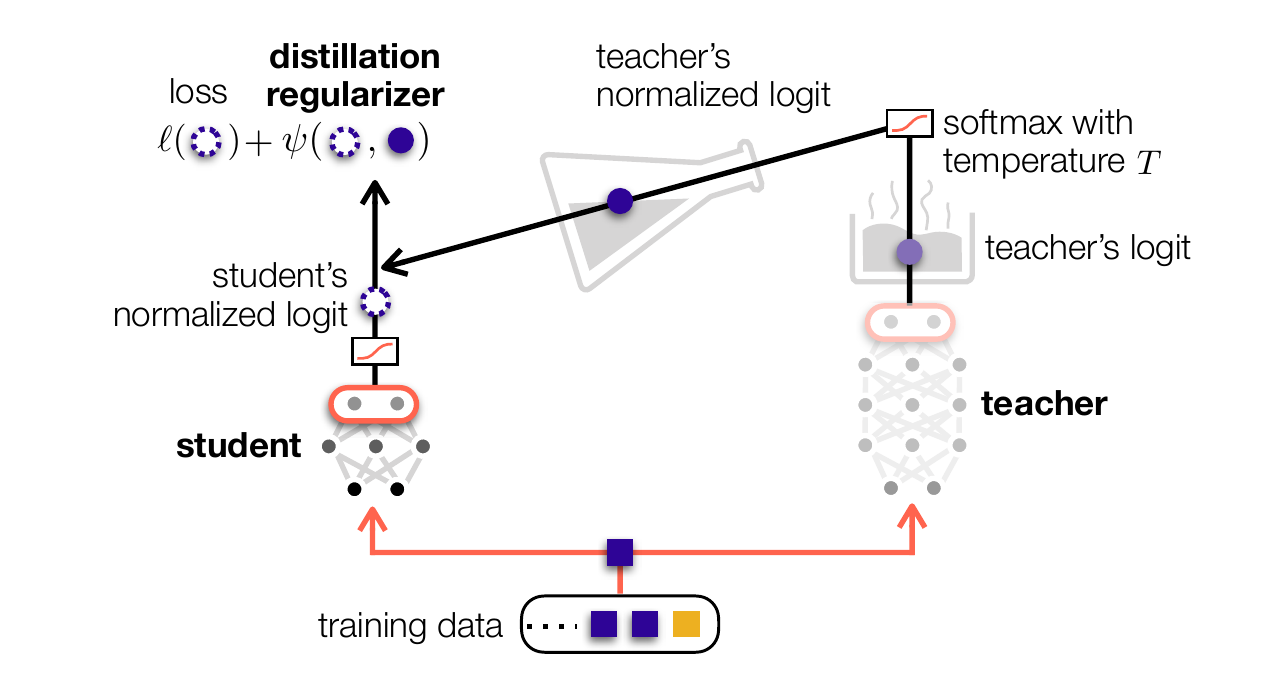}
\caption{Knowledge distillation (KD) that transfers a teacher model's knowledge to its student model.}
\label{fig:knowledge_dist}
\end{figure}

\noindent3) \textbf{Knowledge Distillation (KD)} \quad
Compared to pruning that gradually compresses the model size, KD first constructs an empty NN, referred to as a student NN, with a target compressed model size, and  fills in its weight parameters \cite{HintonKD:14}. KD focuses on training the student NN, while both student and another pre-trained teacher NN observe an identical data sample, as illustrated in Fig.~10. Each NN's prediction of the sample quantifies its current knowledge, and KD transfers the knowledge from the teacher to the student. For the knowledge measurement, instead of the final prediction output, e.g., `dog' in a dog-or-cat problem, KD utilizes the model output prior to the final activation, called \emph{logit} values containing the distributional information of the prediction, e.g., $\{\text{dog},\text{cat}\}=\{80,20\}$. 

The key idea of KD is adding a \emph{distillation regularizer} to the student's loss function. This regularizer inserts a random noise into the student's locally calculated weight, more distorting the weight calculation when the teacher's prediction is more dissimilar to the student's prediction that is likely to be wrong. In this respect, the distillation regularizer is proportional to the knowledge gap between the teacher and the student, and the gap is commonly measured using cross entropy. Since cross entropy only accepts $0$-$1$ ranged values, the original logit values are fed to the distillation regularizer, after being normalized via a softmax function with temperature $T\geq 1$, i.e., $\exp(z_{i}/T)/\sum_{j} \exp(z_{j}/T)$ for the $i$-th label logit~$z_{i}$. The increase in $T$ smoothens the logit values, helping more information being transferred from the teacher particularly when the teacher's logit distribution is peaky. Afterwards, the student's final activation is set as another softmax function without temperature, i.e., $T=1$, which resembles a distilling process as visualized in Fig.~\ref{fig:knowledge_dist}. By default, pruning and KD are performed separately after training completes. The resultant extra latency and memory usage may not well correspond to edge ML, calling for different techniques to be discussed in Sect.~\ref{sec:tech_enablers}.

\subsection{Grand Challenges}
From a theoretical standpoint, unraveling the black-box operations of ML has focused primarily on centralized ML architectures. Investigating edge ML architectures is therefore a daunting task, notably with the model split, where the analysis becomes extremely cumbersome due to the data flow dependency among the split model segments. Moreover, as each device feeds the data samples that can be non-IID, most of the analytical frameworks built upon IID data samples fall short in measuring the reliability of edge ML systems. 

From a technical perspective, the aforementioned training and model compression approaches cannot be straightforwardly applicable for edge ML, but encounter key challenges due to devices' characteristics  and their inter-device communication links, as detailed next.

\subsubsection{On-Device Constraints}
\noindent1) \textbf{Energy Limitation}\quad
In a large-scale ML system, each device is likely to be mobile, equipped with a capacity limited battery. Thus, the finite energy needs to be efficiently utilized for the computation process during training and inference operations, as well as the communication process. In this respect, one may prefer to slightly decrease the inference accuracy in order to save the training energy consumption that is proportional to the model size \cite{Zhang:Energy}. Unfortunately, traditional ML architectures cannot dynamically change the model size in real time, and are thus unable to flexibly optimize the energy consumption. Next, with the helper-device split, the device may want to offload its computation to the helper with  higher computation power without energy limitation. The bottleneck is the offloading communication overhead, and optimizing such a computation-communication trade-off is a large field of research in mobile edge computing (MEC) \cite{wang:2016:MEC,Elbamby:2017:eucnu,chenfeng:2017:GC,Elbamby:2019}.
\vspace{5pt}

\noindent2) \textbf{Memory Limitation}\quad
The optimal NN model size for inference is determined by the devices' memory sizes of SRAM and DRAM. This may conflict with the optimal model size for the training operations, which is mainly determined by the communication overhead induced by the MSI exchanges. Furthermore, the limited size of the devices' storages such as hard disk drive (HDD) and solid-state drive (SSD) constrain the input data sample sizes as well as the number of intermediate calculation values stored during the training phase. Therefore, the training and inference operations need to be jointly optimized under the constraints of communication links, as well as of various types of devices' memories, respectively.
\vspace{5pt}

\noindent3) \textbf{Privacy Guarantees}\quad
Some of the data samples owned by the devices may be privacy sensitive, e.g., medical records. Exchanging MSIs instead of data samples can partly preserve privacy, yet is still vulnerable to being reversely traced by eavesdroppers. Extra coding such as homomorphic encryption \cite{C.-Gentry-and-D.-Boneh:2009aa} may guarantee privacy, but its processing delay may exceed the low latency deadline. Exchanging redundant information may hide the private information, but unfortunately results in extra communication delay. Inserting noise has a potential to protect privacy while playing a role as regularizer, as long as the noise level is properly adjusted; otherwise, it decreases the inference accuracy significantly.

\subsubsection{Communication Bottlenecks}
\noindent1) \textbf{Wireless Capacity Dynamics}\quad
In centralized ML, the communication links are implicitly assumed to be wired; for instance, PCI Express between controller-GPU connections or Ethernet between cloud-device connections under the master-device split. Compared to this, the communication links in edge ML are mostly capacity-limited wireless channels. Furthermore, the wireless channel capacity changes more frequently due to intermittent channel conditions and network congestion. To cope with this, the communication payload, i.e., MSI, needs to be properly compressed. With whom and how often to exchange the MSIs should be dynamically optimized. 
\vspace{5pt}

\noindent2) \textbf{Uplink-Downlink Asymmetry}\quad
Wireless cellular systems follow the helper-device split, where each edge BS becomes the helper. 
In this case, due to the device's lower transmission power, the uplink communication from the device to the helper is much slower than the downlink communication~\cite{JHParkTWC:15}. 
Such characteristics are not utilized in the aforementioned MSI exchanging methods in Sect.~\ref{subsec:BB_MSI_xchange}, where both uplink local MSI and downlink global MSI are of the same type with  identical payload size.
\vspace{5pt}

\section{Theoretical Enablers}\label{sec:theory_enablers}

In this section, we provide theoretical principles to characterize inference reliability and training dynamics in edge ML under the aforementioned communication and on-device constraints. 
Their related applications and techniques in the context of edge ML are exemplified in blue boxes.

\subsection{ML Reliability Guarantees}\label{subsec:T_E_reliability}

Traditional ML focuses heavily on minimizing the average loss and/or average training latency. These approaches omit the credibility intervals of the calculations, and therefore are insufficient for supporting URLLC applications. Instead, it is mandatory to evaluate the loss and latency that guarantee a target reliability. With this end, we first focus on the connection between inference reliability and training latency, followed by the relationship between communication latency and reliability during the training process for a given end application.

The training latency is determined primarily by the number of required training data samples, often referred to as \emph{sample complexity}, until reaching (i) a target inference accuracy with (ii) a target inference reliability. Unfortunately, traditional ML only outputs (i) but not (ii). This missing block can be addressed by Bayesian learning theory \cite{R.-Meir-and-T.-Zhang:2003aa} and the probably approximately correct (PAC) framework~\cite{L.-G.-Valiant:1984aa}. 
\vspace{5pt}

\noindent1) \textbf{Bayesian Learning}\quad
In contrast to traditional ML approaches, an NN is described not directly by the weights of the NN but by their stochastic distribution. The weight distributions before and after the training process are called as prior and posterior distributions, respectively. The prior distribution can be initialized by an arbitrary distribution or by accounting for the training data characteristics. For a given prior distribution, training an NN is recast as estimating the posterior distribution by maximizing a likelihood between the training data samples and the weights, as per the Bayes theorem:
\begin{align}
\underbrace{\Pr(w\mid \mathcal{D}_n)}_\text{posterior} = {\underbrace{\Pr(w)}_\text{prior}\cdot \underbrace{\Pr(\mathcal{D}_n\mid w)}_\text{likelihood}}\Big\slash\Pr(\mathcal{D}_n).
\end{align}
where $w$ is the set of NN's weights and $\mathcal{D}_n\subset\mathcal{D}$ is the set of training data samples. With the estimated posterior distribution, the interference with a set $\mathcal{D}'\subset \mathcal{D}\backslash\mathcal{D}_n$ of test data samples is described as
\begin{align}
\underbrace{\Pr(y \mid \mathcal{D}')}_\text{inference} = \int \Pr(y\mid w,\mathcal{D}')\underbrace{\Pr(w \mid \mathcal{D}')}_\text{posterior} dw. \label{Eq:BayesInf}
\end{align}
Calculating this by averaging over the posterior distribution is cumbersome. Instead, one can approximate the posterior distribution with its generated weights, e.g., via Markov chain Monte Carlo (MCMC) methods.

\small\begin{bluebox}{\textbf{Bayesian Learning} Techniques}
\begin{itemize}[leftmargin=5pt]
	\item \textbf{Gaussian Process (GP)}\quad 
	 Gaussian-distributed priors with infinitely many perceptrons in a single hidden layer result in GP via the central limit theorem \cite{R.-M.-Neal:1995aa}. GP enables regression  using the second-order statistics of Gaussian processes, thereby achieving low complexity in terms of sample and computation. GP has been used for controlling robots~\cite{J.-Fink-A.-Ribeiro-and-V.-Kumar:2013aa} and Google's Internet balloons~\cite{Wired:17}.
	\vspace{5pt}

	\item \textbf{Stochastic Gradient Langevin Dynamics (SGLD)}\quad 
	SGLD is the mini-batched Bayesian learning where the posterior distribution is approximated as the weights being generated via a form of MCMC \cite{Welling:ICML11}, i.e., Langevin dynamics given as $\Delta w_k = \frac{\eta_k}{2}(\nabla \log \Pr(w_k) + \frac{N}{n} \sum_{i=1}^n \log \Pr(y_t\mid w_t ) + \varepsilon_t$ for $\varepsilon_t\sim \mathcal{N}(0,\eta_t)$. This approximates SGD's weight update dynamics with an additional noise $\varepsilon_t$. Inserting $\varepsilon_t$ is useful not only for reducing the generalization error, but also for connecting SGD with the differential privacy framework~\cite{Dziugaite:18} (to be elaborated in Sect. 4.1 and 4.3) and with Entropy SGD~\cite{Chaudhari:ICLR17} (Sect. 5.1).
	\end{itemize}
\end{bluebox}\normalsize
\vspace{10pt}

\noindent2) \textbf{PAC Framework}\quad 
Inference uncertainty mainly comes from the unseen data samples at the training phase. 
To quantify this, for $n=|\mathcal{D}_n|$  training data samples, one can compare the trained NN's (i) empirical average training loss, $\hat{L}(w) = \frac{1}{n}\sum_{i=1}^{n} \ell(y_i,w)$ with (ii) the expected inference loss, $L(w)=\E_{\mu}[\ell(y, w)]$ averaged over all seen/unseen data samples following a distribution $\mu$. 
PAC framework focuses on the bound of the difference between $L(w)$ and $\hat{L}(w)$, i.e., generalization error of the approximation-generalization trade-off shown in Fig.~\ref{fig:avg_risk_tradeoff} in Sect.~\ref{subsec:BB_training_prin_pract}, such that
\begin{align}
\Pr\(L(w) - {\hat{L}(w)} \leq   \text{GE}( \mu, \varepsilon)\) \geq 1-\varepsilon, \label{Eq:PACvar}
\end{align}
where $\text{GE}( \mu, \varepsilon)$ is the achievable generalization error with probability at least $1-\varepsilon$. All terms in \eqref{Eq:PACvar} depend on the NN's hypothesis~$h\in\mathcal{H}$ with the entire hypothesis space $\mathcal{H}$~\cite{D.-Haussler:1992aa}. Treating an NN as an approximated function in regression, $\mathcal{H}$ implies the set of functions that the NN is allowed to select as being the solution \cite{GoodfellowBook:16}. When $\mathcal{H}$ is finite, applying Hoeffding's inequality and union bound, one can derive the GE as $\sqrt{(\log|\mathcal{H}|+\log(1/\varepsilon))/(2 n) }$. 
Here, $n$ is the sample complexity to satisfy a target generalization error with a target reliability $1-\varepsilon$.

\small\begin{bluebox}{\textbf{PAC Framework} Applications}
\begin{itemize}[leftmargin=5pt]
\item \textbf{PAC-VC Bound}\quad
If the hypothesis space $\mathcal{H}$ is infinite, e.g., in deep NNs or non-parametric models such as GP, one can derive the GE  using the Vapnik-Chervonenkis (VC) dimension $\text{VC}(\mathcal{H})$. This yields the PAC-VC bound's GE $\sqrt{(\text{VC}(\mathcal{H})+\log(4/\varepsilon) )/n}$. Here, $\text{VC}(\mathcal{H})$ is simply calculated as the number of NN's parameters, and so is the PAC-VC bound.
\vspace{5pt}

\item \textbf{PAC-Rademacher Bound}\quad
The PAC-VC bound is quite loose in general, since its VC dimension is determined solely by the NN regardless of the training dataset. To rectify this, one can measure $\mathcal{H}$ using the Rademacher compexity $\text{Rad}_\mathcal{D}(\mathcal{H})$ that depends not only on the NN model but also on the training dataset, yielding the GE of the PAC-Rademancher bound, given as $\text{Rad}_\mathcal{D}(\mathcal{H}) + \sqrt{\log(1/\varepsilon)/n}$. 
\end{itemize}
\end{bluebox}\normalsize
\vspace{10pt}

\noindent3) \textbf{PAC-Bayesian Framework}\quad
In the previous PAC framework, PAC-VC bounds become accurate only in the case of as many training data as model parameters, which is infeasible particularly for deep NNs~\cite{J.-Langford-and-M.-Seeger:2001aa}. PAC-Rademancher bounds are free from the said limitation, yet become  vacuous for modern NN architectures under ReLU activations with an SGD training process \cite{G.-K.-Dziugaite-and-D.-M.-Roy:2017aa}. Utilizing Bayesian learning methods, one can resolve these issues, resulting in the following PAC-Bayes bound~\cite{D.-A.-McAllester:1998aa}.
\begin{align}
\Pr\(L(q) - \hat{L}(q) \leq  \sqrt{\frac{\text{KL}(q||p) + \log(1/\varepsilon) }{2 n} }\) \geq 1-\varepsilon, \label{Eq:McAllester99}
\end{align}
where $L(q)$ and $L(\hat{q})$ denote the empirical and expected average loss values for a posterior $q$, respectively. In \eqref{Eq:McAllester99}, the GE is a function of the Kullback-Leibler (KL) divergence of $q$ and a prior $p$, which is also called as the \emph{complexity} required for mapping $p$ to $q$. It is worth noting that the difference between $L(q)$ and $\hat{L}(q)$ can be measured using their KL divergence, yielding the refined version by \cite{M.-Seeger:2002aa} as stated below.
\begin{align}
\Pr\(\text{KL}\(\hat{L}(q) || L(q) \) \leq \frac{1}{n}\[\text{KL}(q|| p) + \log\frac{n+1}{\varepsilon} \]\) \geq 1-\varepsilon \label{Eq:Seeger}
\end{align}
Both \eqref{Eq:McAllester99} and \eqref{Eq:Seeger} are derived under an IID training dataset, which can be extended to a non-IID dataset as described in the following applications.

\small\begin{bluebox}{\textbf{PAC-Bayes Framework} Application}
\begin{itemize}[leftmargin=5pt]
\item \textbf{Chromatic PAC-Bayes Bound for Non-IID Data}\quad
When the dependency in a training dataset $\mathcal{D}_n$ is modeled as a dependency matrix $\Gamma(\mathcal{D}_n)$ using fractional covers in graph theory, the PAC-Bayes bound becomes

\vspace{-10pt}\footnotesize\begin{align}
\hspace{-7pt}\Pr\(\text{KL}\(\hat{L}(q) || L(q) \) \leq \frac{\chi^*}{n}\[\text{KL}(q|| p) + \log\frac{n/\chi^*+1}{\varepsilon} \]\) \geq 1-\varepsilon, \nn
\end{align}\small
where $\chi^*$ is the fractional chromatic number of the dependency matrix $\Gamma(\mathcal{D}_n)$, defined as the minimum weight over all proper exact fractional covers of all vertices in $\Gamma(\mathcal{D}_n)$. The fractional chromatic number can be obtained using linear programming, and its closed form is available if $\Gamma(\mathcal{D}_n)$ belongs to some special classes of graphs \cite{Weisstein:MathWorld}. The definition of $\Gamma(\mathcal{D}_n)$ is detailed in~\cite{L.-Ralaivola-M.-Szafranski-and-G.-Stempfel:2010aa}.
\vspace{5pt}

\item \textbf{Collective Stable PAC-Bayes Bound for Non-IID Data}\quad
Consider $\mathcal{D}_n$ is divided into  $m$  subsets. When the dependency among $m$ subsets is modeled using the collective stability framework, the PAC-Bayes bound is given as

\vspace{-10pt}\footnotesize\begin{align}
\Pr\(L(q) - \hat{L}(q) \leq 2 \beta || \Gamma||_\infty \sqrt{\frac{\text{KL}(q||p) + \log(2/\varepsilon) }{2 n m} }\) \geq 1-\varepsilon,  \nn
\end{align}\small
where $\beta$ is the Lipschitz constant under the Hamming distance between the training inputs. The term  $\Gamma$ is the training dataset's dependency matrix whose definition is elaborated in~\cite{B.-London-B.-Huang-and-L.-Getoor:2016aa}.
\vspace{5pt}

\item \textbf{PAC-Bayes Bound with Data-Dependent Priors}\quad
The Bayesian prior $p$ of a PAC-Bayes bound should be chosen independently of the training dataset~$\mathcal{D}_n$, yet can still depend on the distribution of the dataset. Utilizing this idea, it is possible to characterize the dependency between~$\mathcal{D}_n$ and $p$ via the differential privacy framework~\cite{G.-K.-Dziugaite-and-D.-M.-Roy:2018aa}, yielding the following PAC-Bayes bound

\vspace{-10pt}\footnotesize\begin{align}
\hspace{-5pt}\Pr\(\text{KL}\(\hat{L}(q) || L(q) \) \leq \frac{1}{n}\[\text{KL}(q|| p) + 2 c(\varepsilon,\epsilon) \]\) \geq 1-\varepsilon, \nn
\end{align}\small
where $c(\varepsilon,\epsilon) = \max\l\{\log (3/\varepsilon),  n\epsilon^2 \r\}$, and $\epsilon$ implies that the data-dependent prior $p$ ensures $\epsilon$-differential privacy whose definition is elaborated in Sect.~\ref{subsec:T_E_latency_scalability}.

\end{itemize}
\end{bluebox}\normalsize
\vspace{10pt}

\noindent4) \textbf{Meta Distribution}\quad
In both PAC and PAC-Bayes frameworks, the obtained generalization error bounds hold for any $n$ number of selected training samples, as they are averaged over the selections of the training data set. This is suitable for centralized ML where the training mini-batched data samples are frequently renewed. For edge ML, such a scenario may not be feasible due to the communication overhead and/or privacy guarantees. In this case, the meta distribution enables to capture the generalization error bound, by accounting for \emph{a given set $\mathcal{D}_n$ of the training samples} as follows:

\vspace{-10pt}\small\begin{align}
\Pr\l\{  \Pr\(L(\mathcal{D}) - \hat{L}(\mathcal{D}_n)\leq GE(\mu_{n},\varepsilon) \mid \mathcal{D}_n \)\geq 1-\varepsilon  \r\} \geq 1-\delta, \label{Eq:Meta}
\end{align}\normalsize
where $\mu_{n}$ is the distribution of the data samples in $\mathcal{D}_n$, and $\delta$ is the target generalization error outage probability for all training samples. In order not to complicate the calculation, one can approximate the meta distribution with the beta distribution, as proposed in \cite{M.-Haenggi:2015aa}. This is the 2nd-order moment approximation, and thus only requires to calculate the mean and variance of the innermost probability in \eqref{Eq:Meta}.

\small\begin{bluebox}{\textbf{Meta Distribution} Applications}
\begin{itemize}[leftmargin=5pt]
	\item \textbf{Signal-to-Interference-Ratio (SIR) Meta Distribution}\quad 
	The idea of meta distributions was originally proposed in the context of stochastic geometry \cite{M.-Haenggi:2015aa}. It focuses on the meta distribution of the SIR coverage probability for \emph{a given large-scale network topology}. One can thereby quantify, for instance, the fraction of  receivers that guarantees a target wireless communication reliability, which in passing is also useful for the latency reliability analysis in large-scale edge ML design.
	\vspace{5pt}

	\item \textbf{Probably Correct Reliability (PCR)}\quad
	Focusing on estimating wireless communication channels, PCR is the meta distribution of outage capacity for \emph{a given set of channel observations} \cite{M.-Angjelichinoski-K.-F.-Trillingsgaard-and-P.-Popovski:2018aa}. This reliability analysis framework is promising in the context of MLC towards enabling URLLC.
	\end{itemize}
\end{bluebox}\normalsize
\vspace{10pt}

\noindent5) {\textbf{Risk Management Framework}}\quad 
The above-mentioned reliability bounds are determined by the difference between the biased loss after training and the ideally averaged loss with the entire data samples, i.e., generalization error. Instead, one may  inquire whether the biased loss reliably achieves a target loss level. 
The right tail of the biased loss distribution can describe this, which has been investigated in mathematical finance  using \emph{Value-at-Risk (VaR)} and \emph{Conditional-Value-at-Risk (CVaR)} \cite{Rockafellar:2002:VAR}. 
VaR focuses on the tail's starting point, hereafter referred to as the \emph{tail threshold}, specifying the minimum target loss $x$ that guarantees a target reliability $1-\varepsilon$ as follows
\begin{align}
\text{VaR}_{1-\varepsilon}(\hat{L}) = \underset{x}{\arg\min} \l\{  \Pr\( \hat{L} \leq x\) \geq 1-\varepsilon  \r\}.
\end{align}
VaR is often defined by a non-convex and/or discontinuous loss function, which requires complicated calculations. Even with such a loss function, CVaR avoids this complication, by ensuring its monotonicity. To this end, CVaR considers the tail's area, providing the expectation of the loss exceedances with the tail threshold set as the VaR.
\begin{align}
\text{CVaR}_{1-\varepsilon}(\hat{L}) = \E\[ \hat{L} \mid \hat{L} > \text{VaR}_{1-\varepsilon}(\hat{L}) \]
\end{align}

\small\begin{bluebox}{\textbf{Risk Management Framework} Applications}
\begin{itemize}[leftmargin=5pt]
	\item \textbf{Distributional RL}\quad
	Traditional RL is trained so as to approximate the expected cumulative return at the current state $x$ and action $a$, e.g., Q-function $Q(x,a) = \E[R(x,a) + \gamma Q(x',a') ]$ for the reward $R(x,a)$ and a discount factor $\gamma$ when $(x,a)\rightarrow (x',a')$. Instead, one can learn to approximate the distribution of $Z = R(x,a) + \gamma Q(x',a')$, known as the value distribution. Such distributional RL outperforms  state-of-the-art RL methods including deep Q-learning and asynchronous actor-critic RL~\cite{M.-G.-Bellemare-W.-Dabney-and-R.-Munos:2017aa,M.-Hessel-et.-al.:2017aa}.
	\vspace{5pt} 
	
	\item \textbf{Risk-Sensitive RL}\quad
	Maximizing the expected return in RL is insufficient for robust control  warranting a predefined minimum return with a target probability. 
	The mean-variance approach is the simplest solution that additionally minimizes the variance of returns while maximizing the expected return~\cite{Markowitz:mean-var}. 
	A better approach is to maximize the ($1-\varepsilon$)-worst return, i.e., CVAR, by observing extra samples as done in~\cite{Tamar:2015:OCV}.
	Furthermore, one can utilize the value distribution of distributional RL, thereby directly computing and maximizing CVaR~\cite{Silvestr:2018:thesis}.	

	\end{itemize}
\end{bluebox}\normalsize
\vspace{10pt}

\noindent6) \textbf{Extreme Value Theory (EVT)}\quad
The right tail of the biased loss distribution can also be described in the context of  two fundamental theorems in EVT~\cite{EVT:18}. To elaborate, for any distribution of the biased loss, the Pickands-Balkema-de Haan theorem states that the distribution of its loss exceedances with the infinitely large tail threshold $\theta$ converges to  the \emph{generalized Pareto distribution (GPD)}, i.e.,
\begin{align}
\Pr(\theta < \hat{L}(\mathcal{D}_n) \leq x) \overset{\theta\rightarrow \infty}{=} \underbrace{1- (1 + \zeta x/ \sigma)^{-1/\zeta} }_{\text{GPD}(x,\zeta)}, \label{Eq:GPD}
\end{align}\vskip -5pt
\noindent which becomes a Pareto (if $\zeta<0$), an exponential ($\zeta=0$), or a uniform distribution ($\zeta=-1$). Next, focusing on the training data set $\mathcal{D}_n\subset \mathcal{D}$, one may need to guarantee the reliability even for the worst-case loss with respect to a number $K$ of the IID training data set selections $\{\hat{L}(\mathcal{D}_{n,k})\}_{k\leq K}$ in centralized ML, or to $K$ devices in edge ML. In this case, the Fisher-Tippett-Gnedenko theorem of EVT is applicable. The theorem describes that the distribution of the maximum loss out out of the loss values obtained by an infinitely large number of the training data set selections converges to the \emph{generalized extreme value distribution (GEV)}, i.e.,
\begin{align}
\Pr\(\hat{L}_K(\mathcal{D}_n) \leq x \) \overset{K\rightarrow \infty}{=} \underbrace{e^{\text{GPD}(x-m,\zeta)-1}}_{\text{GEV}(x,\zeta)}, \label{Eq:GEV}
\end{align}\vskip -5pt
\noindent where $\hat{L}_K(\mathcal{D}_n) = \max\{\hat{L}(\mathcal{D}_{n,k})\}_{k\leq K}$ and $m$ is the mean of $\{\hat{L}(\mathcal{D}_{n,k})\}_{k\leq K}$. The GEV becomes a Gumbel (if $\zeta=0$), a Fr\'{e}chet ($\zeta>0$), or a reversed Weibull distribution ($\zeta <0$). The relationship between GEV and GPD is obtained trivially by applying Taylor's expansion to \eqref{Eq:GEV} for $\theta < x\rightarrow \infty$.

 \small\begin{bluebox}{\textbf{EVT} Applications}
 \begin{itemize}[leftmargin=5pt]
 \item \textbf{Robust Multiclass Classification} \quad 
 	Each data sample in multiclass classification has one ground-truth label out of multiple labels.
 	In this case, the robustness is defined as the maximum noise (both in attributes and labels) that prevents adversarial inputs. 
 	To evaluate the upper bound for the noise, the maximal gradients of the classifier functions for each label need to be evaluated over all training samples. Alternatively, using EVT, one can evaluate the maximum gradients using only a few training samples \cite{Weng2018EvaluatingTR}.
\vspace{5pt}

 \item \textbf{Kernel-Free Open-Set Recognition} \quad
 	Supervised classification in open sets with unknown number of classes has a challenge of classifying inputs belonging to unseen classes at training time due to  under sampling.
 	The solution is a kernel-free recognition technique known as \emph{extreme value machine} \cite{Rudd2018TheEV}.
 	Therein, EVT is used to characterize the decision boundaries of classes in a probabilistic manner that provide an accurate and efficient data partition.

 \end{itemize}
 \end{bluebox}\normalsize
\vspace{10pt}

\begin{figure}\centering
\includegraphics[width=  \columnwidth]{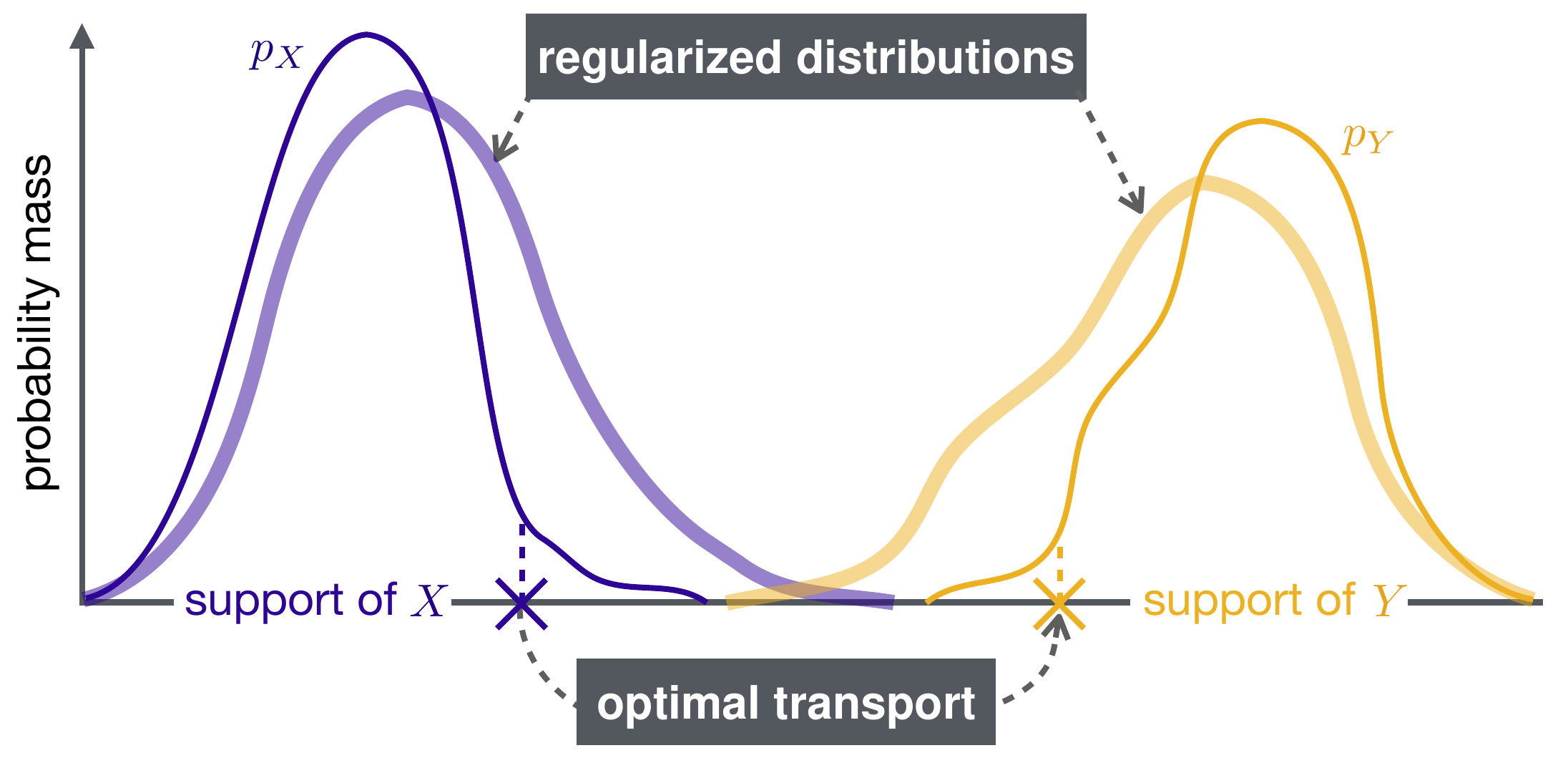}
\caption{Optimal transport (OT), compared to regularized distributions.}
\label{fig:OT}
\end{figure}

\noindent7) \textbf{Optimal Transport Theory (OT)}\quad
Minimizing the KL divergence $\text{KL}(p_Y || p_X)$ between two distributions $p_X$ and $p_Y$ is ubiquitous in NN training and inference processes. 
For training, it is identical to maximizing the likelihood in Bayesian learning. 
It also captures training a generative model such as GAN whose loss function is defined using KL divergence. For inference, it  minimizes the variance term in the PAC-Bayesian bound. 
The caveat is that KL divergence becomes infinite if  two distributions have non-overlapping supports, hindering the reliability of training and inference processes. 
To resolve this, as illustrated in Fig.~\ref{fig:OT}, a na\"ive approach is to insert noise into the distributions so as to secure a common range of support, which however compromises accuracy. 
Instead, replacing the KL divergence minimization with calculating the \emph{Wasserstein distance} in OT~\cite{CuturiBook:18} has recently become a promising solution, which yields a finite value even with non-overlapping distributions. 
The simplest case is calculating the Wasserstein-1 distance $W(p_Y || p_X)$, given as
\begin{align}
W(p_Y || p_X) = \inf_{\gamma\in\Pi_{X,Y}} \E_{X,Y \sim \gamma}\[|| X-Y||\]. \label{Eq:Wasserstein}
\end{align}
This implies the minimum cost to transport the probability mass from $X$ to $Y$ so that $p_X$ equals to $p_Y$, as illustrated in Fig.~\ref{fig:OT}. 
The transported mass amount is characterized by the joint distribution $\gamma$ of $X$ and $Y$. 
The constraint $\gamma\in\Pi_{X,Y}$ ensures that such an optimal transportation $\gamma^*$ has its marginal distributions $p_X$ and $p_Y$. 
The calculation of $W(p_X || p_Y)$ by deriving $\gamma^*$ is challenging, thus commonly relying on approximation and algorithmic techniques.

\small\begin{bluebox}{\textbf{OT} Applications}
\begin{itemize}[leftmargin=5pt]
	\item \textbf{Wasserstein GAN (WGAN)}\quad 
	The training process of WGAN \cite{wassGAN} is equivalent to calculating the Wasserstein distance.
	For its easier calculation, the RHS of \eqref{Eq:Wasserstein} is replaced with its dual formulation $\sup_{||f||_L\leq 1} \E_{X\sim P_X}[f(X)] + \E_{Y\sim P_Y}[f(Y)]$ by applying the Kantorovich-Rubinstein duality. This implies a regression problem with a 1-Lipshitz function $f$, which can be solved via supervised learning.
	\vspace{5pt}

	\item \textbf{Sinkhorn Divergence}\quad 
	Instead of the said dual approach, \cite{Cuturi:2013:SDL} tackles the primal formulation in \eqref{Eq:Wasserstein} with an entropic approximation and the Sinkhorn algorithm \cite{sinkhorn1964}. 
	In this approach, the RHS of \eqref{Eq:Wasserstein} is recast as $\inf_{M\in\mathcal{M}} \text{Tr}(M C^T)$ with the cost matrix $C$. This is a matching problem of the matrix $M$ given by mini-batched $X$ and $Y$, which can be solved via the parallelized Sinkhorn algorithm.
	\end{itemize}
\end{bluebox}\normalsize
\vspace{10pt}

\noindent8) \textbf{R\'{e}nyi Entropy}\quad
The information bottleneck in Sect.~\ref{subsec:BB_training_prin_pract}  describes the information flow across consecutive $L$ layers of an NN during the training process. 
This can be achieved by extending a single bottleneck $\hat{X}$ in \eqref{Eq:IB} into  $L$  bottlenecks $\{X_l\}$'s \cite{Shwartz-Ziv:17}. 
When the bottlenecks are discrete random variables, the flow dynamics are described by the discrete Shannon entropy~\cite{P.-Khadivi-R.-Tandon-and-N.-Ramakrishnan:2016aa}. By contrast, when the bottlenecks are continuous random variables, the dynamics are expressed using continuous entropy, i.e., differential entropy, which unfortunately results in intractable solutions, except when the input $X$ and the output $Y$ are jointly Gaussian \cite{G.-Chechik-A.-Globerson-N.-Tishiby-and-Y.-Weiss:2005aa}. 
In order to derive a tractable solution, one can leverage the \emph{R\'{e}nyi entropy}:
\begin{align}
H_\alpha(X)=\log_2\(\sum_{i=1}^n {p_i}^\alpha \) \big / (1-\alpha),
\end{align}
where its limiting case becomes the Shannon entropy $H_1(X)=-\sum_{i=1}^n p_i \log_2 p_i$ when $\alpha$ approaches $1$. 
With its matrix version expression, the IB formulation with $L$ layers is given as
\begin{align}
\min_{f} \sum_{i=1}^L I_\alpha(X;X^l) + \beta H(Y, f(X)),
\end{align}
where $f$ is a function that the NN tries to learn and $H(Y, f(X))$ is the cross entropy between $Y$ and $f(X)$. The term $I_\alpha(X;X^l)$ is the matrix based R\'{e}nyi mutual information that equals $I_\alpha(X;X^l) = S_\alpha(X) + S_\alpha(X^l) - S_\alpha(X,X^l)$. 
Here, the first term is the matrix based R\'enyi entropy that equals $S_\alpha(X)= \log_2\text{Tr}(X^\alpha)/(1-\alpha)$. 
The last term is the matrix based joint R\'enyi entropy that is given by $S_\alpha(X,X^l)=S_\alpha(X \circ X^l/ \text{Tr}(X \circ X^l))$, where $\circ$ is the Hadamard product.

\subsection{Latency Reduction and Scalability Enhancement}\label{subsec:T_E_latency_scalability}
The performance of data-driven ML approaches rests on how many data samples are utilized. In edge ML in which  data samples are generated by the devices, the problem boils down to  \emph{how many devices can be federated}. Providing a privacy-preserving mechanism during their federation is key to increasing the range of federation, which can be addressed through the lens of differential privacy (DP)~\cite{McSherry:2007:MDV,Liu:2018:GausMech}.

A large range of federation brings to the fore the subsequent question, \emph{how to cope with the MSI communication costs of a large number of edge ML devices} whose NN model sizes may exceed the wireless channel capacity. In this respect, rate-distortion theory establishes a guideline for compressing MSI by balancing the compression rate and its resulting distortion~\cite{Cover:Book,R.-M.-Gray-and-D.-L.-Neuhoff:1998aa}. Furthermore, in a multi-agent RL (MARL) setting,  mean-field game (MFG) theory \cite{larsy:2007:MFG,Bensoussan:2013:MFG} provides an elegant method making full use of local computations, as detailed next.
\vspace{5pt}

\noindent1) \textbf{Differential Privacy (DP)}\quad
With local datasets owned by devices, distributed training operations should preserve the data privacy. To this end, one can apply an auxiliary mechanism, such that its output cannot tell whether a particular local dataset is participated in the training operations, thereby preserving data privacy. DP formalizes this idea, while quantifying the privacy loss of a mechanism $\mathcal{M}(\cdot)$ and its target threshold $\epsilon$ as follows:
\begin{align}
\underbrace{\log\(\frac{\Pr\(\mathcal{M}(\mathcal{D}_i)\in \mathcal{S} \) }{\Pr\(\mathcal{M}(\mathcal{D}_j)\in \mathcal{S}\)} \)}_\text{privacy loss} \leq \epsilon,
\end{align}
which holds for any subsets $\mathcal{D}_i$ and $\mathcal{D}_j$ with $i\neq j$ out of the entire dataset $\mathcal{D}$. When the aforementioned constraint is satisfied, the mechanism achieves $\epsilon$-differential privacy, in which an adversary can only differentiate whether the output is from $\mathcal{D}_i$ or $\mathcal{D}_j$, with an uncertainty inversely proportional to $\epsilon$.
For SGD, this is achieved by simply inserting a Gaussian noise into each data input, i.e., the Gaussian mechanism \cite{Liu:2018:GausMech}. 
For Bayesian learning, one can sample the weights from the noise inserted posterior distribution, i.e., the exponential mechanism \cite{McSherry:2007:MDV}, achieving the $\epsilon$-differential privacy.

\small\begin{bluebox}{\textbf{DP} Application}
\begin{itemize}[leftmargin=5pt]
	\item \textbf{DP Regularizer}. The cost for achieving $\epsilon$-differential privacy is the increase in noise. This is not always a foe but can be a friend. In fact, inserting noise is treated as adding a regularizer. Thus, so long as the noise level is appropriately adjusted, one can improve both accuracy and privacy, as done in \cite{Zhang:2012:FMR}.
	\end{itemize}
\end{bluebox}\normalsize

\noindent2) \textbf{Rate-Distortion Theory}\quad
Quantizing MSI reduces latency by decreasing the communication payload size, at the cost of compromising MSI accuracy due to the distortion induced by quantization. Balancing latency and accuracy can therefore be achieved by optimizing the number $\ell$ of quantization levels. Rate-distortion theory characterizes the optimum $\ell$ under a given channel condition, by stating that the transmitting rate $R(D)$ bits/sample with distortion $D$ should not exceed a given wireless channel capacity $C$ bits/sample, i.e., $R(D)\leq C$. To elaborate, when the MSI is treated as an IID. Gaussian source with variance $\sigma^2$, the Shannon distortion-rate function~\cite{Cover:Book} is given as
\begin{align}
R(D) = 1/2\cdot \log_2(\sigma^2/D). \label{Eq:RD}
\end{align}
Using a uniform scalar quantizer with sufficiently small step size $\Delta$, the mean squared error distortion is approximated as $D\approx \Delta^2/12$ \cite{R.-M.-Gray-and-D.-L.-Neuhoff:1998aa}. Next, assuming the source deviation $\sigma$ equals the difference between the maximum and minimum quantized values, we obtain the quantization levels $\ell= \sigma/\Delta$. Applying these two results to \eqref{Eq:RD} thereby leads to the following bound
\begin{align}
C \geq R(D) \approx \log_2(\ell) + \log_2(12)/2. \label{Eq:RD2}
\end{align}
This shows the upper bound of quantization levels, which is useful when transmitting the MSI as much as possible for a given channel condition. Since the Gaussian source assumption yields the lowest rate, the result provides the worst-case quantizer design guideline in practice.

\small\begin{bluebox}{\textbf{Rate-Distortion Theory} Applications}
\begin{itemize}[leftmargin=5pt]
	\item \textbf{Regularization via Quantization}\quad
	If one is willing to minimize the MSI communication latency, the lower bound of quantization levels is needed. In fact the distortion can contribute positively to the regularization in edge ML \cite{Sau:16}. The MSI can thus be distorted until reaching an optimal regularizing noise level. This maximum distortion yields the lower bound.
	\vspace{5pt}

	\item \textbf{IB under KL Divergence}\quad
	The formulation \eqref{Eq:IB} in IB is a special case for obtaining the rate-distortion function $R(D)$ with a distortion measure given as a KL divergence \cite{N.-Tishby-F.-C.-Pereira-and-W.-Bialek:2000aa}. Namely, \eqref{Eq:IB} originates from $R(D) = \min_{p(\hat{x};x)} I(X;\hat{X})$ s.t. $\sum_{x,\hat{x}} p(x)p(\hat{x}| x) \mathcal{D}(x| \hat{x}) D$, where $\mathcal{D}(x| \hat{x})$ is a distortion measure between the original input $x$ and its compressed information $\hat{x}$. Then, its Lagrangian relaxed formulation with the distortion measure set as the KL divergence between $p(y|x)$ and $p(y|\hat{x})$ becomes \eqref{Eq:IB}.
	\end{itemize}
\end{bluebox}\normalsize
\vspace{5pt}

\noindent3) \textbf{Mean-Field (MF) Control Framework}\quad 
In MARL, $N$~devices are strategically interacting by individually taking actions without a central coordinator, formulated as an $N$-player game. In this game, both computational complexity and the number of communication rounds across devices increase exponentially with $N$, which can be remedied at using \emph{MF game (MFG)} theory \cite{larsy:2007:MFG}. 

To illustrate, consider a $3$-player game with the devices $A,B,C \in\mathcal{P}$. The device $A$ with a given state first takes an action that affects the states of the other devices, i.e., the device $A$ interacting with $\mathcal{P}\backslash A$. Likewise, the device $B$ interacts with $\mathcal{P} \backslash B$, while the device $C$ does with $\mathcal{P} \backslash C$. All these problems are coupled, and calculating their Nash equilibrium induces the undesirable complexity. However, if $N$ becomes extremely large, in the aforementioned example, it becomes $\mathcal{P} \backslash A \approx \mathcal{P} \backslash B\approx \mathcal{P} \backslash C\approx \mathcal{P}$, since each action is likely to affect a negligibly small fraction of the entire population. This implies that each device plays a $2$-player game with a virtual device, i.e., the entire population, which is called an MFG. The originally coupled $N$-player game thereby becomes a number $N$ of individual device's $2$-player games that can be locally solved for a given state of the entire population, referred to as the \emph{MF distribution}. 
The MF distribution is obtained by solving the \emph{Fokker-Planck-Kolmogorov (FPK)} equation of a continuous Markov process \cite{Bensoussan:2013:MFG}. 
For the given MF distribution, the optimal action of each device is then taken by solving the \emph{Hamilton-Jacobi-Bellman (HJB)} equation, a continuous version of the Bellman backward equation in Markov Decision Process (MDP) \cite{Bensoussan:2013:MFG}. An MFG theoretic communication-efficient UAV control example will be elaborated in Sec~6.7.

\begin{figure}\centering
\includegraphics[width= \columnwidth]{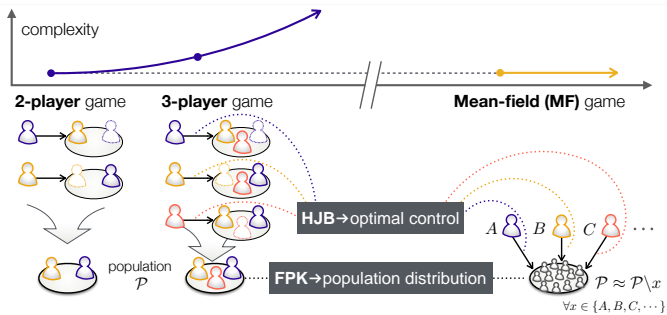}
\caption{Mean-field game (MFG) where each device locally plays a 2-player game with a virtual device, i.e., the entire population whose state is given by the MF distribution.}
\end{figure}

\small\begin{bluebox}{\textbf{MF Control Framework} Applications}
\begin{itemize}[leftmargin=5pt]	
	\item \textbf{MFG in Wireless Systems}\quad 
	With the aforementioned wind dynamics, the work~\cite{KimSPAWC:18}  investigated the UAV mobility control that avoids  inter-UAV collisions while maximizing their air-to-ground communication performance. With the wireless channel dynamics, MFG has shown its effectiveness in transmission power control and resource management particularly under ultra-dense cellular networks (UDNs)~\cite{MobilMFGSG:GC16}. 
	Moreover,  the spatio-temporal content popularity dynamics were modeled in \cite{kim:2017:MFCA}, and an optimal file caching strategy was found using MFG.
	\vspace{5pt}

	\item \textbf{MF Control in ML}\quad
	With the feed forward dynamics of the deep NN training process, the optimal weights are determined by an HJB that can be solved by the Euler approximation method \cite{E:2017:HJB}, to be detailed in Sec.~5.4.2. 
	Note that all the said examples assume that the initial state of the population is given by a Gaussian distribution that is not always realistic. 
	To fill this gap, in the context of large-scale MARL, the population dynamics of the MFG has been inferred  using an inverse RL method in \cite{yang:2018:MF-MARL}.
	\end{itemize}
\end{bluebox}\normalsize

\section{Technical Enablers}\label{sec:tech_enablers}

In this section, we propose technical solutions that enable low-latency decentralized training as well as reliable and accurate decentralized inference, under communication and on-device constraints. Their relationships with the theoretical principles presented in Sec.~3 are elaborated  upon in yellow boxes.

\subsection{ML Reliability Improvement}

Generalization errors can be reduced by designing an NN training algorithm for finding flat minima in the EL, as exemplified by Entropy SGD~\cite{Chaudhari:ICLR17}. For multiple different tasks, one can reduce generalization errors by training NNs based on task correlations~\cite{jnl:smith17,Finn:2017:MMF:3305381.3305498}. On another level, the training process can become robust against malicions and/or malfunctioning devices by the aid of blockchain technologies~\cite{KimCL:19}, as elaborated next.
\vspace{5pt}

\noindent1) \textbf{Entropy SGD}\quad
The goal of entropy SGD is to obtain a flat minimum solution~\cite{Chaudhari:ICLR17}. To this end, for a given original loss function $\mathcal{L}(w)$, entropy SGD minimizes its modified loss function, referred to as local entropy, given as
\begin{align}
u(w_k,\gamma)&= -\log \E_{g}\[e^{-(\mathcal{L}(w_k)+g)}\],
\end{align}
where $g$ follows a Gaussian distribution with variance $\gamma$. The local entropy loss is designed by Gaussian sampling from the $\mathcal{L}(w)$'s Gibbs entropy that is proportional to the number of local maxima within $\gamma$ in the EL. The local entropy can be minimized using an MCMC algorithm~\cite{Chaudhari:17}. 

\small\begin{yellowbox}{\textbf{Entropy SGD} Related Theory}
\begin{itemize}[leftmargin=5pt]
	\item \textbf{Serial-to-Parallel Conversion via~SGLD}\quad 
	It is remarkable that entropy SGD for a single device is identical to elastic SGD operated with multiple devices, under an ergodicity condition $\nabla^2 L(w) + 1/\gamma I \succ 0$. Edge ML can hence be analyzed by reusing most of the theoretical principles that were originally applicable for a single device. The said conversion is validated by first exploiting the Hope-Cole transformation \cite{Chaudhari:ICLR17,Chaudhari:17} that shows the local entropy loss of entropy SGD is the solution of a viscous Hamilton-Jacobi partial differential equation (PDE). Solving this PDE  using a homogenization technique \cite{Chaudhari:ICLR17,Chaudhari:17} yields the loss dynamics that is identical to the elastic SGD's loss dynamics.
	
	\vspace{5pt}
	\item \textbf{PAC-Bayes Bound for Entropy SGD}\quad 
	A recent work \cite{Dziugaite:18} verifies that Entropy SGD works by optimizing the Bayesian prior, i.e. the distribution of $w_k + g$. This clarifies the difficulty of deriving the PAC-Bayes bound for Entropy SGD, as the prior of the original PAC-Bayes framework is constrained to be chosen independently of the training data. The work \cite{Dziugaite:18} resolves this problem, by first adding an extra random noise to Entropy SGD's weight updates, as done in SGLD \cite{Welling:ICML11}. The resulting algorithm is referred to as Entropy SGLD that is interpreted as a mechanism achieving $\epsilon$-differential privacy. Then, utilizing the PAC-Bayes with the data-dependent prior satisfying the $\epsilon$-differentially privacy (see Sect. 4.1) yields the PAC-Bayes bound for Entropy SGD, thereby quantifying its generalization capability.
	\end{itemize}
\end{yellowbox}\normalsize
\vspace{10pt}

\noindent2) \textbf{Task-Aware Training}\quad
Generalization errors result not only from unseen dispersed samples (see Sect. 4.1), but also from distinct tasks of devices. For a given set of tasks, \emph{multi-task learning (MTL)}~\cite{Zhang2013ARA,jnl:smith17} trains an NN so as to ensure high accuracy for multiple tasks, by inserting a task correlation regularizer into the original loss function~\cite{Zhang2013ARA}. When the task correlation is unknown, the correlation can also be trained by alternating (i) optimizing the NN weights while fixing the correlation matrix and (ii) optimizing the correlation matrix while fixing the NN weights~\cite{jnl:smith17}. On the other hand, if the tasks are not given a priori, MTL is ill-suited because task correlations cannot be specified. Alternatively, \emph{meta learning} is still effective in this case, which aims to train a meta-learner that is capable of quickly learning various types of tasks~\cite{Finn:2017:MMF:3305381.3305498}. This training objective can be achieved by randomly sampling the loss function out of all possible loss functions corresponding to different known tasks, thereby guaranteeing robustness against unseen tasks. 
\vspace{5pt}

\noindent3) \textbf{Blockchained Training}\quad
In edge ML, malicious devices may participate and disrupt the training process. Furthermore, selfish devices may not contribute to the local training procedures, while only receiving the global training results computed by the other devices. Keeping a record of the training process using distributed ledger technology (DLT) is useful to mitigate these problems. For example, when exchanging local MSIs, each device cross-validates the MSIs, and stores the accepted MSIs in its local distributed ledger. The local distributed ledger is synchronized with the other devices' ledgers via DLT such as a blockchain algorithm~\cite{KimCL:19} or a directed acyclic graph (DAG) based Byzantine fault tolerance (BFT) algorithm~\cite{H.-Seo-J.-Park-M.-Bennis-and-W.-Choi:2018aa,Park:2019:RC}. Thereby, only the legitimate local MSIs contribute to the global MSI calculation, ensuring the reliability of the training process.

\subsection{Communication-Aware Latency Reduction}\label{subsec:TEC_E_latency}

In what follows we introduce and propose MSI exchange schemes to reduce the communication latency during distributed training operations. We hereafter focus on the h-d architectural split where training data samples are generated by devices, unless otherwise specified.

\subsubsection{Periodic Model MSI Exchange}
\noindent1) \textbf{Federated Averaging (FAvg)}\quad Too frequent MSI exchanges in CSGD incur significant communication overhead. In order to mitigate this, the devices in FAvg exchange the local MSIs at an interval of $\tau$ epochs~\cite{Brendan17}. Following the notations defined in Sect.~3.2.2, at the $k$-th epoch, the $i$-th device's weight $w_{k}^{(i)}$ is described as below:
\begin{align}
w_{k+1}^{(i)} = \begin{cases}w_k^{(i)} - \eta \bar{g}_k\quad \text{if $k$ mod $\tau = 0$,} \\ w_k^{(i)} - \eta g(w_k^{(i)})\quad \text{otherwise.} \end{cases}
\end{align}
Similar to CSGD, the learning rate $\eta$ needs to decrease with $k$ in order to reduce the weight update variance induced by the randomness of the user-generated training samples.
\vspace{5pt}

\begin{figure}\centering
\includegraphics[width= \columnwidth]{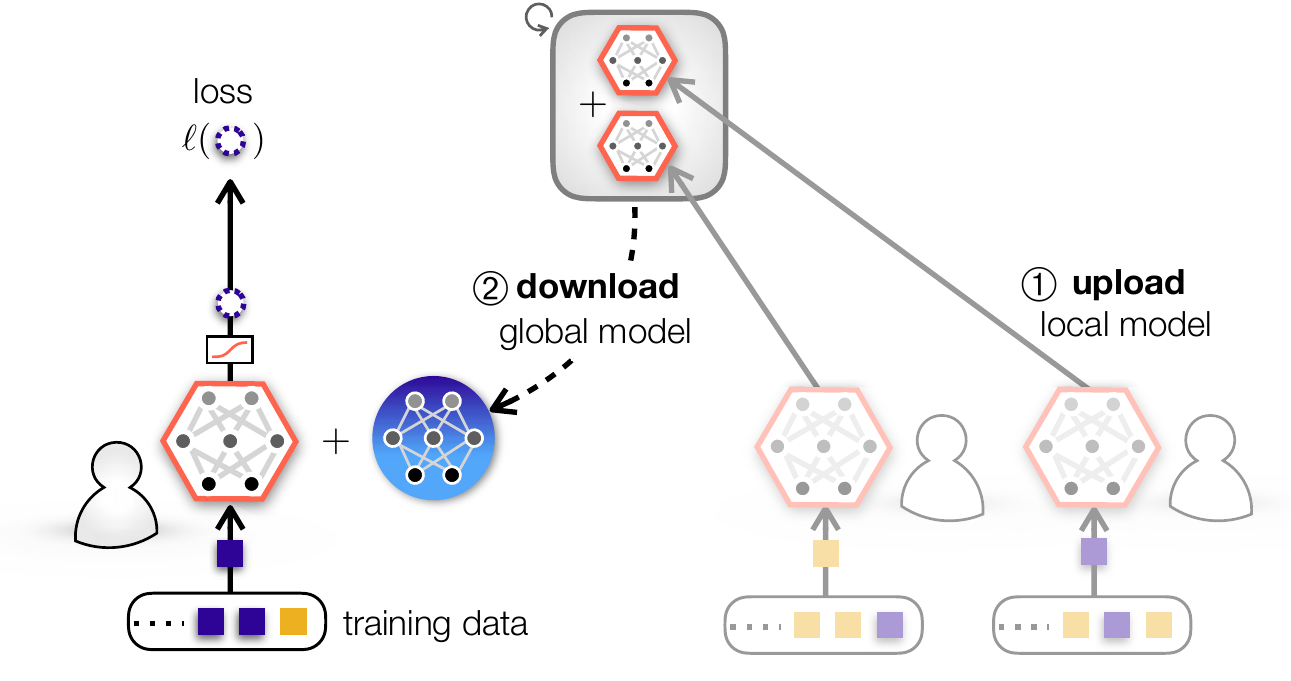}
\caption{Operational structure of federated learning (FL).}
\end{figure}

\noindent2) \textbf{Federated SVRG (FSVRG)}\quad
Allowing a constant learning rate can reduce the training latency compared to FAvg whose learning rate decreases with time. As a variant of FAvg, FSVRG applies the stochastic variance reduced gradient (SVRG) \cite{SVRG_NIPS13} that minimizes the weight update variance by additionally utilizing the difference between the local and global gradients,  allowing a constant learning rate~\cite{pap:jakub16}. 
Besides, similar to ESGD, FSVRG keeps track of the global MSI, and updates it based on the distributed approximate Newton (DANE)~\cite{shamir:2014:dane} that also ensures a constant learning rate. 
This yields the following local and global weight update rules:
\begin{align}
\hspace{-2pt}&w_{k+1}^{(i)} \!=\! \begin{cases} w_k^{(i)} - \eta\(\bar{g}(\hat{w}_k) +  g(w_k^{(i)}) - g(\hat{w}_k) \)  \text{if $k$ mod $\tau \!=\! 0$,}\\
w_k^{(i)} - \eta g(w_k^{(i)})  \quad \text{otherwise,} 
\end{cases}\\[-11pt]
&\text{where } \hat{w}_k = \hat{w}_{k-1} + \sum_{i=1}^M \frac{n_i}{n} \(w_k^{(i)} - \hat{w}_{k-1} \) \text{if $k$ mod $\tau \!=\! 0$}.\nn
\end{align}
 $\bar{g}(\hat{w}_k) = 1/M \sum_{i=1}^{M} g(\hat{w}_k^{(i)})$. The notation $n_i$ is the local training dataset size, and $n$ is the size of the entire devices' aggregate dataset, i.e., global dataset. Like FAvg, FSVRG applies periodic MSI exchanges at an interval of $\tau$, reducing the communication overhead. Nonetheless, FSVRG requires extra MSI weight exchanges, in addition to the gradients of FAvg. Thus, the resulting communication payload size is doubled from FAvg. 

Both FAvg and FSVRG are often referred to as FL that work under non-IID training datasets in practice~\cite{Brendan17,pap:jakub16}, though the accuracy is degraded compared to the best case performance under IID datasets~\cite{ARM18,Jeong:18}.
\vspace{5pt}

\noindent3) \textbf{Co-Distillation (CD)}\quad
While keeping communication overhead the same as in FAvg, CD has the potential to obtain a more accurate model by exploiting extra computation and memory resources during the training phase~\cite{OnlineKD}. In fact, all the aforementioned training methods directly apply the globally averaged MSI to the local MSI update calculation via an ensembling method. Alternatively, one may focus on the fact that after downloading the global average weight MSI evaluated by exchanging local weight MSI, each device can have two separate NN models: its local model and the globally averaged model.~\footnote{The original CD in~\cite{OnlineKD} considers that each device has: (1) its local model and (2) all copies of the other devices' models, which may incur  huge communication overhead. Instead, we replace (2) with the globally averaged model, without loss of generality.} In the next local weight update, the device feeds a training data sample to both models. Then, from the KD point of view, the global model is interpreted as a teacher whose inference output, i.e., knowledge, can be transferred to a student, i.e., the local model. This is enabled by the following weight update rule:
\begin{align}
\hspace{-4pt}w_{k+1}^{(i)}\!=\! \begin{cases}
w_k^{(i)} - \eta\(g(w_k^{(i)}) + \psi(F^{(i)}_k,\hat{F}^{(i)}_k )\) \; \text{if $k$ mod $\tau = 0$,}\\
w_k^{(i)} - \eta g(w_k^{(i)}) \; \text{otherwise}.
\end{cases}
\end{align}
The local model's term $F^{(i)}_k$ is a set of  normalized  logits using a modified softmax function with a temperature hyperparameter $T$, henceforth denoted as a \emph{logit vector}. To illustrate, for the $\ell$-th label's logit $x_\ell$ with  $L$  labels, the normalized value equals $\text{softmax}(x_\ell) = \exp(x_\ell/T)/\sum_{j=1}^{L} \exp(x_j/T)$. Adjusting the temperature $T$ helps the logits to be closely mapped into the per-label inference probabilities, such that the normalized values become identical across all labels or the maximum logit for $T\rightarrow \infty$ or $0$, respectively. Likewise, the global model's logit vector is given as $\hat{F}^{(i)}$. With these local and global logit vectors, the distillation regularizer $\psi(F^{(i)}_k,\hat{F}^{(i)}_k )$ is given as the gradient of their mean squared error~\cite{Ba:NIPS16} or of the cross entropy~\cite{HintonKD:14,OnlineKD}. As such, when the local model's output $F^{(i)}_k$ is close to the global model's output $\hat{F}^{(i)}_k$, i.e. small distillation regularizer, the local weight is determined mostly by the locally calculated gradient, and otherwise perturbed in order not to follow the local bias. In CD, thanks to the periodic communication interval~$\tau$, also known as a checkpoint interval, the communication overhead becomes as small as FAvg. Its downside is consuming extra memory and computation resources for separately storing the global model and performing  inference  using the global model.
\vspace{5pt}

\begin{figure}\centering
\includegraphics[width= \columnwidth]{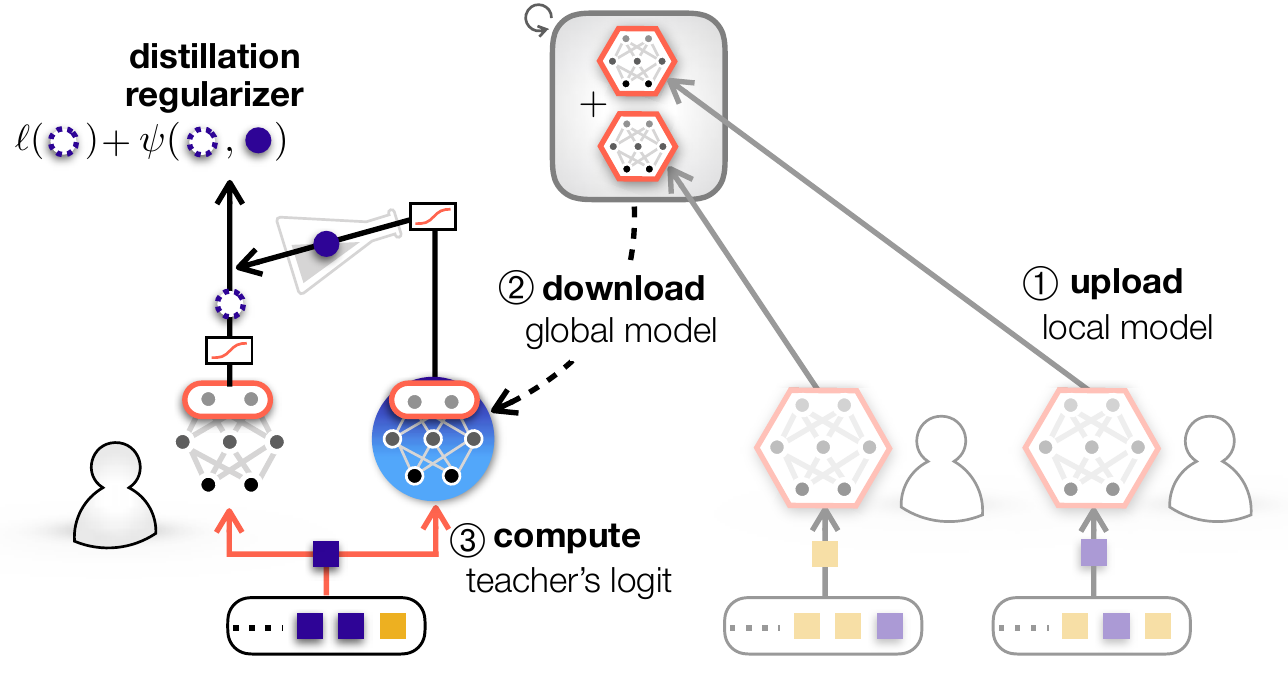}
\caption{Operational structure of co-distillation (CD).}
\end{figure}

\subsubsection{ Output MSI Exchange}
\noindent1) \textbf{Federated Distillation (FD)}\quad
Exchanging the model parameter MSI, i.e., weights and/or gradients, may induce  large communication payload size particularly for a deep NN, since the number of parameters is proportional to the model size. To resolve this, we propose FD that exchanges model output MSI, i.e., normalized logits whose payload size depends only on the output dimension, i.e., the number of labels~\cite{Jeong:18}. The weight update rule is then implemented using KD. The key challenge is that a set of normalized logits, henceforth denoted as a logit vector, is associated with its input training data sample. Therefore, to operate KD between the exchanged global average logit vector and the local model's logit, both logit vectors should be evaluated using an identical training data sample. Unfortunately, synchronous logit vector exchanges as many as the training dataset size brings about significant memory and communication costs. Such communication cost may even exceed the model parameter MSI payload, which is the reason why CD resorts to parameter MSI exchanges. 

To rectify this, each device in FD exchanges a set of mean logits per label, each of which is locally averaged over epochs until a checkpoint. This \emph{local average logit MSI} is associated not with individual data samples but with the accumulated training dataset, enabling periodic local MSI exchanges. The exchanged local MSIs are then averaged across devices, yielding a set of global mean logits per label, i.e., \emph{global average logit MSI}. At the next local weight update phase, each device selects  distillation regularizers which are synchronous with its training data samples. This weight update rule is represented as follows:
\begin{align}
w_{k+1}^{(i)}\!=\! \begin{cases}
w_k^{(i)} - \eta\(g(w_k^{(i)}) + \psi(\bar{F}^{(i)}_{k,\ell},\breve{F}^{(i)}_{k,\ell} )\) \; \text{if $k$ mod $\tau = 0$,}\\
w_k^{(i)} - \eta g(w_k^{(i)}) \; \text{otherwise},
\end{cases}
\end{align}
where $\bar{F}_{k,\ell}^{(i)}$ is the local average logit vector when the training sample belongs to the $\ell$-th ground-truth label. The global average logit vector equals $\breve{F}_{k,\ell}^{(i)}=\sum_{j\neq i}\bar{F}^{(j)}_k/(M-1)$. The exchanged local average logit MSI is the set of per-label local average logit vector for all labels, i.e., $\{\bar{F}_{k,\ell}^{(i)}\}_{\ell=1}^L$, so does the global average logit MSI $\{\breve{F}_{k,\ell}^{(i)}\}_{\ell=1}^L$.

\begin{figure}\centering
\includegraphics[width= \columnwidth]{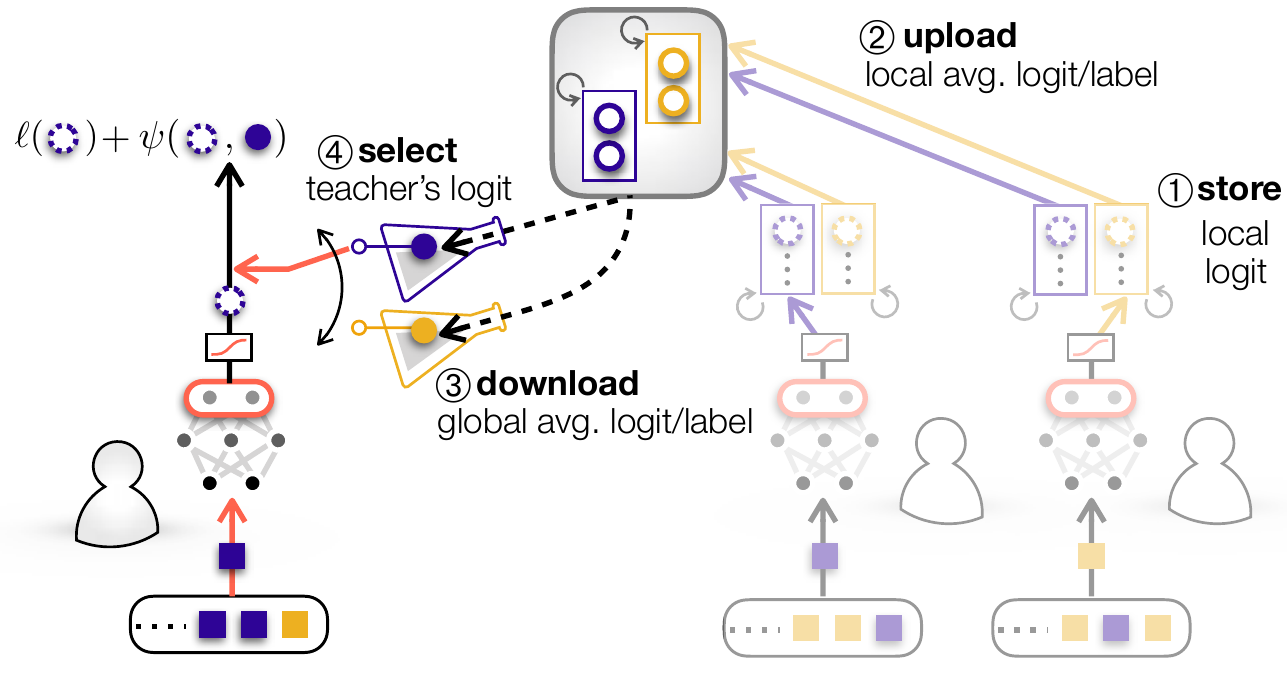}
\caption{Operational structure of federated distillation (FD).}
\end{figure}

\begin{table*}[]
\caption{Comparison of MSI exchanges in centralized ML and edge ML. Compared to CSGD as a baseline scheme, the advantages of each method appear {blodfaced}, whereas the disadvantages are {italicized}.}
\centering
\renewcommand{\arraystretch}{1.2}
\resizebox{\textwidth}{!}{\begin{tabular}{l  c c c c c c c c c c c}
\toprule
\multicolumn{3}{c}{\quad\quad\quad\quad\quad Centralized ML}& \multicolumn{6}{c}{\textbf{Edge ML}}& \multicolumn{2}{c}{MARL via \textbf{Edge ML}}\\\cmidrule(r){2-3} \cmidrule(r){4-10} \cmidrule(r){11-12}
& {CSGD~\cite{Synch_sgd,Zinkevich:10,Jin:16,Dean:12}} & {ESGD~\cite{Zhang:15}} & \textbf{FAvg}~\cite{Brendan17} & \textbf{FSVRG}~\cite{pap:jakub16}  & \textbf{CD}~\cite{OnlineKD} & \textbf{FD}~\cite{Jeong:18}  & \textbf{FJD} & \textbf{DSGD}~\cite{Lian:17} & \textbf{GADMM}~\cite{Anis:NeurIPS19} & \textbf{FRL}~\cite{WangHan:18} & \textbf{FRD}  \\\midrule
Split  &  {m-d}       & {m-d}   &  h-d    &  h-d     & h-d   &  h-d  &  h-d  & d-d & d-d    & d-d & d-d   \\[5pt]

Tx. MSI      & grad. & weight& \begin{tabular}[c]{@{}c@{}}grad.\\ (\textbf{periodic})\end{tabular}& \begin{tabular}[r]{@{}c@{}}grad., \emph{weight}\\ (periodic)\end{tabular} &  weight  &  \begin{tabular}[c]{@{}c@{}}local avg.\\ \textbf{logit/label} \end{tabular}  & \begin{tabular}[c]{@{}c@{}}local avg.\\ \textbf{Jacob./label}\end{tabular} & \begin{tabular}[c]{@{}c@{}}weight\\ (\textbf{neighbor})\end{tabular} & \begin{tabular}[c]{@{}c@{}}weight\\ (\textbf{neighbor})\end{tabular}  &  grad.  &  \begin{tabular}[c]{@{}c@{}}local\\ Q/\emph{state}\end{tabular}\\[10pt]

Rx. MSI      & \begin{tabular}[c]{@{}c@{}}sample,\\avg. grad.\end{tabular}& \begin{tabular}[c]{@{}c@{}}sample,\\avg. weight\end{tabular}     & \begin{tabular}[c]{@{}c@{}}avg. grad.\\(\textbf{periodic})\end{tabular}     &  \begin{tabular}[c]{@{}c@{}}avg. grad., \emph{weight}\\(periodic)\end{tabular}  & \begin{tabular}[c]{@{}c@{}}avg. weight\\(periodic)\end{tabular}   & \begin{tabular}[c]{@{}c@{}}global avg.\\\textbf{logit/label}\end{tabular}   & \begin{tabular}[c]{@{}c@{}}global avg.\\\textbf{Jacob./label}\end{tabular}   & \begin{tabular}[c]{@{}c@{}}weight\\(\textbf{neighbor})\end{tabular} & \begin{tabular}[c]{@{}c@{}}weight\\ (\textbf{neighbor})\end{tabular} & \begin{tabular}[c]{@{}c@{}} avg. grad.\\(periodic)\end{tabular} & \begin{tabular}[c]{@{}c@{}}global avg.\\Q/\emph{state}\end{tabular}\\[10pt]

Comp.     &    -     &  -    &   -        &  -     &  \emph{inference}  & -   & \emph{Jacob.}   & -  & \emph{optimization} &  -  & -\\[5pt]

Mem.    & -        &  \emph{previous weight}    &  -    &  \emph{previous  weight}      & -  &  \textbf{logit/label, $\downarrow\downarrow$}  & \textbf{Jacob./label, $\downarrow$}  &- &  -
 & - &  Q/\emph{state}\\
\bottomrule
\end{tabular}}
\end{table*}

The performance of FD can further be improved with a slight modification. In fact, model output accuracy increases as training progresses. Thus, it is better to use a \emph{local weighted average logit MSI}, where the weight increases with time. Alternatively, one can implement FD by only exchanging the most mature knowledge. For instance, consider that  devices share a knowledge test dataset a priori. Just before each checkpoint, using the test set, each device measures its latest inference knowledge. Then, FD is enabled by exchanging \emph{local checkpoint logit MSI}, a set of local logit vectors obtained by the test dataset, each of which corresponds to a single ground-truth label. Exchanging the local MSI.
\vspace{5pt}

\noindent2) \textbf{Federated Jacobian Distillation (FJD)}\quad
Model input-output Jacobian matching is interpreted as a KD operation that inserts noise into logits~\cite{Srinivas:ICML18}. Since noisy logit exchanges improves the KD performance when the noise level is properly adjusted~\cite{Sau:16}, we expect exchanging Jacobian matrices to improve FD. With this motivation, the weight update rule of FJD is given as:
\begin{align}
w_{k+1}^{(i)}\!=\! \begin{cases}
w_k^{(i)} - \eta\(g(w_k^{(i)}) + \psi(\bar{G}^{(i)}_{k,\ell}, \breve{G}^{(i)}_{k,\ell}) \) \; \text{if $k$ mod $\tau \!=\! 0$,}\\
w_k^{(i)} - \eta g(w_k^{(i)}) \; \text{otherwise}, \label{Eq:CJD}
\end{cases}
\end{align}
In the distillation regularizer, $\bar{G}^{(i)}_{k,\ell}$ is the local average of the logit vector's Jacobian when the input sample's ground-truth belongs to the $\ell$-th label, and $\breve{G}^{(i)}_{k,\ell}$ is the global average of local average logit Jacobian vectors for the $\ell$-th label. Following FD, the local and global MSIs are given as $\{\bar{G}_{k,\ell}^{(i)}\}_{\ell=1}^L$ and $\{\breve{G}_{k,\ell}^{(i)}\}_{\ell=1}^L$, respectively. Compared to FD, the disadvantage is that FJD requires extra memory and computation resources for storing and computing the Jacobian matrices. Another burden is that Jacobian matrix communication leads to the payload size being proportional to the output dimension multiplied by the input dimension, which is nonetheless still independent of the model size.
\vspace{5pt}

\subsubsection{Uplink-Downlink Asymmetric MSI Exchange}
Due to the device's limited transmission power, the uplink communication is likely to be slower than the downlink communication speed~\cite{JHParkTWC:15}. To reflect this difference, as demonstrated in \cite{Park:2019:FLlet}, the local MSI to be uploaded from each device to the helper can be FD that minimizes the payload size. By contrast, the global MSI to be downloaded to the devices can be the entire model parameters, as used in FSVRG, which may lead to the largest payload size that contains the largest knowledge. Since this global model parameter MSI is not consistent with the uploaded model output MSI, one needs to reconstruct a global model from the model outputs, which can be done via the KD operations with a test dataset as exemplified in FD. Consuming the helper's extra computation resource and time can be justified so long as the computation cost is cheaper than the communication cost.

\subsubsection{Device-to-Device MSI Exchange}
\noindent1) \textbf{Distributed Parallel SGD (DSGD)}\quad
Exchanging only with with neighboring devices can reduce the communication overhead. Namely, with the d-d split, the local MSI of DSGD is each device's model weights~\cite{Lian:17,Jiang:2017_7172}, and its weight update is represented~as:
\begin{align}
w_{k+1}^{(i)} =  \tilde{w}_k - \eta g(w_k^{(i)}),
\end{align}
where the global MSI $\tilde{w}_k$ is the average weight among the communicating devices, i.e., $\tilde{w}_k= \sum_{j=1}^M a_{ji} w_k^{(j)} /{\sum_{j=1}^M w_{ji}}$. A pre-defined mixing matrix determines with whom to exchange the local MSIs, which has the element $w_{ji}=1$ if the $j$-th device communicates with the $i$-th device, and otherwise we obtain $w_{ji}=0$. When the device indices follow their physical locations, neighboring communication topology is characterized by a diagonally clustered weight matrix.
\vspace{5pt}

\noindent2) \textbf{Group ADMM (GADMM)}\quad
Without the aid of any central entity, GADMM exchanges model weights with neighboring devices, and achieves fast training convergence with much less communication rounds~\cite{Anis:NeurIPS19,Anis:2019:GC}. The key idea is to apply the Alternating Direction Method of Multiplier (ADMM) algorithm after grouping devices into head and tail devices, such that each device in the head group $\mathcal{N}_h$ is connected to two neighboring devices in the tail group $\mathcal{N}_h$. In GADMM, the weights of devices belonging to the same group are updated in parallel, but the weights of devices belonging to different groups are updated in an alternating fashion. When odd and even superscripts denote head and tail devices respectively, for a loss function $\ell(\cdot)$, GADMM updates primal variables (i.e., weights) and dual variables (\small$\lambda_{n-1}$\normalsize~and \small$\lambda_n$\normalsize) as follows.

\begin{itemize}[leftmargin=12pt]
	\item Each head device updates its primal variables~as:

\vspace{-10pt}{\small\begin{align}
&\hspace{-8pt}{w}^{(i \in {\cal N}_h)}_{k+1}\!\! = \arg\!\min_{w^{(i)}_k} {\lambda_k^{(i-1)}} (w_{k}^{(i-1)} \!\!-\! w_{k}^{(i)})+{\lambda_k^{(i)}}(w_{k}^{(i)} \!\!-\! w_{k}^{(i+1)}) \nonumber\\
&\quad+ \frac{\rho}{2}\( \| w_{k}^{(i-1)} \!\!-\! w_k^{(i)}\|_2^2+  \| w_k^{(i)} \!\!-\! w_k^{(i+1)}\|_2^2\) + \ell(w^{(i)}_k).
\label{headUpdate}
\end{align}\normalsize}
These updates are sent to two tail neighbors.

\item Next, each tail device updates its primal variables~as:

\vspace{-10pt}{\small\begin{align}
&\hspace{-8pt}{w}^{(i \in {\cal N}_t)}_{k+1}\!\! = \arg\!\min_{w^{(i)}_k} {\lambda_k^{(i-1)}} (w_{k+1}^{(i-1)} \!\!-\! w_{k+1}^{(i+1)})+{\lambda_k^{(i)}}(w_{k}^{(i)} \!\!-\! w_{k+1}^{(i+1)}) \nonumber\\
&\quad+ \frac{\rho}{2}\( \| w_{k+1}^{(i-1)} \!\!-\! w_k^{(i)}\|_2^2+  \| w_k^{(i)} \!\!-\! w_{k+1}^{(i+1)}\|_2^2 \) + \ell(w^{(i)}_k).
\label{tailUpdate}
\end{align}\normalsize}
These updates are sent to two head neighbors.

\item Finally, every device updates its dual variables as:

\vspace{-10pt}{\small\begin{align}
{\lambda}_{k+1}^{(i)}={\lambda}_k^{(i)} + \rho({w}_{k+1}^{(i)} - {w}_{k+1}^{(i+1)}).
\label{lambdaUpdateb}
\end{align}\normalsize}
\end{itemize}

\noindent Consequently, each device communicates with only two neighbors to update its own weights, while only half of devices broadcast their weights per communication round.

\vspace{5pt}
\noindent3) \textbf{Federated Reinforcement Learning (FRL)}\quad
The mixing matrix based MSI exchange in DSGD is applicable for the MSI exchange in MARL that  follows the d-d split. 
Then, exploiting FL further improves the communication efficiency, leading to FRL. To illustrate, consider the policy gradient method in MARL, where each agent's policy is trained using SGD with an NN \cite{bansal:2018:emergent}. Following either FAvg or FSVRG, the agents identified in a mixing matrix collaboratively train an ensemble policy by periodically exchanging the NN's model parameters. A similar procedure is applicable to deep Q-learning in MARL, where each agent approximates the Q values using an NN, as demonstrated in~\cite{WangHan:18}.
\vspace{5pt}

\noindent4) \textbf{Federated Reinforcement Distillation (FRD)}\quad
FD is applicable for the MSI exchange in MARL with the d-d split, leading to FRD.
For Q learning in MARL, the agents identified in a mixing matrix collectively predict a set of Q values and their associated states.
The MSI exchange and weight update rule follow FD, by replacing its normalized logits and labels with the Q values and the states, respectively. Similarly, a policy-based method in MARL can be improved in combination with FD. For actor-critic RL, FD is applicable to either one of the actor (policy) NN or the critic (value) NN or to both NNs~\cite{Park:2019:FDR}. However, with large input state and/or output dimensions, these approaches may incur huge communication and memory costs. To resolve this issue, states and/or actions can be compressed based on their correlations. To illustrate its effectiveness, consider the Atari gaming environment \cite{atari:2013} whose output action dimension is only $\{\text{up, down, left, right}\}$, whereas the input state dimension is the entire pixels per frame. Since neighboring pixels are highly correlated, one can reduce the input dimension by grouping multiple neighboring raw states as a single proxy state that is mapped into the average action of the raw states.

All the aforementioned MSI exchange methods are summarized in Table~1. On top of these methods, one can further enhance communication efficiency, by additionally quantizing gradients \cite{NIPS2017_Alistarh}, removing insignificant gradients, i.e., sparsification \cite{NIPS2018_Sebastian}, opportunistic uploading based on gradient magnitudes \cite{NIPS2018_Tianyi}, and adaptively adjusting the communication intervals \cite{Wang:2019:AFL,FL_Nishio}. For more details on state dimensionality reduction and advanced FL and FD frameworks in both supervised learning and RL, the readers are encouraged to check~\cite{Park:2019:FLlet,Smith:FLSurvey,Yang:FLSurvey}.

\vspace{5pt}

\small\begin{yellowbox}{\textbf{MSI Exchange} Related Theory}
\begin{itemize}[leftmargin=5pt]
	\item \textbf{Optimal Regularization via SGLD}. 
	Distillation-based MSI exchange methods rely on inserting a noise proportionally to the knowledge gap between the teacher and student NNs. The noise amount is adjustable via the temperature parameter $T$ when normalizing the teacher's logit values, and can further be optimized via SGLD. For a fixed learning rate $\eta$, a recent work \cite{Smith:2017:Bayes} shows that the noise amount maximizing the test accuracy in centralized ML is characterized by the optimal noise scale $g=\eta(n/B-1)$ of the SGLD's noise $\varepsilon_t$ (see Sect. 4,1), for $n$ training samples and batch size $B$. The definition of $g$ comes from the autocorrelation of $\varepsilon_t$, given as $\E[\varepsilon_{t+\tau} \varepsilon_t] = g F(w)\delta_\tau$, where $F(w)$ is a matrix describing the covariance of gradients, and $\delta_\tau$ is the Dirac delta function of $\tau$.
	\vspace{5pt}

	\item \textbf{Wasserstein Distillation Loss}. Cross entropy is widely utilized as the distillation regularizer, which is decomposed into entropy and KL divergence terms. Due to the KL divergence, the distillation regularizer may diverge, particularly when the teacher and students' logits are too peaky to have an overlapping support. A quick fix is to increase the teacher's temperature $T$ for smoothing its logits, at the cost of compromising the accuracy of measuring the knowledge gap (see. Fig.~11). Instead, Wasserstein distance (see Sect. 4.1) can be a suitable regularizing function for such cases.
	\end{itemize}
\end{yellowbox}\normalsize

\subsection{Computation-Aware Latency Reduction}

Compressing model parameters during training operations is effective in the latency reduction, as demonstrated by high-accuracy low-precision training (HALP) that adjusts the arithmetic precision based on the training dynamics~\cite{C.-D.-Sa-et.-al.:2018aa}. Processing the training operations together with other incumbent applications is another viable solution, in which their operation scheduling is optimized as done in MEC~\cite{Cao:2018:MEC} and in the context of exploration-exploitation trade-off~\cite{McInerney:2018:EEE}. 

Furthermore, compressing the NN model after completing the training process reduces the inference latency. Such compression can be co-designed with hardware and computational characteristics such as the energy consumption, compression ratio, and the model parameters' frequency of use, as studied in energy-based pruning \cite{T.-J.-Yang-Y-H.-Chen-and-V.-Sze:2017aa}, Viterbi-based compression~\cite{D.-Lee-D.-Ahm-T.-Kim-P.-I.-Chuang-and.-J-J.-Kim:2018aa}, and Deep compression~\cite{Han:16}, as detailed~next.
\vspace{5pt}

\noindent1) \textbf{Adaptive-Precision Training}\quad
 In mixed-precision training \cite{S.-Narang-et.-al.:2018aa}, a low-precision representation consists of an exponent $\delta$ and mantissa $b$, and the tuple $(\delta,b)$ can express a numerical value within the range $\{-\delta 2^{{b-1}},\cdots, -\delta, 0, \delta, \cdots, \delta(2^{{b-1}}-1)\}$ by only using $(\delta + b + 1)$ bits, where the last $1$-bit is allocated for the sign. Here, the lower precision (i.e., the more compression), the higher quantization noise that increases gradient variance during the training process. To optimize the compression-distortion trade-off, 
high-accuracy low-precision training (HALP) is a viable solution~\cite{C.-D.-Sa-et.-al.:2018aa}. This firstly utilizes SVRG~\cite{SVRG_NIPS13} for reducing the gradient variance. Besides, HALP applies a bit-centering technique that dynamically re-centers and re-scales the low-precision numbers by setting $\delta= g(\hat{w}_{k})/[\mu (2^{{b-1}}-1)]$ for a $\mu$-strongly convex loss function where its gradient $g(\hat{w}_{k})$ is defined in Sect. 5.2.1. This lowers the quantization noise asymptotically as the training converges, thereby achieving the accuracy of a full-precision SVRG. Furthermore, while slightly compromising accuracy, one can represents the gradients only using their ternary directions $\{-1,0,1\}$ \cite{wen:2017_6749}, minimizing the memory usage.
\vspace{5pt}

\noindent2) \textbf{Application-Training Co-Processing}\quad
In edge ML, a device is likely to perform both the NN training and its end application processing simultaneously. 
Let us recall the real-time AR/VR application example in Sect.~\ref{sec:significance}, where each headset device predicts the future gaze direction, thereby pre-rendering  future visual frames. 
In this case, the NN training and the rendering processes are simultaneously performed at the device, and  the device's computing energy allocation needs to be optimized under the widely-known exploration-exploitation trade-off \cite{McInerney:2018:EEE}. 
Furthermore, with the helper-device split, a part of the demanding rendering processes can be offloaded from the device to the helper that also participates in the NN training process. 
At the helper side, its computation energy has  to be optimized, as investigated in the context of multi-use MEC \cite{Cao:2018:MEC}.
\vspace{5pt}

\noindent3) \textbf{Hardware-Efficient Compression}\quad
After a training process, compressing the model reduces the memory usage. In this respect, one can prune the model via DropOut and/or DropConnect based on e.g., Fisher information of each node~\cite{Theis:18}. Alternatively, one can optimize the pruning process based on energy consumption \cite{T.-J.-Yang-Y-H.-Chen-and-V.-Sze:2017aa}, as illustrated in~Fig.~16 In energy-based pruning, it first estimates the energy consumption per layer, and then prune the weights within each layer in order of energy.

After these pruning processes, the resulting weight parameters are expressed as a sparse weight matrix where there exist only a few non-zero values in a large-sized matrix, which can be compressed using, e.g., the compressed sparse row (CSR) format. The compression rate depends on the pruning process, and therefore a compression-rate optimal pruning is needed. Exploiting the Viterbi algorithm is useful for this purpose, guaranteeing a constant maximum compression rate~\cite{D.-Lee-D.-Ahm-T.-Kim-P.-I.-Chuang-and.-J-J.-Kim:2018aa}. On the other hand, in Deep compression~\cite{Han:16}, the weights after pruning are quantized and clustered, yielding a set of shared weights as shown in Fig.~16. Afterwards, the shared weights are further compressed using Huffman coding that allocates more bits, i.e., longer codeword length, to the shared weights that appear more frequently.

\begin{figure}\centering
\includegraphics[width= .7\columnwidth]{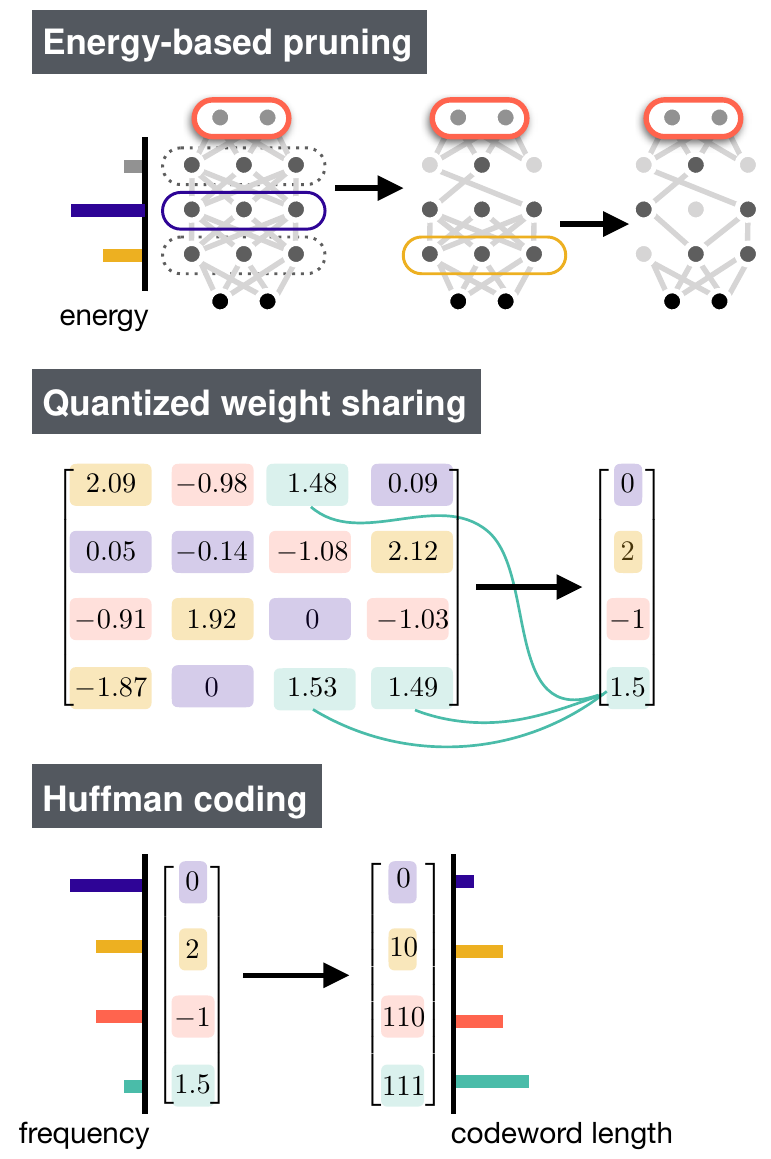}
\caption{Illustration of hardware-efficient model compression via energy-based pruning, quantized weight sharing, and Huffman coding.}
\end{figure}

\subsection{Scalability Enhancement}

Modern deep NN architectures often have too large depths to be stored at mobile devices. This calls for splitting a NN into segments distributed over multiple devices, as shown in Fig.~9(d). Furthermore, the range of federation in edge~ML is constrained by the battery levels and privacy requirements of mobile devices, and the effectiveness of federation is delimited by its non-IID training dataset. These challenges and their suitable solutions are described as follows.
\vspace{5pt}

\noindent1) \textbf{Stacked Model Split}\quad
In the model split, the data may flow back and forth between the split model segments. The resulting dependency among the segments stored in multiple devices obstructs the parallelism of local training. To minimize such dependency, one can start from the original model comprising multiple stacks of components that can be easily parallelized. 
A suitable example can be a discriminator distributed GAN \cite{hardy:2018:MDGAN} whose discriminators can be distributed over multiple devices. 
In the opposite way, multiple generators can be distributed over the devices that share a single discriminator \cite{Hoang:2017:MultiGeneratorGA}. Both cases are illustrated in Fig.~17.
Another example is an DBN that comprises multiple stacked RBMs  distributed over several devices. More general model split based edge ML frameworks are discussed in \cite{Vepakomma:2018:Splita,Vepakomma:2019}, in the context of health applications with private patient data.
\vspace{5pt}

\begin{figure}\centering
\hspace{-10pt}\subfigure[Multiple generators.]{\includegraphics[width=.4\columnwidth]{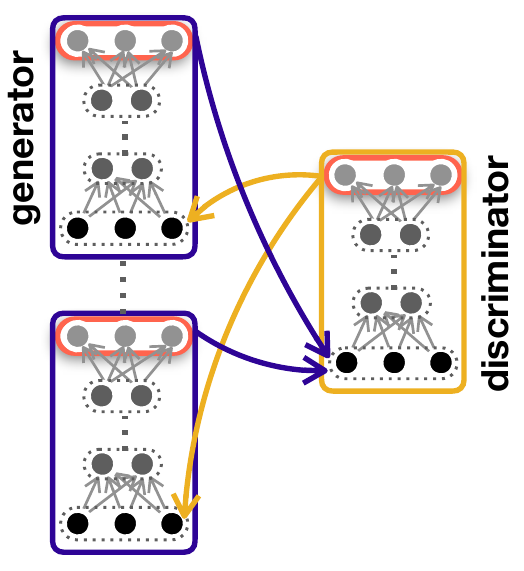}}\hspace{10pt}
\subfigure[Multiple discriminators.]{\includegraphics[width=.4\columnwidth]{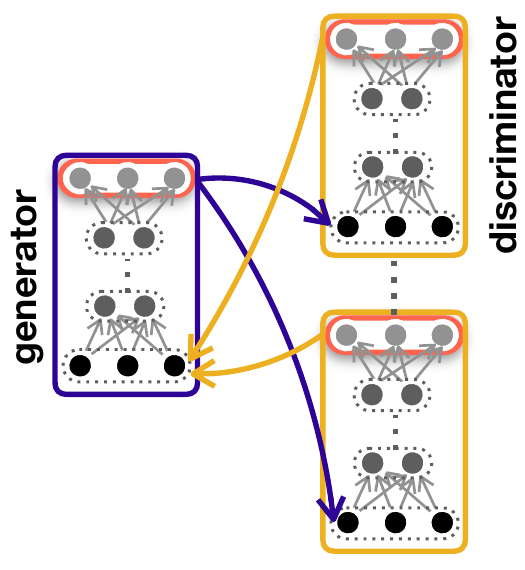}}
\caption{Illustration of a distributed GAN with: (a) multiple generators and (b) multiple discriminators.}
\end{figure}

\noindent2) \textbf{ODE-based Training}\quad
Traditional ML cannot dynamically change the model size during the training process, so is difficult to flexibly adjust the energy consumption.
This is critical for mobile devices whose battery level is limited and fluctuate over time, which hinders adopting edge ML into more devices. One promising solution for this bottleneck is to train an NN by solving an ordinary differential equation (ODE)~\cite{Chen:18}. 
In this framework, adjusting the model size is recast as changing the number of evaluations that can be easily feasible in practice.
Furthermore, compared to SGD rooted in convex optimization methods, ODE-based trainin  is able to directly solve a non-convex optimization, avoiding local minima issues as addressed in Sect.~\ref{subsec:BB_training_prin_pract}. 
To this end, one can utilize an algorithmic approach by exploiting the feed forward dynamics from the $l$-th layer to the next layer with a generic loss function~$\mathcal{L}$.
\begin{align}
x_{l+1} &= x_l + \mathcal{L}(x_l, w_l), \quad l\in\{0,\cdots, L-1\} \label{Eq:FFdynamics}
\end{align}
If the number $L$ of layers is sufficiently large, \eqref{Eq:FFdynamics} is approximated as the following ODE
\begin{align}
d x(l)/dl &= \mathcal{L}(x(l), w(l)), \quad l\in[0,L]. \label{Eq:ODE}
\end{align}
This ODE can be numerically solved via the Euler's approximation method with the approximation accuracy proportional to the number of evaluations \cite{E:2017:HJB}. 
\vspace{5pt}

\small\begin{yellowbox}{\textbf{ODE-based Training} Related Theory}
\begin{itemize}[leftmargin=5pt]
	\item \textbf{MF Controlled Training}. 
	One can analytically solve \eqref{Eq:ODE} with the empirical loss function with a regularizer~$R$.
\begin{align}
\mathcal{L}(w) &= \frac{1}{n} \sum_{i=1}^n \[\Phi(x_L^i,y_o^i) + \int_0^L R(x_l^i,w_l)dt \]
\end{align}
Here, the weight parameters $w_l$ is shared by a number $n$ of the training data samples. So long as $n$ is sufficiently large, according to mean-field control theory, the minimized loss, i.e., value function, satisfies \emph{Pontryagin's maximum principle (PMP)}. The optimal solution thereby guarantees a necessary condition that is recast as maximizing the Hamiltonian:
\begin{align}
H(x_o,w) = -\nabla H(x_o,w) \mathcal{L}(x,w) - \mathcal{R}(x,w). \label{Eq:Hamilton}
\end{align}
Particularly when $H(x_o,w)$ is strongly concave, by setting $T\rightarrow 0$, this solution guarantees the global optimum.
	\end{itemize}
\end{yellowbox}\normalsize
\vspace{10pt}

\noindent3) \textbf{Private Data Augmentation}\quad
Due to the user-generated data samples, training dataset across devices can be non-IID in edge ML, which severely degrades the benefit of distributed training. A simple example is the situation when all devices have identical data samples, i.e., fully correlated. In this case, the global and local MSI become identical, negating the diversity gain from distributed training. At the opposite extreme, if all the datasets are entirely not correlated, then reflecting the global MSI in the local weight update is no more than inserting a randomly noisy regularizer. Data augmentation can render such a non-IID dataset amenable to distributed learning. One possible implementation is \emph{partially exchanging} the other devices' data samples. In fact, an experimental study \cite{ARM18} has shown that FL under a non-IID dataset achieves only 50\% inference accuracy compared to the case under an IID dataset, which can be restored by up to 20\% via randomly exchanging only 5\% of the devices' local training data samples. Another way is \emph{locally augmenting} data samples. This is viable, for instance, by a generative model that is capable of generating all samples, which can rectify the Non-IID dataset towards achieving an IID dataset across devices.

Both approaches require access to data samples owned by other devices, thus necessitating privacy guarantee. Data samples can be exchanged while preserving privacy via DP by partially inserting noise and/or redundant data samples. Local data oversampling needs to exchange the local data sample distribution to collectively construct the entire dataset distribution that is to be compared with the local data sample distributions. Such distribution information including any excess or shortage of data samples per label, e.g., per medical checkup item, may easily reveal private sensitive information, e.g, diagnosis result. A GAN-based solution \cite{Jeong:18} for this case is elaborated in Sect.~\ref{CS_FAug}.

\section{Case Studies}\label{sec:case_studies}

From the standpoint of CML, this section aims at demonstrating the effectiveness of the proposed theoretical and technical solutions in edge ML. Several use cases that follow MLC are also introduced at the end, while addressing their connections to MLC.

\subsection{Federated Learning with EVT for Vehicular URLLC}\label{CS_ExtFL}

EVT-based FL enables URLLC in vehicular communication networks as discussed in our preliminary study~\cite{FL_v2x}. 
EVT provides an analytical parametric model to study the tail distribution of queue lengths at vehicular transmitters over the whole network.
By combining the parametric model from EVT with FL, referred to as ExtFL, the individual vehicles learn the tail distribution of queue lengths over the network without a need of exchanging the local queue length samples.
Therein, the key advantage of FL is the reduction of communication payload during the model training compared to a centralized training model relying on exchanging the training samples.
In this regard, the impact of communication latency for training on the vehicular-to-vehicular (V2V) communication is reduced.

\begin{figure}\centering
	\includegraphics[width= \columnwidth]{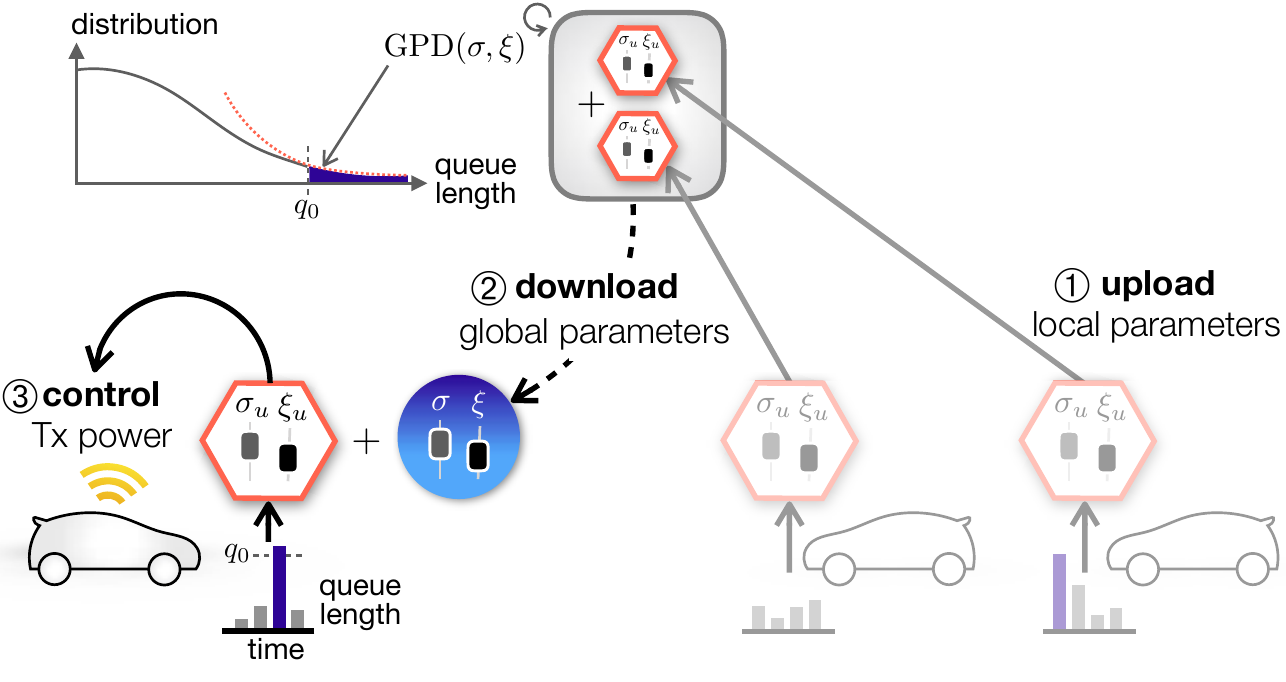}
	\caption{Operational structure of EVT parametric FL (ExtFL).}
	\label{fig:v2v_fl_mle}
\end{figure}

The objective is to minimize the network-wide power consumption of a set $\vueSet$ of vehicular users (VUEs) while ensuring low queuing latencies with high reliability. 
Yet there exists worst-case VUEs who are experiencing high latencies with a low probability.
In this regard, extreme events pertaining to vehicles’ queue lengths exceeding a predefined threshold with non-negligible probability is considered to capture the performance losses of worst-case VUEs.
Using the principles of EVT, the tail distribution of the queue lengths exceeding a predefined threshold are characterized by a generalized Pareto distribution $\gpdML(\cdot)$ with two parameters $\gpdCombined = [\gpdScale,\gpdShape]$ scale and shape, respectively.
The knowledge of the tail distribution over the network is utilized to optimize the transmit power of each VUE to reduce the worst-case queuing delays.

To estimate the queue tail distribution using the queue length samples $\{\sampleSet_{\vue}\}_{\vue\in\vueSet}$ observed at each VUE $\vue$, using the concepts in maximum likelihood estimation (MLE), a cost function is defined as follows:
\begin{align}\label{eqn:likelihood_local}
\loglikelihood(\sampleSet)
= \frac{1}{\sum_{\vue} \setSize{\sampleSet_{\vue}}} \sum_{\vue\in\vueSet}\sum_{\sample\in\sampleSet_{\vue}} \log \gpdML(\sample)
= \sum_{\vue\in\vueSet} \sampleSizeRatio \loglikelihood(\sampleSet_{\vue}),
\end{align}
where $\sampleSizeRatio = \frac{\setSize{\sampleSet_{\vue}}}{\sum_{\vue} \setSize{\sampleSet_{\vue}}}$.
The operation of ExtFL is visualized in Fig. \ref{fig:v2v_fl_mle}, and summarized as below.
\begin{enumerate}
	\item VUE $\vue$ uses $\loglikelihood_{\vue} = \loglikelihood(\sampleSet_{\vue})$ to evaluate $\gpdCombined_{\vue}$ and $\grad{\gpdCombined}{\loglikelihoodSP{\gpdCombined_{\vue}}_{\vue}}$ locally, where $\gpdCombined_{\vue}$ is the local estimate of $\gpdCombined$ at VUE $\vue$.
	Then, the local learning \emph{model} $\localmodel$ is uploaded to the road-side unit (RSU).
	\item  RSU does the model averaging and shares the global model $\globalmodel$ with the VUEs.
	\item VUEs use the global parameters to model the tail distribution of queue lengths and utilize it to control their transmit powers.
\end{enumerate}

\begin{figure}
	\centering
	\subfigure[The amount of data exchanged between RSU and VUEs (left) and the achieved reliability (right).]{
		\includegraphics[width=.46\columnwidth]{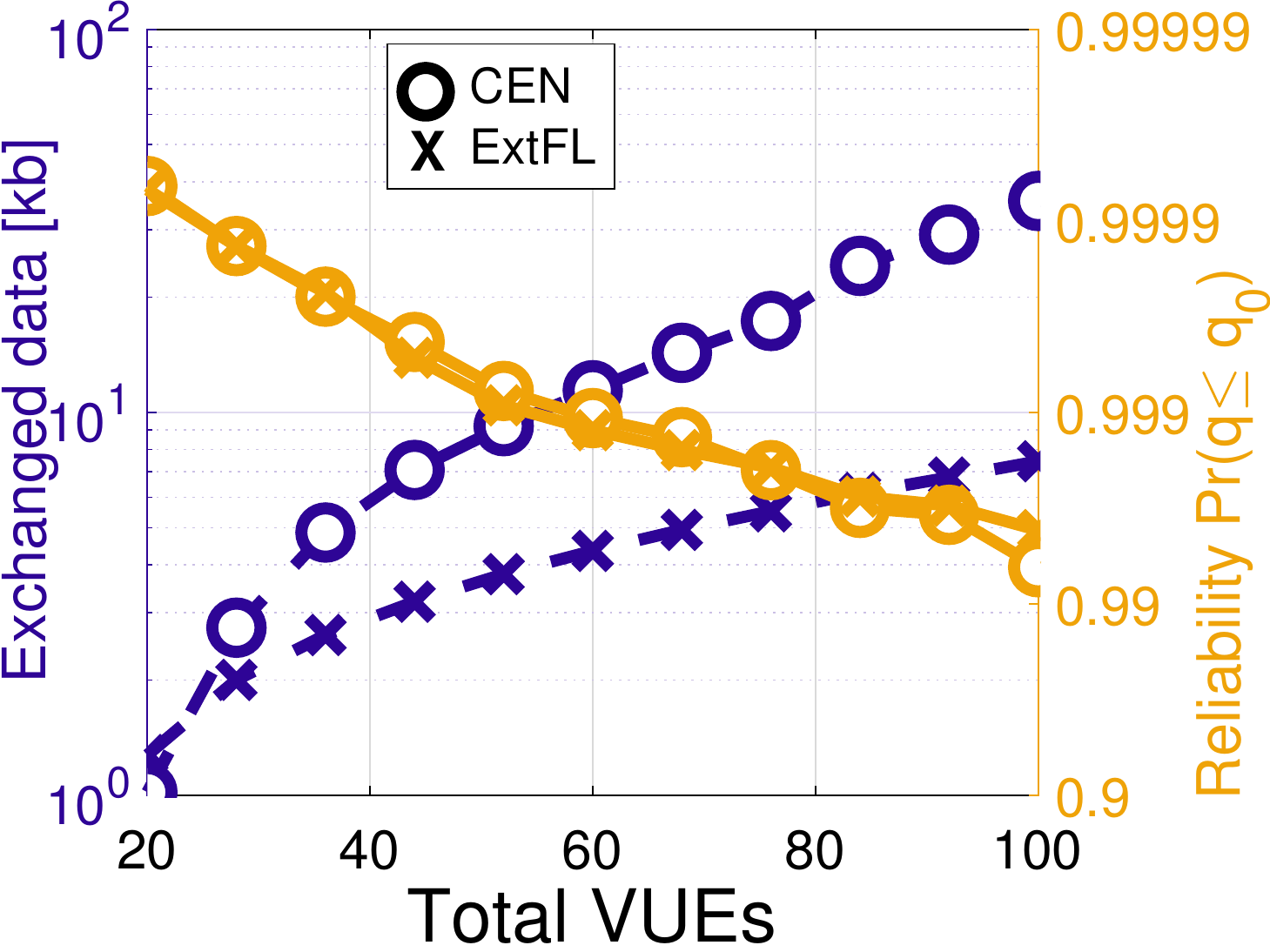}
		\label{fig:FLvsCEN}}
	\subfigure[The mean and the variance of the worst-case VUE queuelengths.]{
		\includegraphics[width=.46\columnwidth]{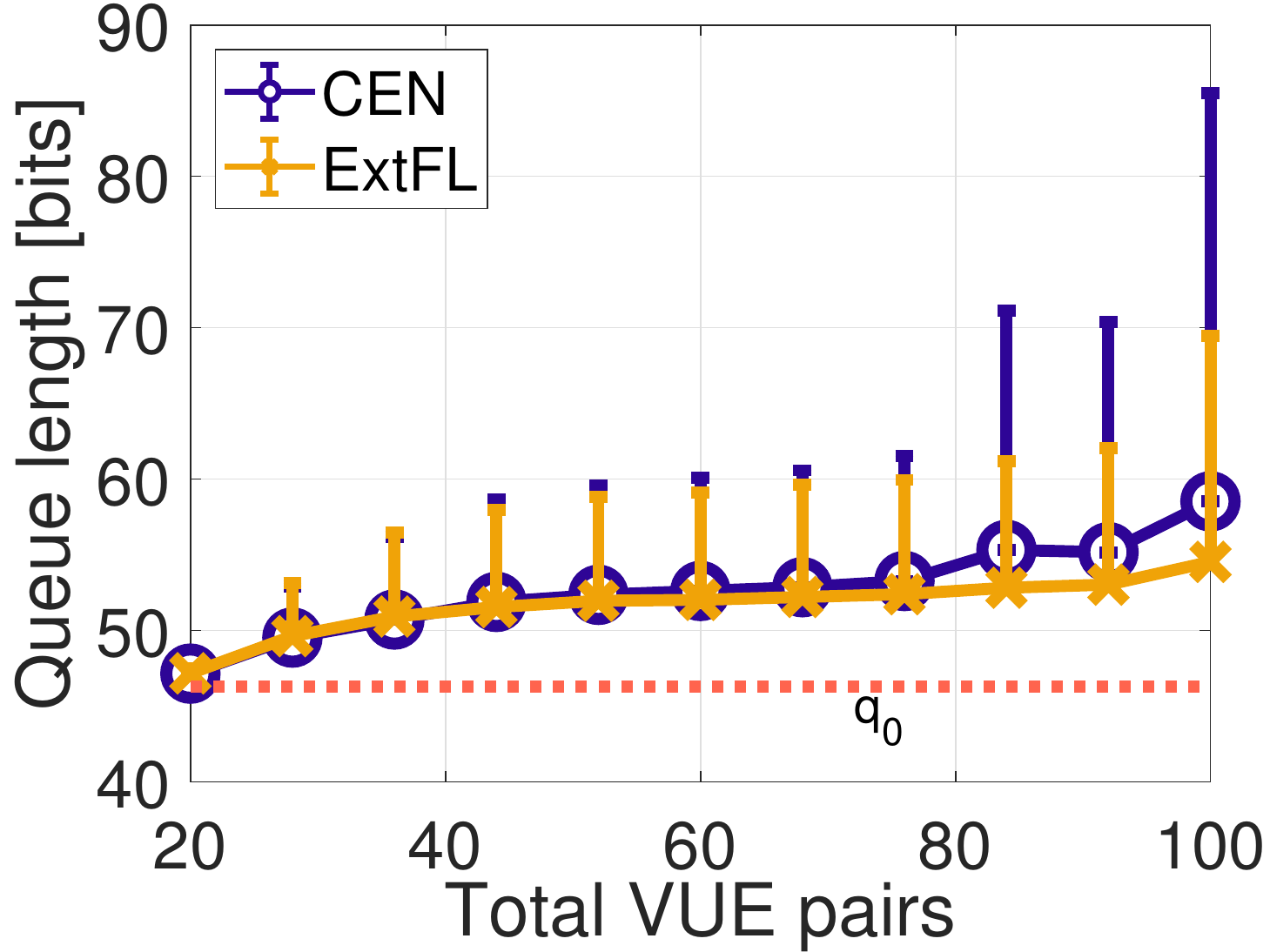}
		\label{fig:v2v_queues}}
	\caption{Comparison between CEN and ExtFL.}	
	\label{fig:accuracy}
\end{figure}

Fig. \ref{fig:FLvsCEN} compares the amount of data exchanged and the achieved V2V communication reliability of ExtFL with a centralized tail distribution estimation model, denoted as CEN. 
%
%
Fig. \ref{fig:FLvsCEN} illustrates that VUEs in ExtFL achieve slightly lower reliability compared to the ones in CEN approach for $\VUE<72$, while outperforming CEN when $\VUE>72$.
Note that the CEN method requires all VUEs to upload all their queue length samples to the RSU and to receive the estimated GPD parameters.
In contrast, in ExtFL, VUEs upload their locally estimated learning models $\localmodel$ and receive the global estimation of the model.
As a result, ExtFL yields equivalent or better end user reliability compared to CEN for denser networks while reducing the amount of data exchange among VUEs and RSU by 79\% when $\VUE=100$.

The worst-case VUE queue lengths, i.e., queue lengths exceeding $\queueTH$, are compared in Fig. \ref{fig:v2v_queues}.
Here, the mean and variance of the tail distribution for CEN and ExtFL are plotted for different numbers of VUEs.
The mean indicates the average queuing latency of the worst-case VUEs while the variance highlights the uncertainty of the latency.
As the number of VUEs increases, it can be noted that both the mean and variance in ExtFL are lower than the ones in CEN.
The reason for above improvement is the reduced training latency in ExtFL over CEN.

\subsection{Federated Learning with Wasserstein Distances} \label{CS_FL_wass}

The Wasserstein distance can precisely measure the similarity between two distributions, even when they have non-overlapping support. To show its effectiveness, consider the same application in Sect. \ref{CS_ExtFL} while replacing its MLE based tail distribution estimation with the Wasserstein distance based estimation, i.e., \emph{Wasserstein-based FL}.

\begin{figure}
	\centering
	\subfigure[Scale parameter.]{
		\includegraphics[width=.47\columnwidth]{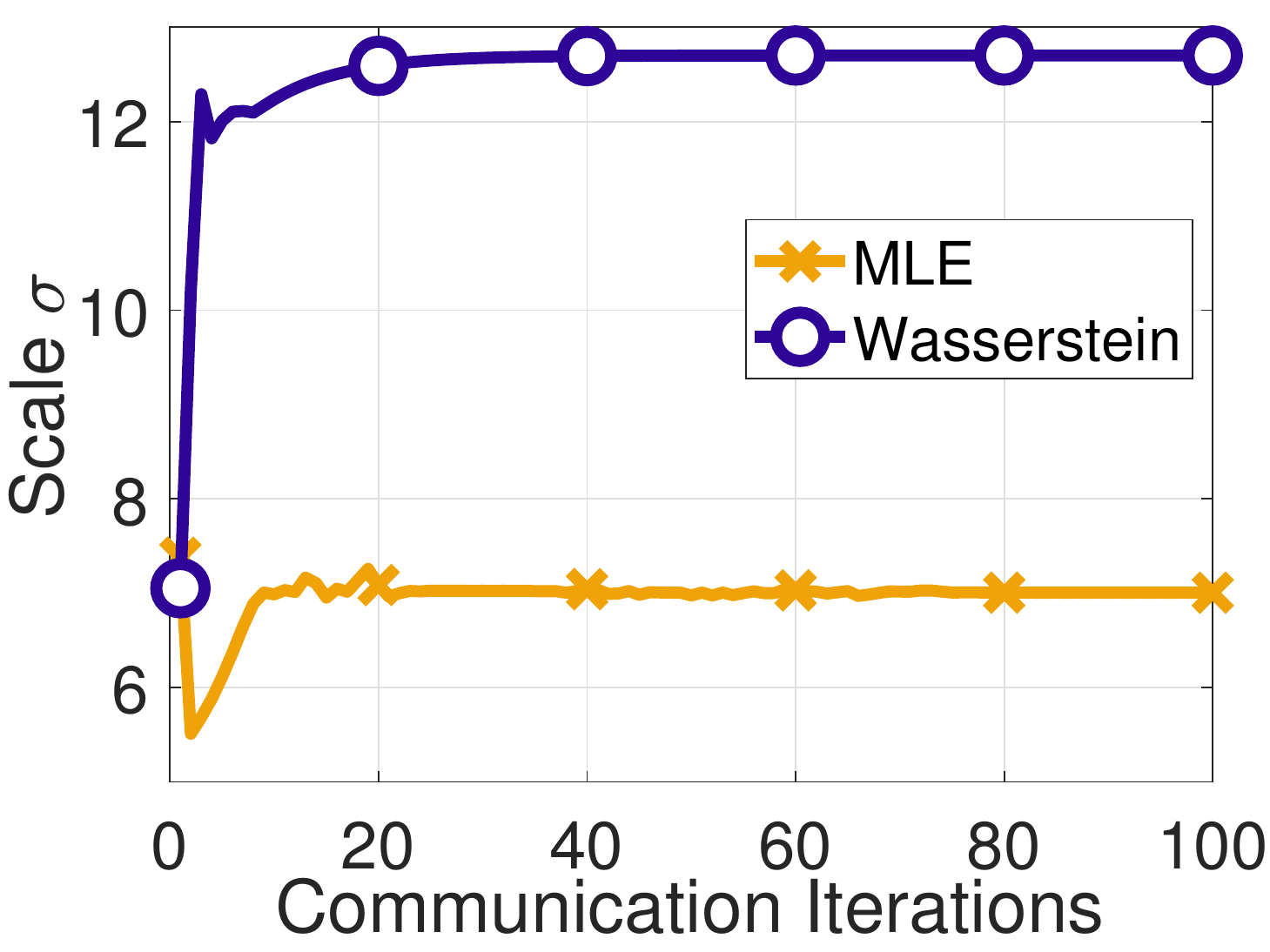}
		\label{fig:wass_scale_comp}}

	\subfigure[Empirical and estimated tail distributions.]{
		\includegraphics[width=.47\columnwidth]{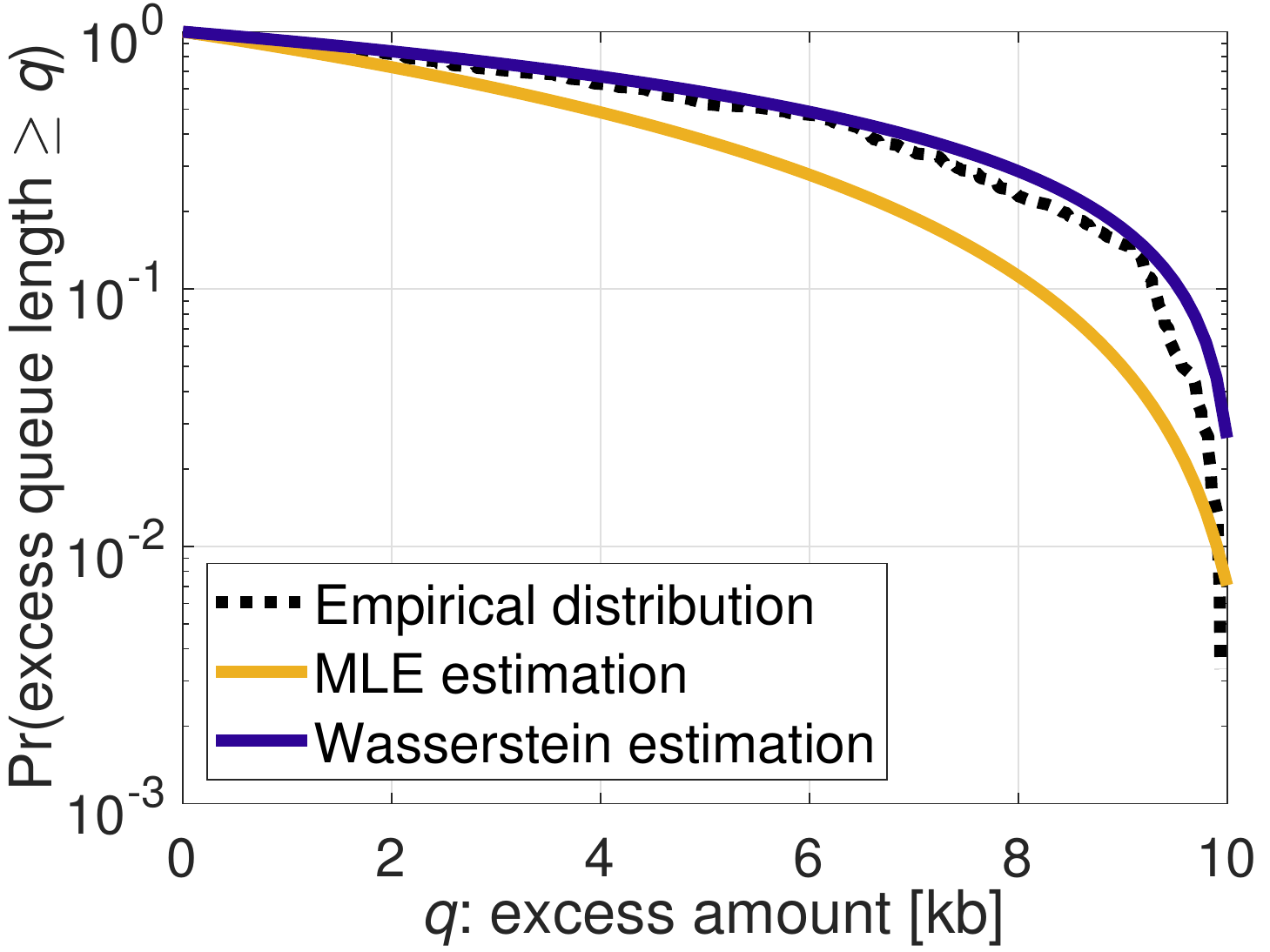}
		\label{fig:ccdf_tail}}
	\caption{ Comparison of GPD parameter estimation and tail distribution estimation between MLE and Wasserstein distance.}
	\label{fig:wass_vs_mle}
\end{figure}

Accordingly, we redefine the cost function $\loglikelihood(\sampleSet)$ in \eqref{eqn:likelihood_local} with the Wasserstein distance given in \eqref{Eq:Wasserstein}.
In this regard, the local and global models defined for FL require to compute the gradient as follows~\cite{Montavon2016WassersteinTO}: 
\begin{equation}\label{eqn:wass_gradient}
\grad{\gpdCombined}{\loglikelihoodSP{\gpdCombined_{\vue}}_{\vue}}(\sample)
= \textstyle -\frac{1}{\sampleSizeRatio} 
\alpha\optimal(\sample)
\grad{\gpdCombined}{ F^{\gpdCombined_{\vue}}_{\vue}(\sample) },
\end{equation}	
where $\alpha\optimal(\sample)$ is the dual function of the corresponding primal problem defined in \eqref{Eq:Wasserstein}.
Here, the function $F^{\gpdCombined_{\vue}}_{\vue}$ satisfies $\gpdML(x') = e^{-F^{\gpdCombined_{\vue}}_{\vue}(x')}/ Z$ with the partition sum $Z = \sum_{x'} e^{-F^{\gpdCombined_{\vue}}_{\vue}(x')}$. 
By utilizing the knowledge of the parametric representation of $\gpdML(x')$, the function $F^{\gpdCombined_{\vue}}_{\vue}$ can be derived while $\alpha\optimal(\sample)$ is calculated using the Sinkhorn algorithm \cite{sinkhorn1964}.

Fig. \ref{fig:wass_vs_mle} corroborates the advantage of Wasserstein-based FL, compared to MLE-based FL in Sect. \ref{CS_ExtFL}. Compared to MLE-based FL, Fig.~\ref{fig:wass_scale_comp} shows that Wasserstein-based FL converges as fast as MLE-based FL, but with a different converging point compared to MLE-based FL's. Fig.~\ref{fig:ccdf_tail} validates that the converging point of Wasserstein-based FL is closer to the optimum, thereby more accurately estimating the tail distribution of the queue lengths exceeding a target threshold. The higher accuracy of Wasserstein-based FL results from the fact that the Wasserstein distance counts the differences between empirical and parametric distributions over the entire supports, whereas the KL divergence in MLE ignores the differences over only the points at which the parametric distributional values are sufficiently large~\cite{GoodfellowBook:16}.

\begin{figure}
	\centering
	\subfigure[Scale parameter.]{
		\includegraphics[width=.47\columnwidth]{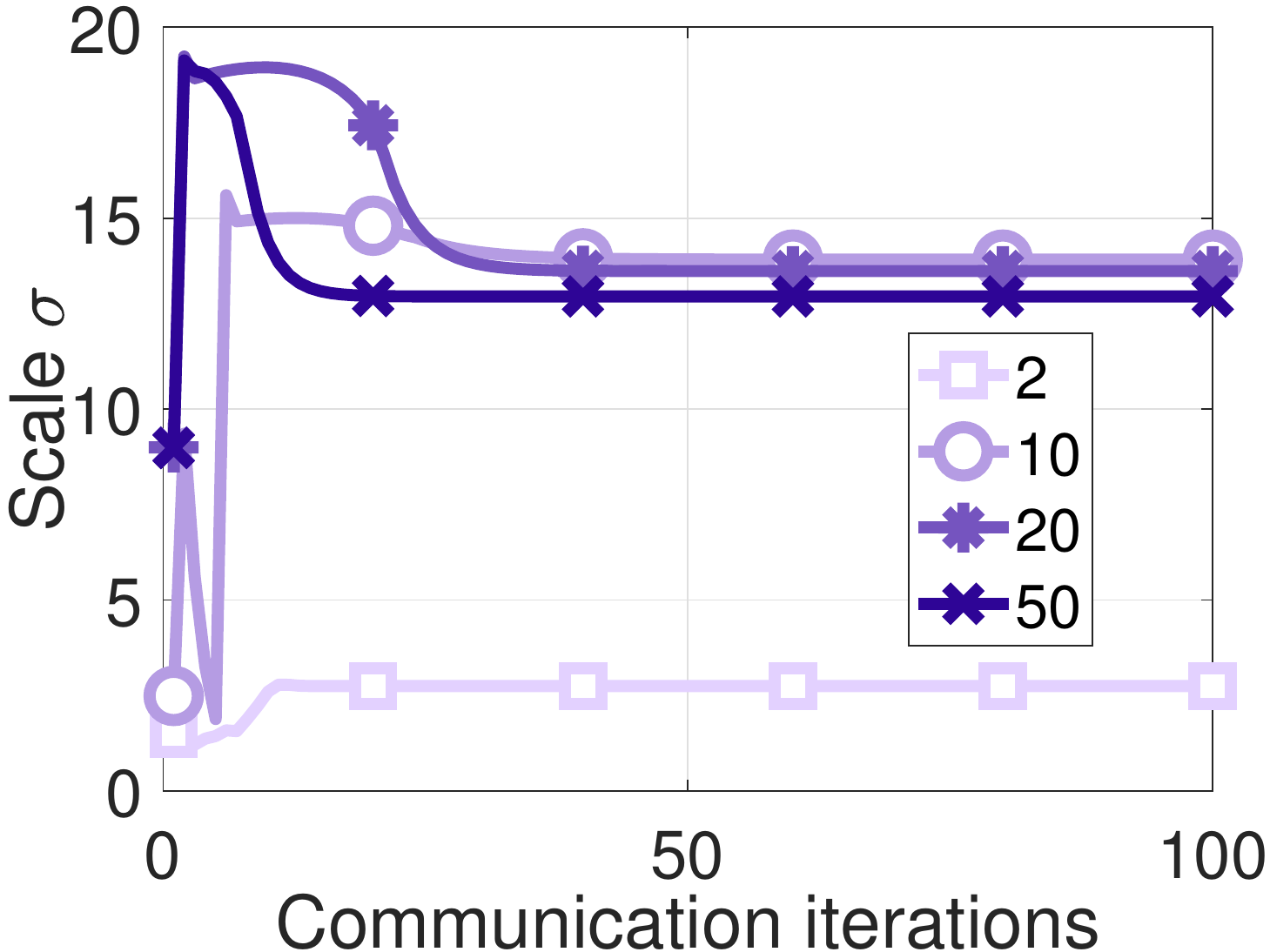}
		\label{fig:wass_scale_sgd}}
	\subfigure[Shape parameter.]{
		\includegraphics[width=.47\columnwidth]{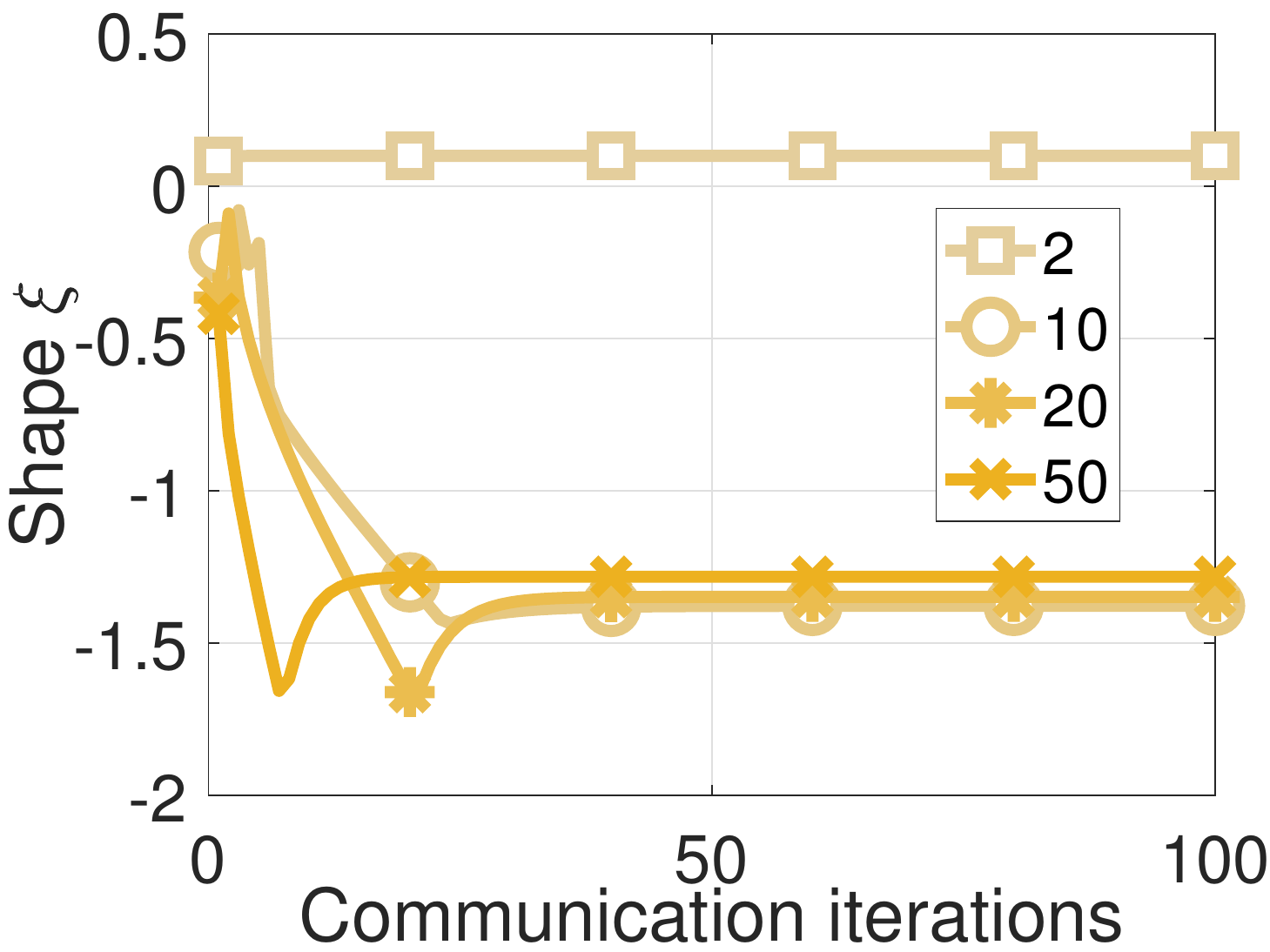}
		\label{fig:wass_shape_sgd}}
	\caption{Convergence speed for different local SVRGD iterations. }	
	\label{fig:wass_sgd}
\end{figure}

Finally, Fig. \ref{fig:wass_sgd} illustrates the impact of local computing iterations on the training convergence. In both Figs. \ref{fig:wass_scale_sgd} and \ref{fig:wass_shape_sgd}, reducing local SVRGD iterations yields faster training convergence, thanks to exchanging the model parameters more frequently. However, as it exchanges the parameters of less trained models, the converging points fall farther away, lowering the tail distribution estimation accuracy. This relationship highlights the importance of optimizing local computing and global communication iterations.

\subsection{Federated Learning with Blockchain}\label{CS_FL_blockchain}

The reliability and scalability of FL can further be improved by adopting blockchain~\cite{Bitcoin,info_prop}, as exemplified by \emph{block-chained FL (BlockFL)} in our preliminary study~\cite{KimCL:19}. 
BlockFL provides incentives to devices who own a larger number of local training samples and consume more computing power, which promotes federation with more devices. 
In addition, local training results in BlockFL are mutually validated, thereby extending the range of federation to untrustworthy devices in a public network. 
All these operations as well as local MSI exchanges are fully decentralized, which is more robust against malfunctions and attacks compared to the original FL ~\cite{Brendan17,pap:jakub16} that relies on a single helper entity.

\begin{figure}
	\centering
	\includegraphics[width=\columnwidth]{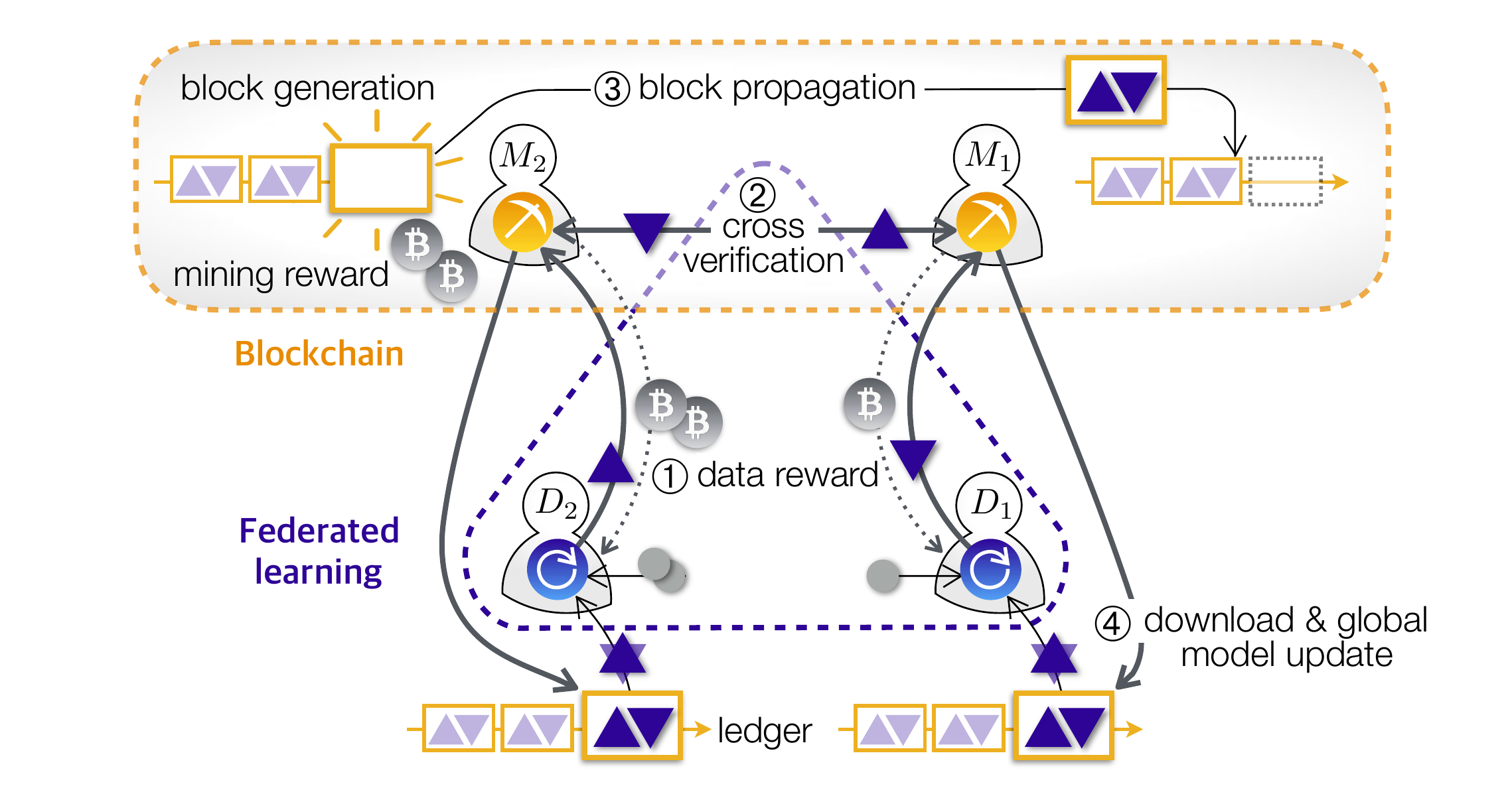}
	\caption{Operational structure of of block-chained FL (BlockFL).} \label{Fig:BlockFL}
\end{figure}

As shown in Fig.~\ref{Fig:BlockFL}, the logical structure of BlockFL consists of devices and miners. Miners can physically be either randomly selected devices or separate nodes like a conventional blockchain network \cite{Bitcoin}. The operation of BlockFL is summarized as follows.
\begin{enumerate}
	\item Each device computes and uploads the local MSI to its associated miner in the blockchain network, while in return receiving the data reward proportional to the number of its data samples from the miner. 
	\item Miners exchange and verify all the MSIs, and then run the Proof-of-Work (PoW) \cite{Bitcoin}.
	\item Once a miner completes the PoW, it generates a block where the verified local MSIs are recorded, and receives the mining reward from the blockchain network. The generated block is propagated and added to every miner's ledger.
	\item Finally, the ledgers are downloaded to the miners' associated devices. Each device locally computes the global MSI from the freshest block, which becomes an input of the next local model update.
\end{enumerate}

At step 3), every miner is enforced to stop its PoW process once it receives a propagated block that should be the earliest generated block, thereby synchronizing the distributed ledgers. However, if a miner generates a block during the earliest generated block's propagation delay, this miner unknowingly adds its own generated block to the ledger that becomes different from the other legitimate ledgers. This forking event incurs extra delays for rolling the unsynchronized ledger back. When the block generation rate $\lambda$ of each miner is centrally controlled by adjusting the PoW difficulty, the optimal block generation rate $\lambda^*$ is thus obtained by balancing between forking occurrences and block propagation delays, which is approximated as
\begin{align}
\lambda^* \approx 2\( T_{\text{bp} }\[ 1 + \sqrt{1 + 4 N_M(1 + T_{\text{wait}}/T_\text{bp})} \] \)^{-1},
\end{align} \label{Eq:OptBlockRate}
where $N_M$ is the number of devices, $T_{\text{bp}}$ is the longest propagation delay of the legitimate block, and $T_\text{wait}$ is the maximum waiting time before starting the PoW process.

\begin{figure}
	\centering
	\subfigure[For different SNRs.]{
		\includegraphics[width=.47\columnwidth]{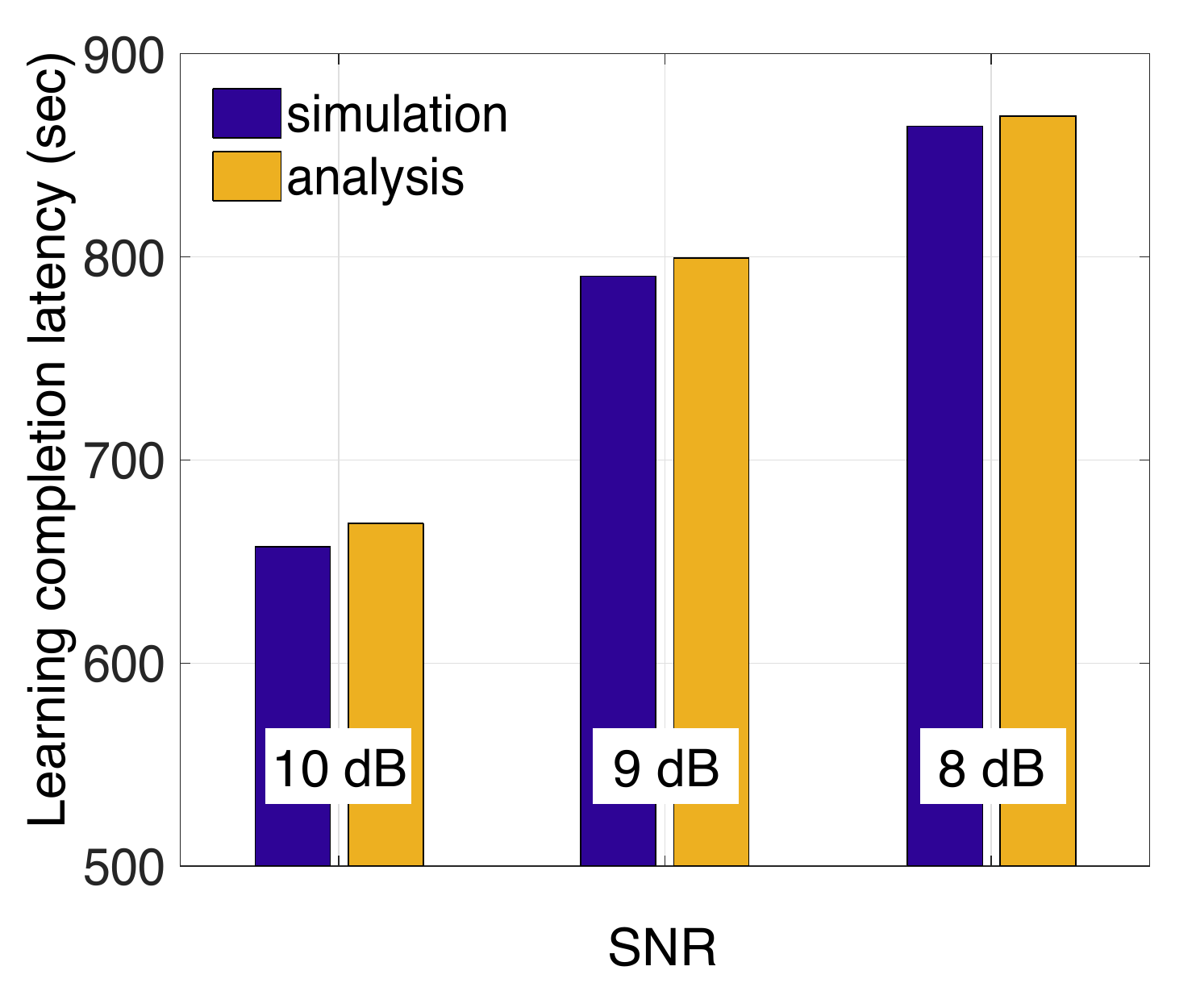}
		\label{Fig:BlockFL_result1}} 
	\subfigure[With/without malfunction.]{
		\includegraphics[width=.47\columnwidth]{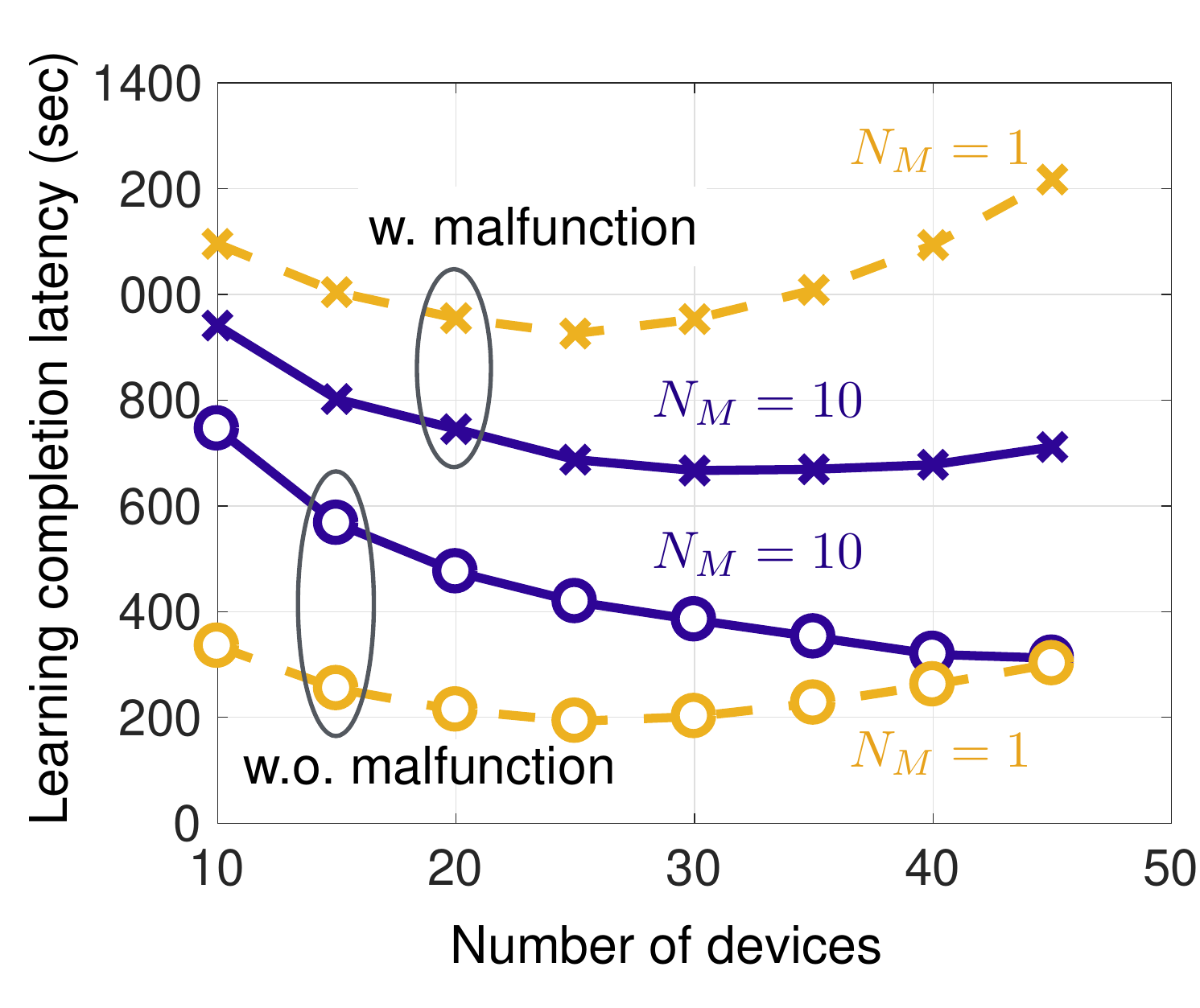}
		\label{Fig:BlockFL_result2}}
	\caption{Average learning completion latency with the optimum block generation rate $\lambda^*$.}	
	\label{Fig:BlockFL_result}
\end{figure}

Fig.~\ref{Fig:BlockFL_result1} plots the E2E latency of BlockFL with $\lambda^*$ until the local model parameters converge, which involves all delays incurred by FL and blockchain operations. 
The simulation parameters follow the 3GPP LTE Cat. M1 specification~\cite{sierra:2017:catM1} with $N_M = 10$. 
As the received signal-to-noise ratio (SNR) decreases from 10 dB to 8 dB, both uplink/downlink MSI delays and block propagation delays increase, and we therefore observe the increased E2E latency. Compared to the simulated optimum, the E2E latency with \eqref{Eq:OptBlockRate} shows only up to only 1.5\% difference.

Fig.~\ref{Fig:BlockFL_result2} demonstrates the scalability and robustness of BlockFL. Without any malfunction, a larger $N_M$ increases the latency due to the increase in their cross-verification and block propagation delays. This does not always hold under the miners' malfunctions, which are captured by adding a Gaussian noise $\mathcal{N}(-0.1,0.01)$ to each miner's aggregate MSIs with probability $0.5$. In BlockFL, global MSI is locally calculated at each device, and each miner's malfunction thus only distorts its associated device's MSI. Such distortion can be restored by federating with other devices that associate with the miners operating normally. For this reason, a larger $N_M$ may achieve even shorter latency, as observed for $N_M=10$ with malfunctions.

\begin{figure}[t]\centering
	\includegraphics[width= \columnwidth]{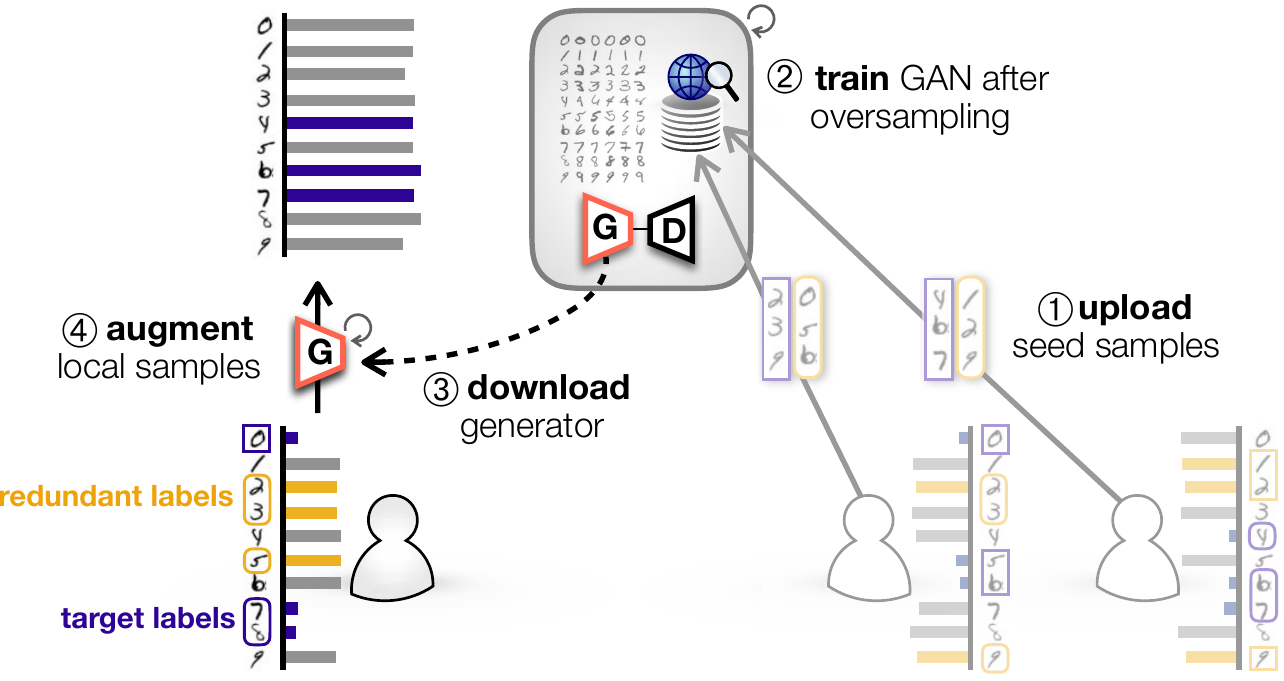}
	\caption{Operational structure of federated augmentation (FAug).} \label{Fig:FAug}
\end{figure}

\subsection{Federated Augmentation Rectifying Non-IID Data}\label{CS_FAug}

The non-IID training dataset of on-device ML can be corrected by obtaining the missing local data samples at each device from the other devices \cite{ARM18}. 
Such a sample exchange may however induce significant communication overhead, especially with a large number of devices, and may violate the privacy requirement of data samples. 
Instead, federated training and exchanging a generative model can rectify the non-IID dataset by enabling each device to locally augment the missing data samples, while abiding by a target privacy requirement. This method and its scalability are demonstrated by \emph{federated augmentation (FAug)} in our previous study~\cite{Jeong:18}.

The procedure of FAug is illustrated in Fig.~\ref{Fig:FAug}, described as follows. At first, each device recognizes the labels being lacking in data samples, referred to as \emph{target labels}, and uploads a few samples of these target labels, referred to as seed samples, to the helper over wireless links. Then, the generative model, a part of a conditional GAN~\cite{CondGAN14}, is trained at a helper with high computing power and a fast connection to the Internet. GAN training commonly requires a large number of training samples. With this end, the helper oversamples the uploaded seed samples, e.g., via Google's image search for visual data, and thereby train the GAN. Finally, downloading the trained GAN's generator empowers each device to replenish the target labels until reaching an IID training dataset.

The operation of FAug needs to guarantee privacy of the user-generated data. In fact, each device's data generation bias, i.e., target labels, may easily reveal its privacy sensitive information, e.g., patients' medical checkup items revealing the diagnosis result. To keep these target labels private from the helper, the device additionally uploads redundant data samples from the labels other than the target labels. The privacy leakage from each device to the helper, denoted as \emph{device-helper privacy leakage (PL)}, is thereby reduced at the cost of extra uplink communication overhead. At the $i$-th device, its device-helper PL is measured as $|\mathbb{L}^{(i)}_t|/(|\mathbb{L}^{(i)}_t| + |\mathbb{L}^{(i)}_r|)$, where $|\mathbb{L}^{(i)}_t|$ and $|\mathbb{L}^{(i)}_r|$ denote the numbers of target and redundant labels, respectively.

The target label information of a device can also be leaked to the other devices since they share a collectively trained generator. Indeed, a device can infer the others' target labels by identifying the generable labels of its downloaded generator. This privacy leakage is quantified by \emph{inter-device PL}. Provided that the GAN is always perfectly trained for all target and redundant labels, the inter-device~PL of the $i$-th device is defined as $|\mathbb{L}^{(i)}_t|/\bigcup_{j=1}^M (|\mathbb{L}^{(j)}_t| + |\mathbb{L}^{(j)}_r|)$. Note that the inter-device~PL is minimized when its denominator equals to the maximum value, i.e., the number of the entire labels. This minimum leakage can be achieved so long as the number of devices is sufficiently large, regardless of the sizes of the target and redundant labels.

Table \ref{tab:1} provides the test accuracy and communication cost of FD and FL, with or without FAug. For FD, the communication cost is defined as the number of exchanged logits, whereas the cost for FL is given as the number of exchanged model parameters. With FAug, the communication cost comprises the number of uploading data samples and the number of downloading model parameters of the trained generator. We observe that FAug is effective for both FL and FD, improving the test accuracy by 0.8-2.7\% for FL and by 7-22\% for FD. Such a gap between the improvements implies that FD is more vulnerable to the non-IID dataset than FL. In FD, even if a device obtains the full teacher's knowledge across all labels, the distillation operation cannot be performed when the device has no local training sample in target labels, which is undesirable.

Overall, we observe that FL achieves the highest test accuracy, while consuming significant communication cost due to exchanging a large number of model parameters. In combination with FAug, FD can cope with the non-IID dataset, and achieve 92-97\% accuracy of FL, In this case, the aggregate communication cost of FD and FAug is up to 25.6x smaller than FL, highlighting the communication-efficiency of FD.

\begin{figure}
	\centering
	\subfigure[Accuracy and inter-device PL.\label{acc_ueue_p}]{\includegraphics[width=.495\columnwidth]{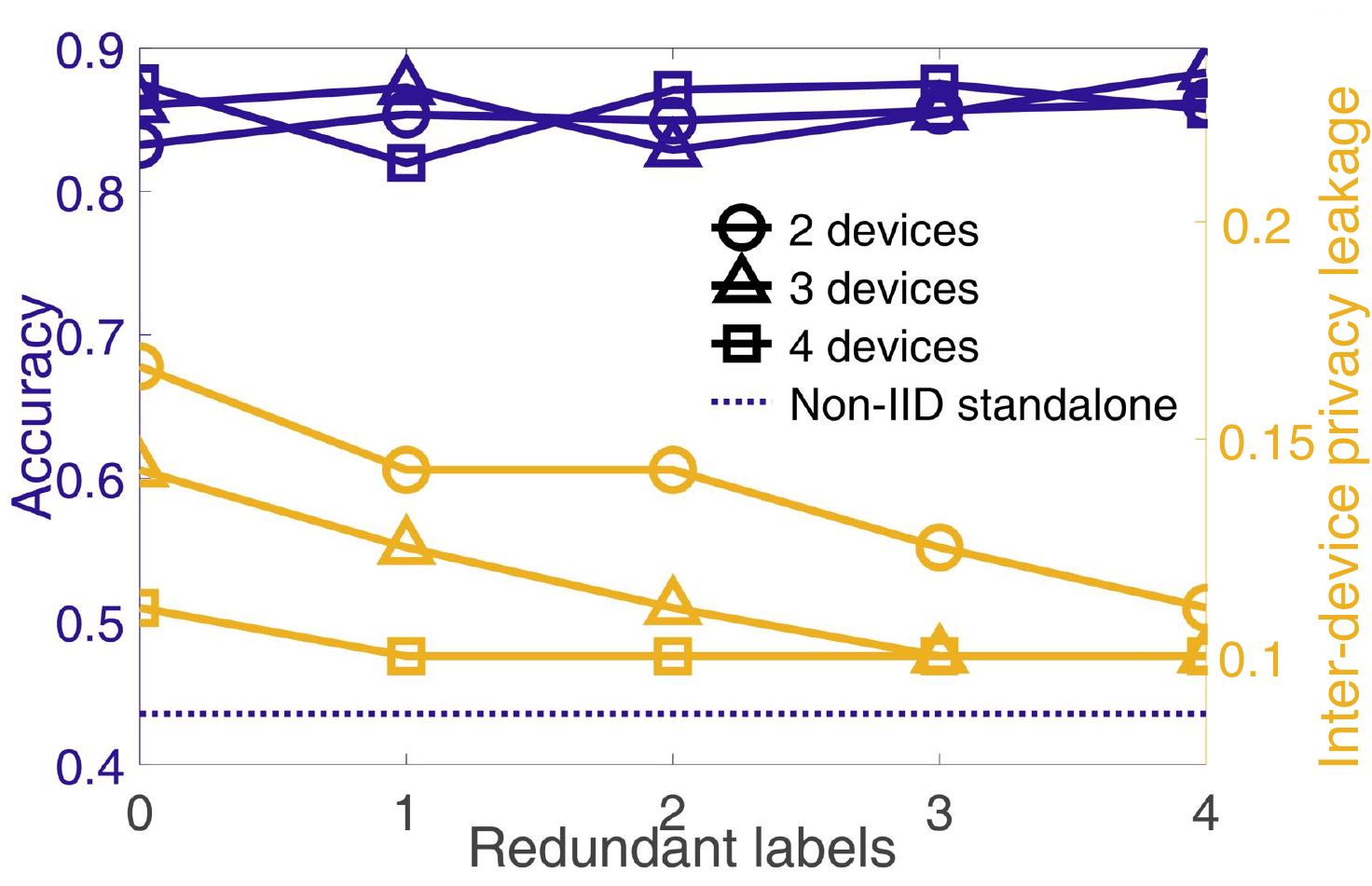}}   
	\subfigure[Device-helper PL.\label{acc_bsue_p}]{\includegraphics[width=.495\columnwidth]{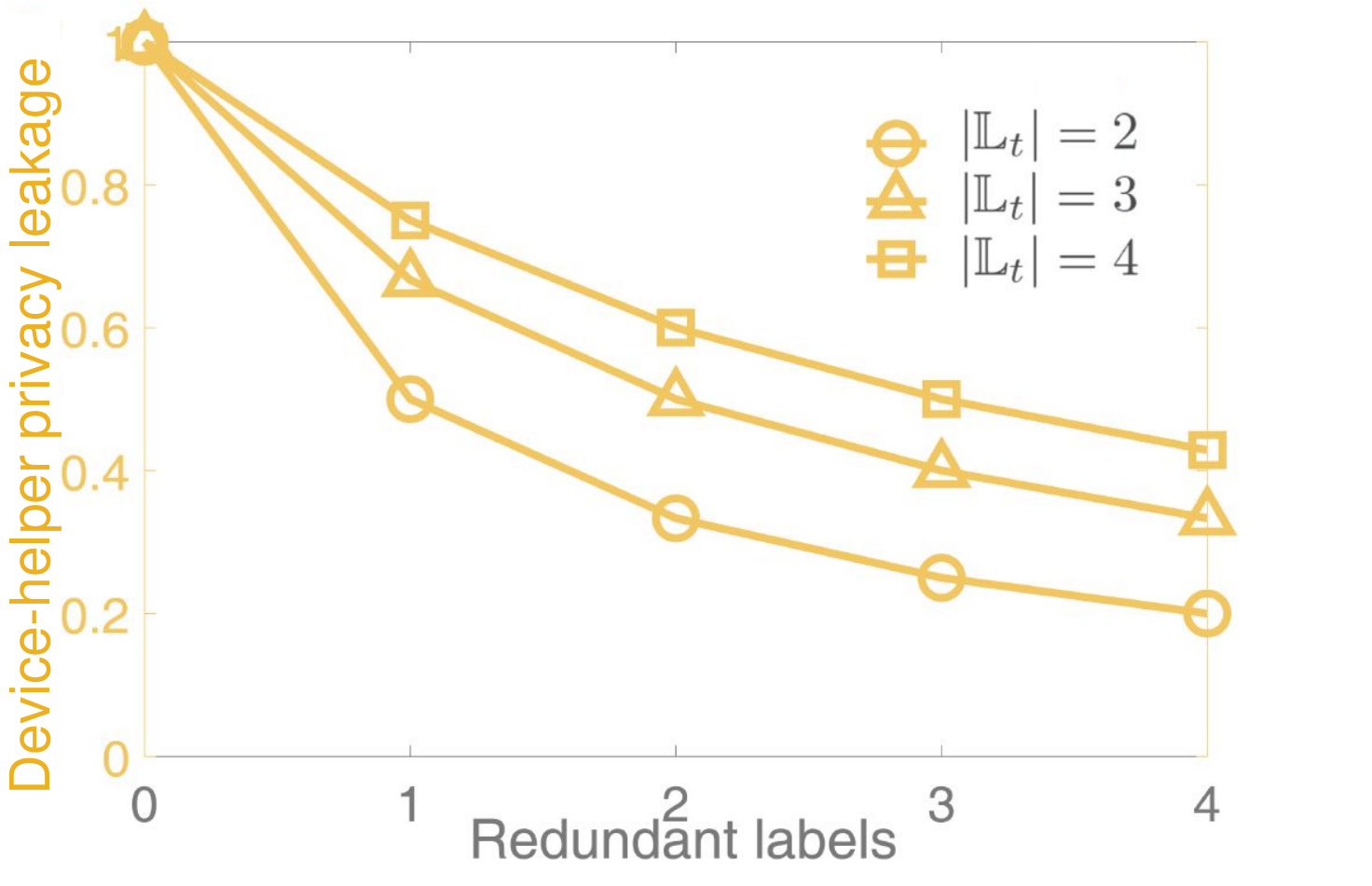}}   
	\caption{\small{Test accuracy and privacy leakage (PL) under a non-IID MNIST dataset: (a) accuracy and inter-device PL; and (b) device-server PL w.r.t. the number of redundant labels.}} \label{acc_per_label}
\end{figure}

Fig. \ref{acc_ueue_p} shows that increasing the number of devices makes the target label uploaders anonymous, thereby reducing the inter-device PL while preserving the test accuracy. More redundant labels allows the uploaders to hide their target labels, which also decreases the inter-device PL. Likewise, the device-helper PL decreases with the number of redundant labels, as shown by Fig. \ref{acc_bsue_p}. Alternatively, multi-hop communication from each device to the server is also effective in hiding target label privacy in the crowds of preceding hops' devices. Its impact on FAug is elaborated in~\cite{Park:2019:MultFAug}.

\begin{table}{\caption{Test accuracy and communication cost of FAug (single target label, no redundant label).}\label{tab:1}} 
	\centering
	\resizebox{\columnwidth}{!}{\begin{tabular}{c c c c c c c c c}
			\toprule
			\multirow{2}[4]{*}{Methods} & \multicolumn{5}{c}{Accuracy w.r.t. the number of  devices}&\multicolumn{3}{c}{Communication cost}\\ 
			\cmidrule(rl){2-6}  \cmidrule(rl){7-9}
			& 2  & 4 & 6 & 8 & 10 & logits &model param. & samples \\ 
			\cmidrule(r){1-1}\cmidrule(l){2-6}   \cmidrule(l){7-9}
			\multicolumn{1}{l}{FD + FAug}&0.8464& 0.8526& 0.8498&  0.8480& \textbf{0.8642}  &3,200&1,493,520 & 5 \\
			\multicolumn{1}{l}{FD (non-IID)}&0.7250& 0.7428 & 0.6951&  \textbf{0.7891}& 0.7524&3,200&-&- \\ \cmidrule(l){1-9} 
			\multicolumn{1}{l}{FL + FAug} & 0.9110& 0.8654& 0.8974&  0.9247& \textbf{0.9259} &-&39,882,256&5 \\
			\multicolumn{1}{l}{FL (non-IID)} &0.9032& 0.8982 & 0.8738&  \textbf{0.9171}& 0.9060 &-&38,388,736&-  \\
			\bottomrule
	\end{tabular}}
\end{table}

\subsection{Field-of-View Prediction for Multicast VR Streaming}\label{CS_RNN}

Gated recurrent units (GRUs), a gating mechanism in RNNs, are commonly used for modeling speech signals and musics.
Their pattern recognition capability can be utilized for predictions that allow proactive control for latency sensitive applications.
In this aspect, proactive content quality adaptation for multi-user  $360^{\circ}$ VR video streaming based on field-of-view (FoV) prediction using GRUs is studied in \cite{MohammedWCNC:18}.

The scenario is a VR theater consisting of a network of VR users watching different HD $360^{\circ}$ VR videos 
streamed over a set of distributed small cell base stations (SBSs).
SBSs operate in the mmWave band and are multi-beam beamforming capable to improve multicast transmission of shared video content to groups of users.
For user grouping, upcoming tiled FoV predictions obtained via a deep neural network architecture based on GRUs are utilized.
By optimizing video frame admission and user scheduling, the goal is to provide highly reliable broadband service for VR users that deliver HD videos with low latency.

Fig. \ref{fig:vr_rate_case} evaluates the impact of the video quality corresponding to the delivered frame rates.
Therein, the proposed FoV prediction based scheme is compared with two benchmark methods: \emph{Baseline 1} and \emph{Baseline 2} that schedule video chunk requests in real time over mmWave unicast and multicast transmissions, respectively.  
Figs. \ref{fig:vr_rate_case}(a) and (b) show that the FoV based predictions allow the proposed proactive scheduler to decrease both average and 99th percentile delays over the baseline methods.
While both baseline schemes reduce the HD delivery rates for increased offered data rates as shown in Fig. \ref{fig:vr_rate_case}(c), the proposed approach maintains about 100\% success HD delivery rate by predicting video frames in advance.

As the predictions become accurate, the need of retransmissions reduces and a surplus of wireless resources can be smartly reused to feed back  prediction event errors.

\begin{figure}
	\centering
	\includegraphics[width=\columnwidth]{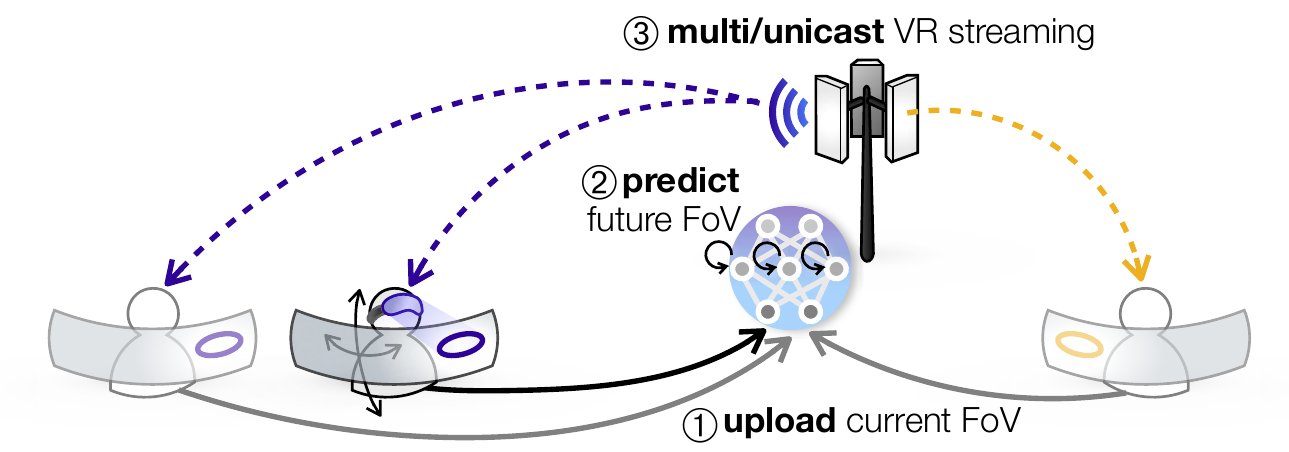}
	\caption{Operational structure of the $360^{\circ}$ VR video streaming via mmWave multicast (blue) and unicst (yellow) transmissions.}
\end{figure}

\begin{figure}
	\centering
	\includegraphics[width=\columnwidth]{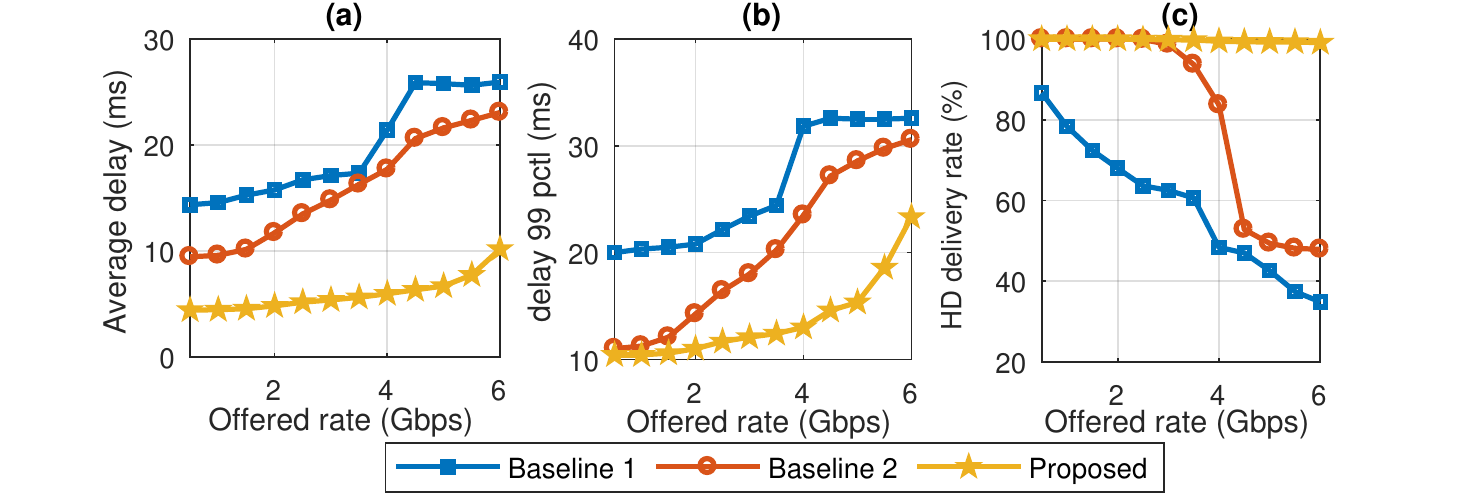}
	\caption{Performance comparison with respect to: (a) average delay, (b) 99th-percentile delay, and (c) HD delivery rate.}
	\label{fig:vr_rate_case}
\end{figure}

\subsection{Actor-Critic RL for Optimizing Age of Information}\label{CS_RL}

\begin{figure}[t]\centering
	\includegraphics[width= \columnwidth]{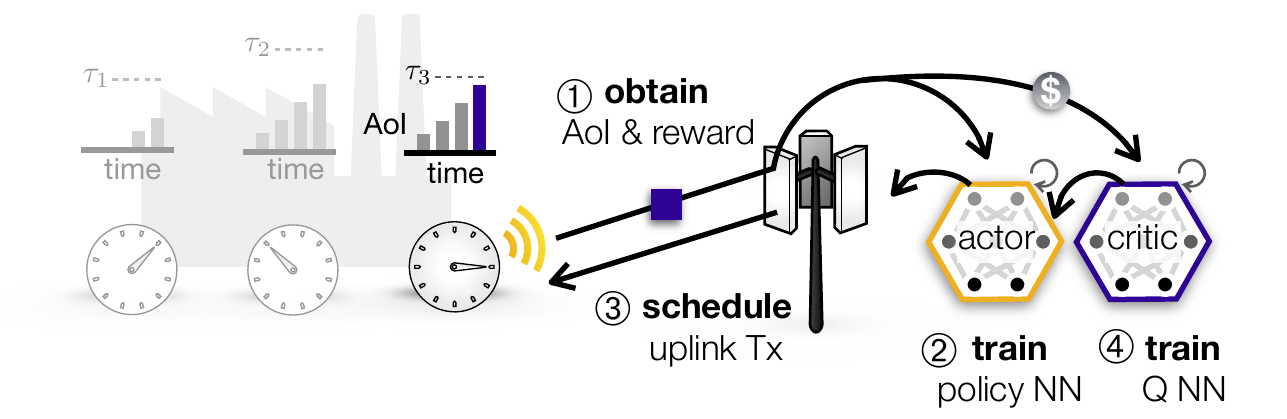}
	\caption{Operational structure of the actor-critic RL for optimizing the age of information (AoI).} \label{Fig:AoI}
\end{figure}

Age of information (AoI) is a measure of the freshness of data to characterize the end-to-end communication latency.
While minimizing AoI enables URLLC, the performance highly depends on the availability of the system state knowledge such as physical resources, channel conditions, packet drops, sampling, etc. 
As a remedy, RL can be adopted to explore and learn the system state dynamics and improve the process of decision making over time.
In this regard, minimizing AoI of remote sensors with the aid of a RL-based scheduler is presented in \cite{anis:2018:RL}.
As illustrated in Fig.~\ref{Fig:AoI}, the scenario is focused on a set of sensors in a factory randomly generating data packets and remotely monitored. 
To ensure high reliability and low latency of the received sensor data at the remote monitor, the controller schedules sensors to report their data.
Due to the lack of knowledge of the conditions of the communication links, packet generation at the sensors and losses during communication, the monitoring unit resorts to a RL-based scheduler.
Here, the states observed at the remote monitoring unit are the AoI of each sensor ($\tau_i$), previous data rates, and the time spent to download the last data packet. 
Based on that, RL scheduler builds a probability distribution over the sensors that is used for scheduling the sensors.
A cost for each action is defined in terms of the average AoI over the sensors and aggregated penalties when sensors' AoI exceed their predefined thresholds.
The RL-scheduler is trained using asynchronous advantage actor-critic (A3C) algorithm \cite{pmlr-v48-mniha16} in an offline manner.
In A3C, several copies of the actor agent are trained in parallel (\emph{asynchronous}) to improve the efficiency using discount functions that indicate gains/losses over average q-values (\emph{advantage}) while the critic NN estimates the cost function.
Here, the trained actor NN is used as the scheduler. 

\begin{figure}
	\centering
	\subfigure[Average AoI.]{
		\includegraphics[width=.47\columnwidth]{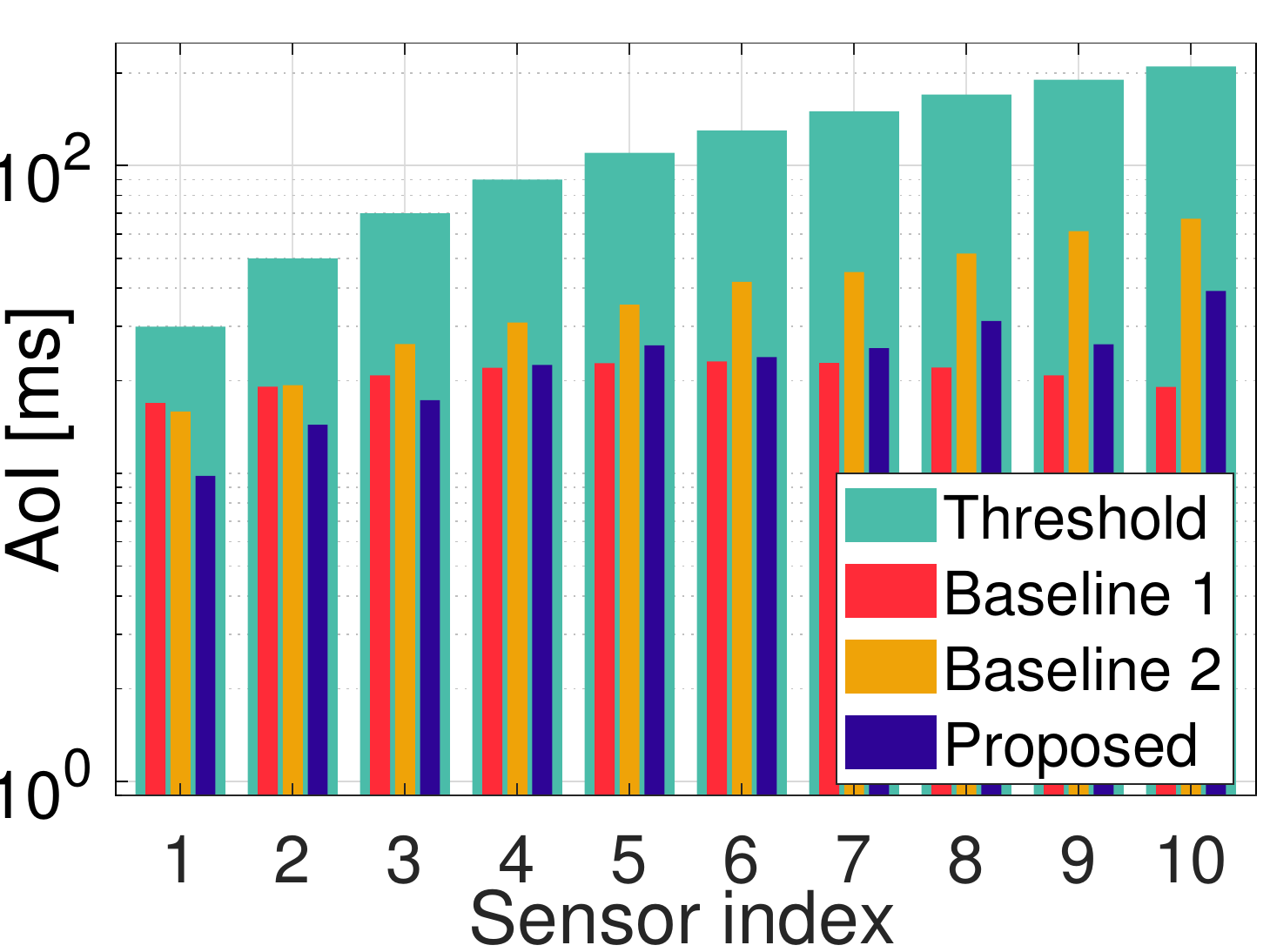}
		\label{fig:rl_aoi}}
	\subfigure[AoI violation probability.]{
		\includegraphics[width=.47\columnwidth]{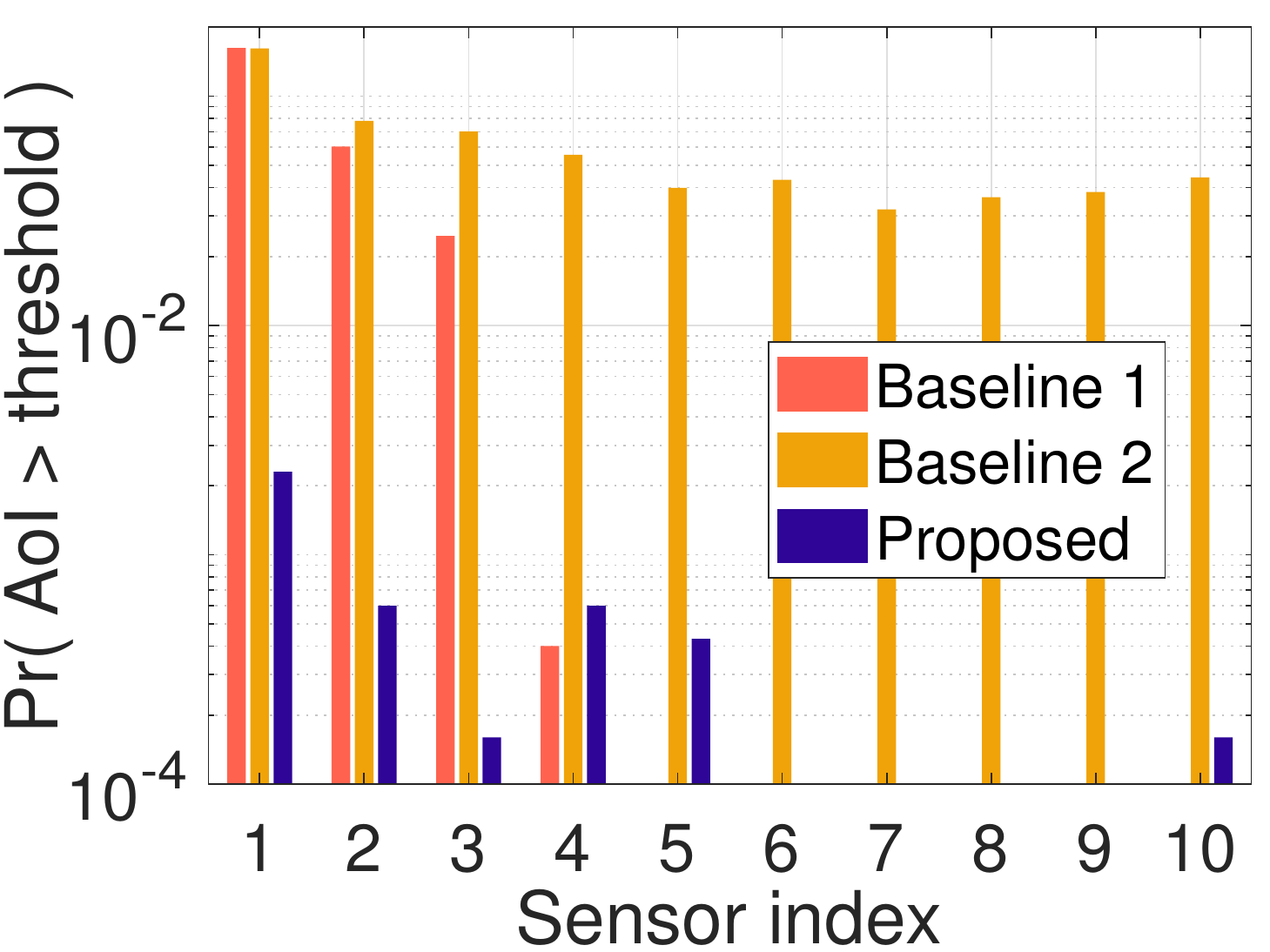}
		\label{fig:rl_rel}}
	
	\caption{Per-sensor performance comparison with respect to: (a) average AoI and (b) AoI violation probability.}	
	\label{fig:rl_comp}
\end{figure}

Fig. \ref{fig:rl_comp} compares the per-sensor performance of the  RL-based scheduler with two baselines: \emph{Baseline~1} schedules a sensor with maximum AoI at a given time while \emph{Baseline~2} randomly schedules sensors proportional to the inverse of their AoI thresholds. 
Here, the system consists of ten sensors indexed by $i= 1,2,\ldots,10$.
Fig. \ref{fig:rl_aoi} indicates that all three methods perform almost same in terms of average AoI.
While all sensors in Baseline~2 exhibit higher average AoI compared to the rest, Baseline~1 and Proposed methods display lowest AoI for sensors with loose and tight thresholds, respectively.
The probability of AoI exceeding the threshold for each sensors are illustrated in Fig. \ref{fig:rl_rel}.
Therein, it can be noted that the proposed method maintains much lower AoI violation probability for sensors with tight AoI thresholds compared to both baseline methods.
For sensors with large thresholds both the proposed and Baseline~1 exhibit no AoI violations.
From Figs. \ref{fig:rl_aoi} and \ref{fig:rl_rel}, we can conclude that the proposed RL-based scheduler is the most reliable approach for sensors with tight thresholds.

By minimizing the AoI over all  sensors, the proposed solution allows the remote monitor to obtain up-to-date data.
For the applications of monitoring and controlling, these data can be utilized to train control modules and/or to issue real-time commands.

\subsection{MFG Theoretic ML for Massive UAV Control}\label{CS_UAV}

\begin{figure}[t]\centering
	\subfigure[HJB learning control.]{\includegraphics[width= .4\columnwidth]{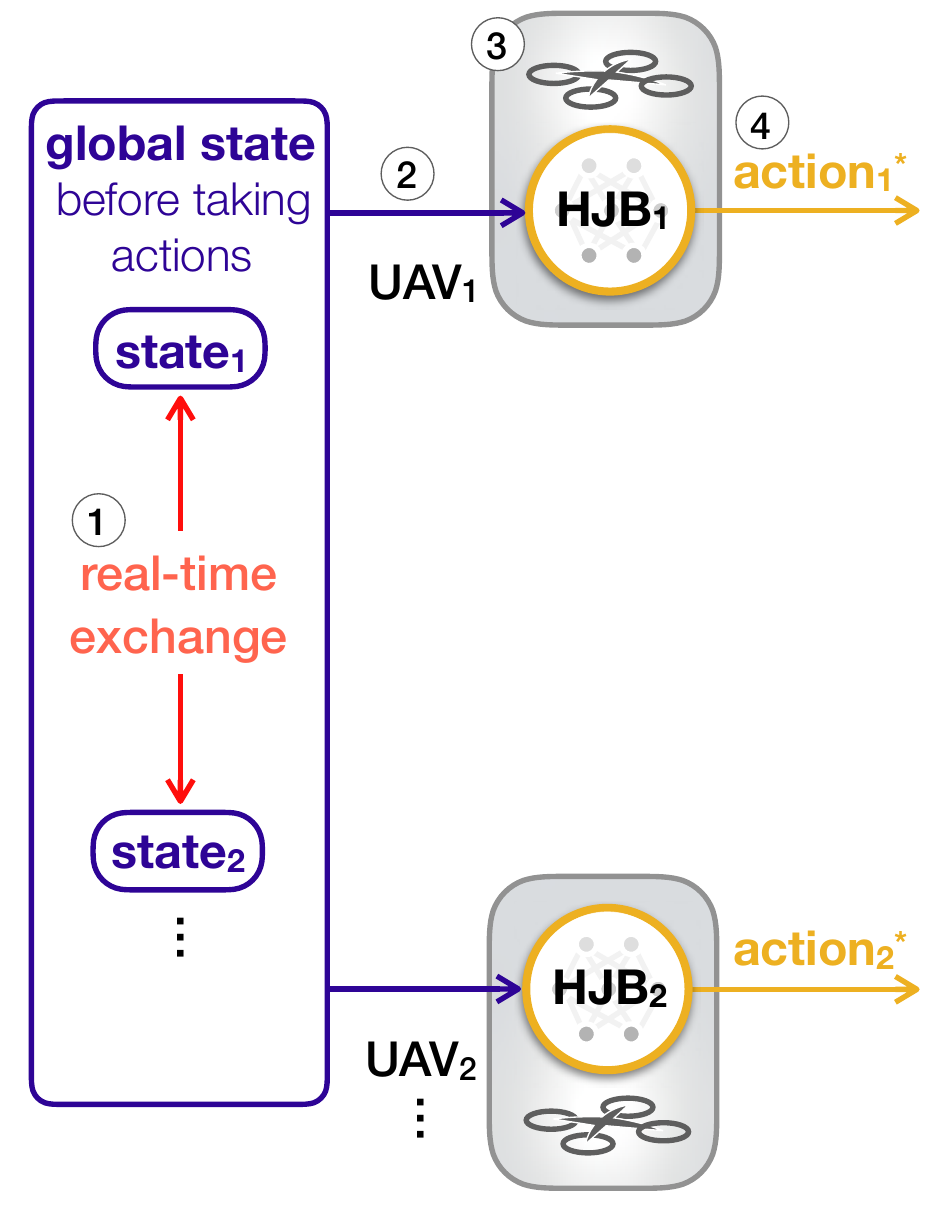}}\hspace{10pt}
	\subfigure[MFG learning control]{\includegraphics[width= .55\columnwidth]{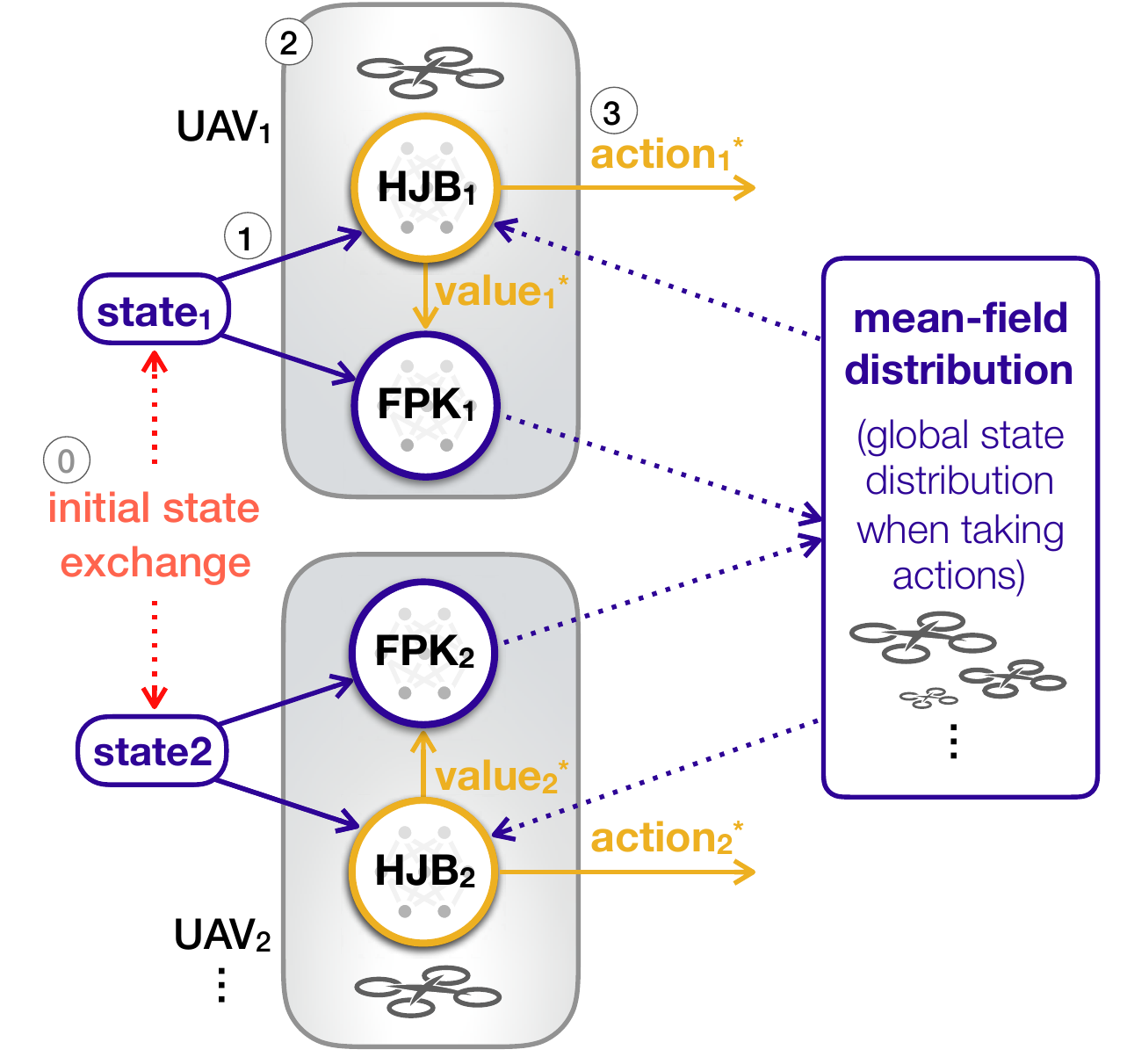}}
	\caption{ Operational structures of (a) HJB learning control and (b) MFG learning control. } \label{Fig:mfgUAV}
\end{figure}

\begin{figure}[t]\centering
	\subfigure[HJB learning control.]{\includegraphics[width= .49\columnwidth]{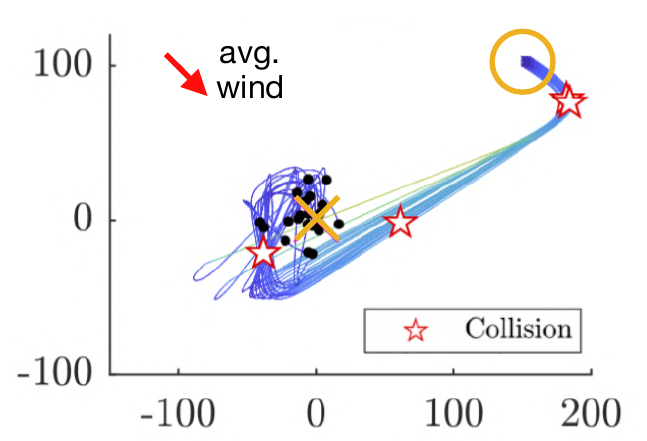}}
	\subfigure[MFG learning control.]{\includegraphics[width= .49\columnwidth]{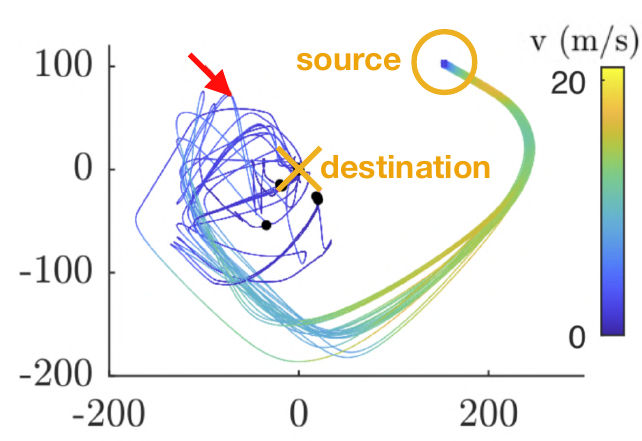}}
	\caption{ Trajectories of 25 UAVs under (a) HJB learning control and (b) MFG learning control. } \label{Fig:mfgUAV_result}
\end{figure}

MFG theoretic control is a key enabler for supporting robust massive autonomous UAV operations against time-varying network connectivity. To illustrate its effectiveness, following \cite{Hamid:GC19}, we consider $N$ UAVs, each of which locally controls its acceleration so as to minimize its travel time from the same source to a common destination, without inter-UAV collisions. To avoid collisions, each UAV communicates with the other $N-1$ UAVs for collecting their real-time states, i.e., coordinates and velocities. To guarantee the optimality of every control, since each control affects the other UAVs' decisions, the states should be recursively exchanged until all controls converge to the Nash equilibrium. This is likely to be infeasible for a large $N$, under the limited communication range of each UAV moving in real time.

This is where MFG theoretic control comes to the rescue, which requires initial state exchanges only once. Afterwards, as elaborated in Sec. 4.2, the control of each UAV is determined by locally solving two partial differential equations, in which the FPK equation ($F=0$) approximates the state distributions of the entire population following optimal controls, i.e., mean-field distribution, while the HJB equation ($H=0$) provides each UAV's optimal control for a given population distribution. However, solving these equations is computationally challenging particularly for high-dimensional states, limiting its adoption commonly within 1-dimensional state cases~\cite{KimSPAWC:18,MobilMFGSG:GC16,kim:2017:MFCA}.

To overcome this limitation, an NN based MFG theoretic control algorithm is developed in~\cite{Hamid:GC19}, denoted as \emph{MFG leaning control} operated by a pair of HJB and FPK NNs, as shown in Fig.~\ref{Fig:mfgUAV}(b). At its core, the HJB and FPK NNs are trained so as to minimize $|H|$ and $|F|$, thereby approximately solving the HJB and FPK equations, respectively. On the one hand, focusing on the relation between HJB and FPK NNs, this ML architecture is similar to actor-critic RL wherein the actor NN (FPK NN or policy NN) yields per-state action distributions, and the critic NN (HJB NN or value NN) evaluates the optimality of controls. On the other hand, focusing on the process of taking actions, MFG learning control resembles DQN whose actions are taken by the Q NN (value NN). Integrating these two RL architectures, MFG learning control addresses the interplay between the population's policy and a single agent's action evaluation, as opposed to traditional RL considering the same agent's policy and action evaluation.

Consequently, the source-to-destination travel trajectory in Fig.~\ref{Fig:mfgUAV_result}(b) corroborates that MFG learning control achieves both goals of collision avoidance and fast travel. This is compared with a baseline \emph{HJB learning control} in Fig.~\ref{Fig:mfgUAV_result}(a), in which UAVs are controlled based not on FPK neural networks but on real-time state exchanges. Due to the limited transmission power, the inter-UAV state exchanges are restricted within the communication range $1$m that is not always sufficient to cover $N-1$ mobile UAVs. This incurs time-varying UAV network connectivity, and makes HJB learning control fail to avoid collision. By contrast, MFG learning control is effective in collision avoidance, by forming a flock of UAVs. In return, the travel distances under MFG learning control become longer. Nevertheless, thanks to flocking behaviors, MFG learning control achieves faster speed during long flights. This compensates the longer travel distances, thereby achieving the travel time almost as fast as the HJB learning control ignoring collision avoidance.

\section{Concluding Remarks}\label{sec:conclusion}

Embracing  recent advances in hardware and communication technologies, edge ML empowers   devices at the network edge by imbuing them with state-of-the art ML techniques. This poses a slew of new research questions centered on E2E latency, reliability, and scalability, under  hardware and privacy constraints. As a first step towards spearheading the edge ML vision and moving beyond centralized and cloud-based ML, this article has explored its key building blocks and theoretical principles warranting a clean-slate design in terms of NN architectures, training and inference operations, and communication. The overarching goal of this article is to foster more fundamental research in edge ML and bridge connections among several communities and mathematical disciplines.

\section{Acknowledgment}
This research was supported in part by the Academy of Finland (Grant Nr. 294128), in part by the 6Genesis Flagship (Grant Nr. 318927), in part by the Kvantum Institute Strategic Project (SAFARI), in part by the Academy of Finland thorough the MISSION Project (Grant Nr. 319759), and in part by the Mobile Edge Intelligence at Scale (ELLIS) project at the University of Oulu.

\bibliographystyle{ieeetr}  

\vfill
\pagebreak
\begin{IEEEbiography}[{\includegraphics[width=1in,height=1.25in,clip,keepaspectratio]{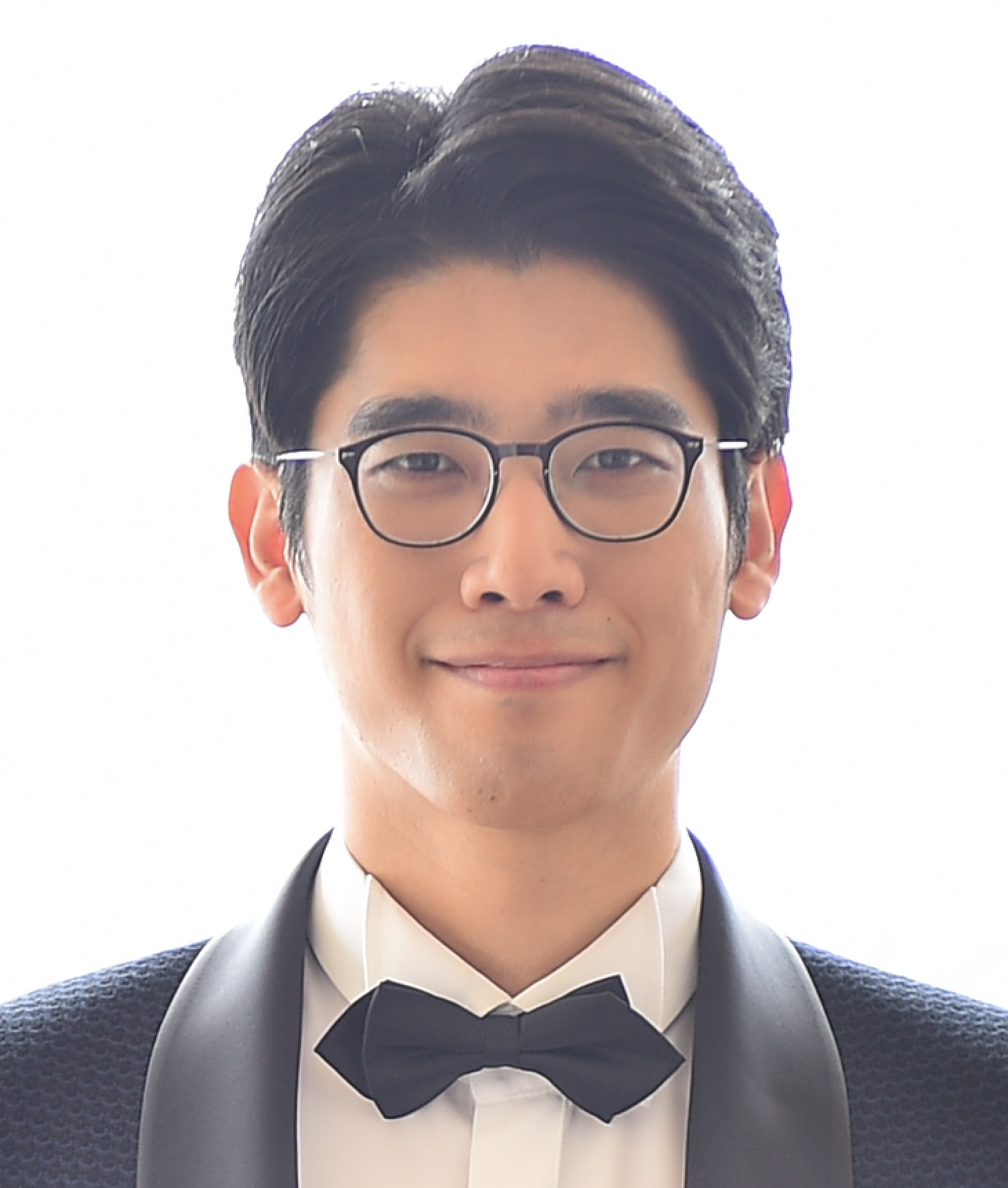}}]{Jihong Park} (S'09-M'16) received the B.S. and Ph.D. degrees from Yonsei University, Seoul, South Korea, in 2009 and 2016, respectively. From 2016 to 2017, he was a Post-Doctoral Researcher with Aalborg University, Denmark. He was a Visiting Researcher with Hong Kong Polytechnic University; KTH, Sweden; Aalborg University, Denmark; and New Jersey Institute of Technology, USA, in 2013, 2015, 2016, and 2017, respectively. He is currently a Post-Doctoral Researcher with the University of Oulu, Finland. From 2020, He will be a Lecturer at the School of Information Technology, Deakin University, Australia. His research interests include distributed machine learning and ultra-reliable designs in 5G communication systems and beyond. He received the 2014 IEEE GLOBECOM Student Travel Grant, the 2014 IEEE Seoul Section Student Paper Contest Bronze Prize, and the 6th IDIS-ETNEWS (The Electronic Times) Paper Contest Award sponsored by the Ministry of Science, ICT, and Future Planning of Korea.
\end{IEEEbiography}

\begin{IEEEbiography}[{\includegraphics[width=1in,height=1.25in,clip,keepaspectratio]{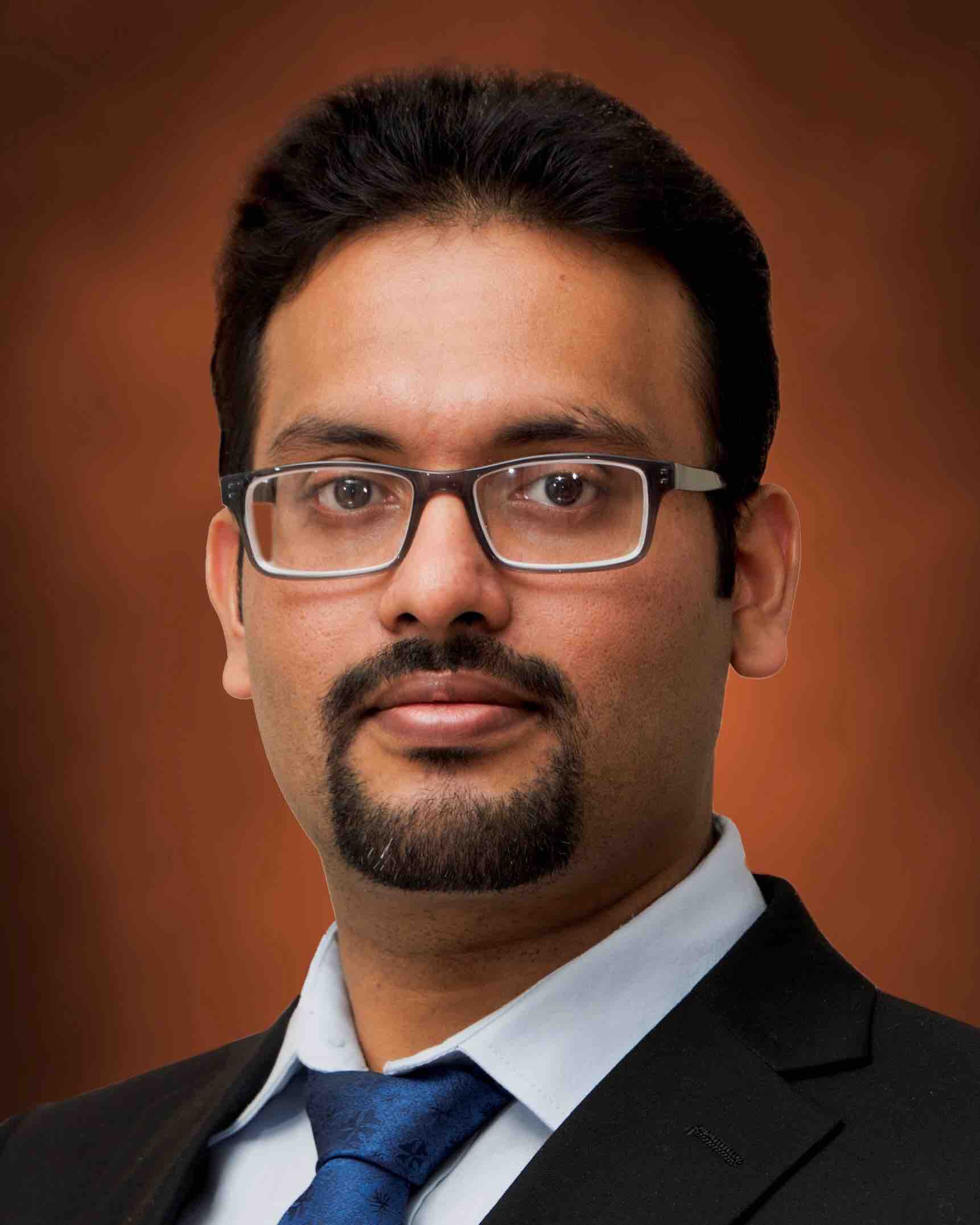}}]{Sumudu Samarakoon} (S'08-AM'18) received his B. Sc. Degree (Hons.) in Electronic and Telecommunication Engineering from the University of Moratuwa, Sri Lanka in 2009, the M. Eng. degree from the Asian Institute of Technology, Thailand in 2011, and Ph. D. degree in Communication Engineering from University of Oulu, Finland in 2017. He is currently working in Centre for Wireless Communications, University of Oulu, Finland as a post doctoral researcher. His main research interests are in heterogeneous networks, small cells, radio resource management, reinforcement learning, and game theory. In 2016, he received the Best Paper Award at the European Wireless Conference and Excellence Awards for innovators and the outstanding doctoral student in the Radio Technology Unit, CWC, University of Oulu.
\end{IEEEbiography}

\begin{IEEEbiography}[{\includegraphics[width=1in,height=1.25in,clip,keepaspectratio]{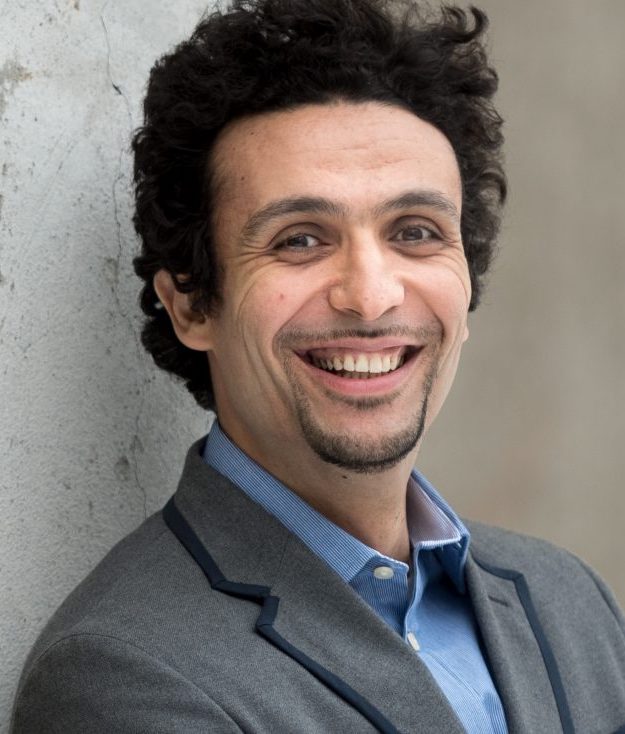}}]{Mehdi Bennis} is an Associate Professor at the Centre for Wireless Communications, University of Oulu, Finland, an Academy of Finland Research Fellow and head of the intelligent connectivity and networks/systems group (ICON). His main research interests are in radio resource management, heterogeneous networks, game theory and machine learning in 5G networks and beyond. He has co-authored one book and published more than 200 research papers in international conferences, journals and book chapters. He has been the recipient of 
several prestigious awards including the 2015 Fred W. Ellersick Prize from the IEEE Communications Society, the 2016 Best Tutorial Prize from the IEEE Communications Society, the 2017 EURASIP Best paper Award for the Journal of Wireless Communications and Networks, the all-University of Oulu award for research and the 2019 IEEE ComSoc Radio Communications Committee Early Achievement Award. Dr Bennis is an editor of IEEE TRANSACTIONS ON COMMUNICATIONS.
\end{IEEEbiography}

\begin{IEEEbiography}[{\includegraphics[width=1in,height=1.25in,clip,keepaspectratio]{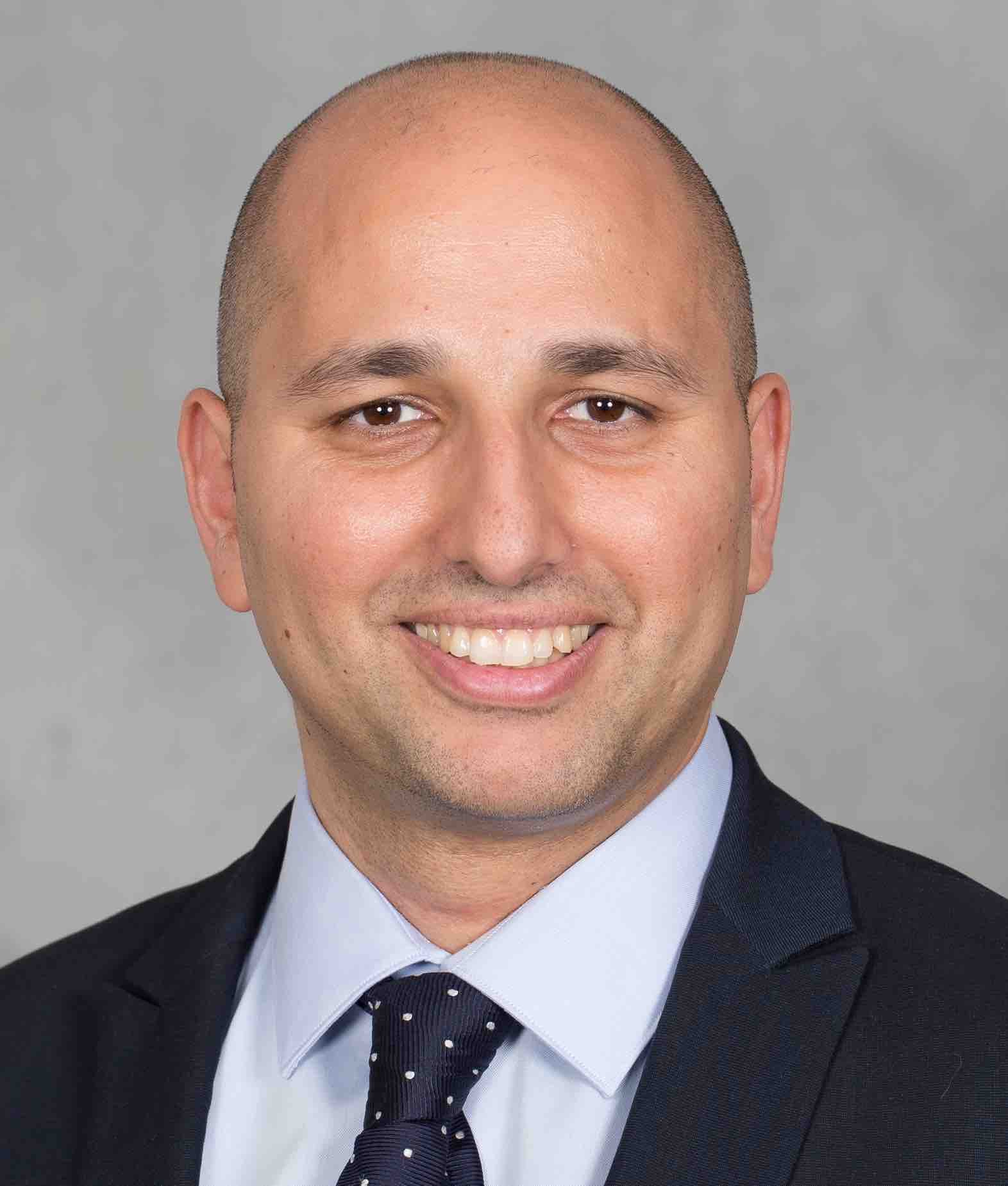}}]{M\'{e}rouane Debbah} (S'01-M'04-SM'08-F'15) received the M.Sc. and Ph.D. degrees from the Ecole Normale Supérieure Paris-Saclay, France. He was with Motorola Labs, Saclay, France, from 1999 to 2002, and also with the Vienna Research Center for Telecommunications, Vienna, Austria, until 2003. From 2003 to 2007, he was an Assistant Professor with the Mobile Communications Department, Institut Eurecom, Sophia Antipolis, France. From 2007 to 2014, he was the Director of the Alcatel-Lucent Chair on Flexible Radio. Since 2007, he has been a Full Professor with CentraleSupelec, Gif-sur-Yvette, France. Since 2014, he has been a Vice-President of the Huawei France Research Center and the Director of the Mathematical and Algorithmic Sciences Lab. He has managed 8 EU projects and more than 24 national and international projects. His research interests lie in fundamental mathematics, algorithms, statistics, information, and communication sciences research. He is an IEEE Fellow, a WWRF Fellow, and a Membre émérite SEE. He was a recipient of the ERC Grant MORE (Advanced Mathematical Tools for Complex Network Engineering) from 2012 to 2017. He was a recipient of the Mario Boella Award in 2005, the IEEE Glavieux Prize Award in 2011, and the Qualcomm Innovation Prize Award in 2012. He received 20 best paper awards, among which the 2007 IEEE GLOBECOM Best Paper Award, the Wi-Opt 2009 Best Paper Award, the 2010 Newcom++ Best Paper Award, the WUN CogCom Best Paper 2012 and 2013 Award, the 2014 WCNC Best Paper Award, the 2015 ICC Best Paper Award, the 2015 IEEE Communications Society Leonard G. Abraham Prize, the 2015 IEEE Communications Society Fred W. Ellersick Prize, the 2016 IEEE Communications Society Best Tutorial Paper Award, the 2016 European Wireless Best Paper Award, the 2017 Eurasip Best Paper Award, the 2018 IEEE Marconi Prize Paper Award, the 2019 IEEE Communications Society Young Author Best Paper Award and the Valuetools 2007, Valuetools 2008, CrownCom 2009, Valuetools 2012, SAM 2014, and 2017 IEEE Sweden VT-COM-IT Joint Chapter best student paper awards. He is an Associate Editor-in-Chief of the journal Random Matrix: Theory and Applications. He was an Associate Area Editor and Senior Area Editor of the IEEE TRANSACTIONS ON SIGNAL PROCESSING from 2011 to 2013 and from 2013 to 2014, respectively.
\end{IEEEbiography}

\end{document}